\documentclass[11pt]{article}
\usepackage[subpreambles=true]{standalone} 

\usepackage[margin=1in]{geometry}

\usepackage{amsmath,amsthm,amssymb,amsfonts,graphicx,dsfont}
\usepackage[maxfloats=100]{morefloats}
\usepackage[utf8]{inputenc}
\usepackage{booktabs}
\usepackage{epsfig}
\usepackage{url}
\usepackage{bm}
\usepackage{rotating}
\usepackage{listings}
\usepackage{verbatim}
\usepackage{xcolor}
\usepackage[title]{appendix}
\usepackage{multirow} 
\usepackage{longtable}
\usepackage{adjustbox}
\usepackage[longnamesfirst]{natbib}
\usepackage{bbm}
\usepackage{float} 
\usepackage{subfigure}
\usepackage[colorlinks,pagebackref=false]{hyperref}
\hypersetup{colorlinks,%
citecolor=black,%
filecolor=black,%
linkcolor=black,%
urlcolor=black
}
\usepackage{todonotes}

\usepackage{setspace} 

\newcommand{\eval}[2][\right]{\relax
	\ifx#1\right\relax \left.\fi#2#1\rvert}

\newcommand{\R}{{\mathbb R}}

\newcommand{\Ical}{{\mathcal I}}

\newcommand{\Ncal}{{\mathcal N}}

\newcommand{\Tcal}{{\mathcal T}}

\newcommand{\Xcal}{{\mathcal X}}

\newtheoremstyle{break}
  {\topsep}{\topsep}%
  {\itshape}{}%
  {\bfseries}{}%
  {\newline}{}%
\theoremstyle{break}

\newtheorem{proposition}{Proposition}
\newtheorem{exampleemph}[proposition]{Example}   

\usepackage{changepage}   
\makeatletter
\newcommand*{\rom}[1]{\expandafter\@slowromancap\romannumeral #1@}
\makeatother

\begin{document}


\title{Empirical Asset Pricing via Ensemble Gaussian Process Regression}

\date{}  

 \author{Damir Filipovi\'c\thanks{\scriptsize \'Ecole Polytechnique F\'ed\'erale de Lausanne and Swiss Finance Institute, Email: damir.filipovic@epfl.ch}
 \and
 Puneet Pasricha\thanks{\scriptsize Indian Institute of Technology Ropar, Email: puneet.pasricha@iitrpr.ac.in}
 }
 
\doublespacing

\begin{titlepage}
\maketitle
\thispagestyle{empty}

\begin{abstract}
We introduce an ensemble learning method based on Gaussian Process Regression (GPR) for predicting conditional expected stock returns given stock-level and macro-economic information. Our ensemble learning approach significantly reduces the computational complexity inherent in GPR inference and lends itself to general online learning tasks. We conduct an empirical analysis on a large cross-section of US stocks from 1962 to 2016. We find that our method dominates existing machine learning models statistically and economically in terms of out-of-sample $R$-squared and Sharpe ratio of prediction-sorted portfolios. Exploiting the Bayesian nature of GPR, we introduce the mean-variance optimal portfolio with respect to the prediction uncertainty distribution of the expected stock returns. It appeals to an uncertainty averse investor and significantly dominates the equal- and value-weighted prediction-sorted portfolios, which outperform the S\&P 500. 
\vspace{1cm}

\end{abstract}
\end{titlepage}

\section{Introduction}
A central problem of empirical asset pricing is the prediction of conditional expected stock returns given the information set of market participants.\footnote{Conditional expected returns in excess of the risk-free rate are also referred to as risk premium.} Conditional expected returns are notoriously hard to predict, for a number of reasons. First, financial markets are very noisy and exhibit a low signal-to-noise ratio when compared to other domains such as computer vision, which is arguably due to the efficiency of the markets. Second, the full information set is not observable and likely too complex to model. The literature has accumulated a non-exhaustive list of predictive stock-level characteristics and macro-economic variables, which continuous to increase. Third, the relation between predictors and returns is evidently non-linear and time-varying due to the dynamically evolving economic conditions, which further complicates the problem.

Over the years, a wide range of methods have been proposed to improve the prediction performance, from traditional statistical methods (\cite{welch2008comprehensive}, \cite{koijen2018carry} etc.) to the modern machine learning methods (\cite{gu_kel_xiu_20}, \cite{chen2019deep}, \cite{gu2021autoencoder} etc.). Interest in machine learning methods for empirical asset pricing has grown tremendously in academic finance and the industry. The leading example in this vein is \cite{gu_kel_xiu_20}, who outline a new research agenda marrying machine learning with empirical asset pricing. They compare several machine learning methods for predicting stock returns, namely linear regression, generalized linear models with penalization, random forests, and neural networks. Other recent studies, such as \cite{chen2019deep}, \cite{gu2021autoencoder}, further explore machine learning methods for empirical asset pricing, which are primarily based on neural networks. These articles provide empirical evidence that neural networks can significantly improve the predictive performance over traditional statistical methods and thus improve our empirical understanding of stock returns. However, focusing exclusively on prediction performance often overlooks an equally important aspect, the epistemic uncertainty inherent in these predictions. Quantifying uncertainty plays a significant role in finance when it comes to making decisions. Capturing uncertainties is crucial and has significant implications because the quality of our predictions directly impacts the quality of applications we build upon these predictions, for example, portfolio selection, hedging, or speculation. 

This paper adopts a broader perspective, stepping back from the traditional ``horse race" to improve prediction accuracy. Instead, it leverages kernel methods in machine learning, particularly Gaussian process regression (GPR), for predicting conditional expected stock returns given observable stock-level characteristics and macroeconomic variables. As a Bayesian non-parametric method, the predictions from a GPR model naturally come along with a predictive distribution. In particular, it gives confidence intervals for the predicted conditional expected returns. As a novel application, we harness these uncertainty estimates by incorporating in portfolio construction, showcasing the methodological advantages of GPR over conventional machine learning models, which often give point estimates of return without any quantification of uncertainty.{\footnote{The standard error estimates of some of the machine learning algorithms are available in the literature. For instance, \cite{giordano2002standard} investigate using the AR-Sieve bootstrap method to estimate the standard error of the sampling distribution of the neural network predictive values in a regression model with dependent errors. \cite{farrell2021deep} use a semi-parametric framework to provide non-asymptotic high-probability bounds for neural network predictions. The estimates of standard error predictions from random forests and LASSO are obtained by \cite{wager2014confidence}, and \cite{casella2010penalized} respectively. Although, these articles attempt to estimate prediction uncertainties, they are either computationally expensive or lack the theoretical rigor of GPR.}}


Our paper makes several methodological and empirical contributions. First, while \cite{gu_kel_xiu_20} study generalized linear models with nonlinear transformations of the original features, they restrict these transformations to second order splines. As such they miss the full range of kernel ridge regression, which comes with a strong mathematical framework and constitutes a powerful machine learning method that works well on medium data sizes. Gaussian processes provide an alternative view on kernel ridge regression, by modeling a distribution over functions and performing inference directly in function space.{\footnote{Gaussian processes are powerful mathematical objects that have enjoyed success in many practical applications. Gaussian processes have very close connections to other regression techniques, such as kernel ridge regression, support vector machines and linear regression with radial basis functions. Gaussian process provide a mathematical framework for many well-known models, including Bayesian linear models, spline models, and large neural networks (under suitable conditions), see \cite{williams2006gaussian}.}} We bridge the gap in \cite{gu_kel_xiu_20} and establish a link between two significant and growing research areas, kernel methods in machine learning and empirical asset pricing in financial economics. Thus, we contribute to the emerging literature on machine learning for empirical asset pricing.

Second, as mentioned earlier, being a Bayesian method, the predictions from a GPR model come along with a predictive distribution. As a novel application, we harness these uncertainty estimates. We find that incorporating uncertainty in portfolio construction leads to substantial statistical and economic improvements in terms of out-of-sample $R$-squared and Sharpe ratio of prediction-sorted portfolios, respectively. More precisely, we first use the predictive covariance matrix to construct a minimum uncertainty-weighted (UW) decile portfolio in the spirit of a global minimum variance portfolio. We find that the UW portfolio delivers a significantly high out-of-sample predictive pooled $R$-squared, $R_{pool}^2$, that outperforms the two traditional portfolios, namely, equal-weighted (EW) and value-weighted (VW). Motivated by this finding, we further exploit the predictive covariance matrix and introduce two new portfolios. The first is a prediction-weighted portfolio, originally proposed by \cite{kaniel2022machine}, which takes advantage of the ranking (to form decile portfolios) and the relative strengths of the predictions. The second is a prediction-uncertainty-weighted (PUW) portfolio in the spirit of the mean-variance optimal portfolio that gives more weight to stocks with higher predicted returns and minimizes uncertainty at the same time, which appeals to an uncertainty averse investor. We find that PW and PUW portfolios generate large economic gains, in terms of Sharpe ratio, compared to EW and VW portfolios.

Third, a well-known issue with implementing GPR is the need to invert the kernel matrix repeatedly for the computation of the marginal log-likelihood function, which has a fundamental time complexity of the order $O(N^3)$, where $N$ is the size of the training sample. This limitation prohibits, both in time and memory space, using GPR for large datasets. To tackle this computational bottleneck of GPR, we introduce an easy-to-implement ensemble learning method in the spirit of the mixture-of-experts approach. Specifically, we partition the large training sample into subsets, apply individual GPRs on all subsets in parallel, and obtain a predictive distribution conditional on the full training data by mixing the predictive distributions over the subsets. This also addresses one of the concerns associated with neural networks in financial applications. Neural networks thrive in data-rich environments. With so many parameters to learn, they require massive training data and are computationally costly to train. Much of the literature on neural networks in empirical asset pricing focuses on monthly returns of a large cross-section of stocks spanning several decades. It is questionable whether neural networks would perform reasonably when restricted to smaller training sets, such as single industry sectors, or the S\&P 500. Such samples only have a few thousand data points, which seems unsuitable for neural networks. What's more, neural networks hardly adapt to online learning, so one has to retrain the whole network when new data arrives.{\footnote{\cite{gu_kel_xiu_20} re-fit the neural network once every year.}} In other words, neural networks are computationally intensive and hence not scalable in scenarios where data comes in sequentially, as with financial data. In contrast, our ensemble learning approach offers several benefits: (i) it allows a straightforward parallel implementation of GPR on small training subsets, thus reducing the computational cost; (ii) it scales well with sample size and naturally lends itself to an online learning framework; and (iii) its data-driven mixing weight scheme takes into account the non-stationarity and heteroscedasticity present in the financial data.

Fourth, our empirical analysis confirms the insights from the growing literature that machine learning methods have excellent potential to predict conditional expected returns. In particular, we show that a simple GPR model with very few hyperparameters outperforms the benchmark models, both statistically and economically, in terms of the out-of-sample $R$-squared and Sharpe ratio of the prediction-sorted portfolios, respectively. Concretely, we conduct an extensive empirical analysis, investigating monthly returns of a large cross-section of US stocks from 1962 to 2016. Our features include 94 time-varying stock-specific characteristics. We compare the performance of our GPR model with a non-linear kernel against linear benchmark models. Specifically, ensemble GPR with an affine kernel, ensemble linear regression, and standard linear regression. We find that our model outperforms, out-of-sample, these benchmarks in predicting individual stock returns. In particular, our model generates an $R^2_{pool}$ of 0.78\% compared to 0.63\% for the ensemble GPR model with affine kernel, 0.61\% for the ensemble linear regression model and 0.37\% for the one-batch linear regression. We also evaluate the predictive performance of our model based on two alternative metrics. The first is the time average of the monthly $R$-squared, $R^2_{avg}$, which gives equal weight to every month, in contrast to $R^2_{pool}$, which places more weight on months with larger cross sections. The second is the information coefficient (IC), which quantifies the model's ability to differentiate the relative performance among stocks disregarding the absolute levels of predictions. We find that $R^2_{avg}$ and IC are 0.39\% and 5.89\% for our model.

We also assess our model's predictive performance at the portfolio level. In particular, we form decile portfolios (bottom $D1$ to top $D10$) sorted on out-of-sample stock return predictions from our model. Our UW portfolios achieve a higher $R$-squared than EW and VW portfolios for each decile. Further, when assessed over the grand panel of all decile portfolios, UW generates a $R_{pool}^2$ of 13.39\% compared to 8.04\% and 3.85\% for EW and VW. The more pronounced predictive power of the UW portfolio shows that reducing prediction uncertainty is economically significant. We also find that the economic gains from the decile portfolios constructed using our predictions are large in terms of Sharpe ratio. For example, the long-short portfolio from PUW with uncertainty aversion value 20 yields an annualized out-of-sample Sharpe ratio of 3.44, which compared to 2.98 from PW, however, at a significantly lower volatility. Moreover, it outperforms the Sharpe ratios of 2.44 (EW) and 0.91 (VW), which confirms that reducing prediction uncertainty matters economically. Further, our prediction uncertainty based portfolios outperform the corresponding portfolios from the linear benchmark models.

We find that the most important features are related to recent price trends, which include variables such as
short-term reversal (mom1m and mom6m), stock momentum (mom12m), momentum change (chmom), long-term reversal (mom36m), and liquidty variables, including bid-ask spread (baspread), dollar volume (dolvol), turnover volatility (SD turn), and Amihud illiquidity (ill). In general, we find that our model is inclusive and extracts predictive information from a wide range of features. We also investigate the cross-sectional heterogeneity in predicted returns and prediction uncertainty. We find that stocks with high predicted returns tend to be less liquid, and stocks with high prediction uncertainty are the ones with limits to arbitrage frictions and that exhibit extreme illiquidity. Stocks with higher predicted returns have a higher 6-month (mom6m) and 12-month (mom12m) momentum and a lower 1-month (mom1m) momentum, which suggests momentum over longer horizons and reversal over short horizons.

In sum, our paper confirms the great potential of machine learning for predicting conditional expected returns. We contribute to this understanding by adding prediction uncertainty, which greatly improves the performance of prediction-sorted portfolios.


Our paper contributes to the fast emerging literature on machine learning for empirical asset pricing.{\footnote{The literature on traditional empirical asset pricing can essentially be divided into two broad categories: time-series and cross-sectional prediction models. For the former see, e.g., \cite{welch2008comprehensive}, \cite{koijen2018carry} and references therein. Cross-sectional models aim at explaining differences in the cross-section of stock returns and do so by regressing returns on stock-level characteristics, e.g., past returns, turnover etc., and macro-economic variables. See, e.g., \cite{fama2008dissecting} and references therein. The main limitation of these traditional regression models is their incapability of incorporating a large number of features and non-linear dependencies.}} In their pioneering work, \cite{gu_kel_xiu_20} conduct a comparative study of several machine learning methods to predict the grand panel of individual stock returns from the US markets and demonstrated the advantages of machine learning methods over traditional approaches. Similar studies are performed for European stock markets, \cite{drobetz2021empirical}, and bond markets, \cite{bianchi2021bond}. \cite{gu2021autoencoder} use an autoencoder neural network to demonstrate that imposing economic structure on a machine learning algorithm can substantially improve the estimation. In the same spirit, \cite{chen2019deep} use deep neural networks with the fundamental no-arbitrage condition as a criterion function to estimate an asset pricing model for individual stock returns. These articles, however, exclude the important class of kernel-based models. We bridge this gap and show that our simple ensemble method based on GPR dominates the performance of their best benchmark models, which are based on neural networks. Our model leads to a better out-of-sample predictive $R$-squared, and further taking into account the estimates of the prediction accuracy leads to better portfolio performance in terms of Sharpe ratio.

The adoption of GPR in finance is only recent, albeit GPR has demonstrated much success outside of finance under the name of kriging. \citet{williams2006gaussian} provide an extensive background on GPR models and highlight its applications in various fields. For instance, they emphasize that Gaussian processes can be viewed as a Bayesian non-parametric generalization of well-known econometrics techniques. In particular, a time series model, $AR(p)$, is a discrete-time equivalent of a Gaussian process model with Mat\'{e}rn covariance functions with an appropriate hyperparameter choice. \cite{han2016gaussian} combine Gaussian process state space models with stochastic volatility models and propose a GPR stochastic volatility (GPRSV) model to predict the volatility of the stocks returns. They also present an adjusted Markov Chain Monte Carlo to estimate their model and demonstrate, through empirical analysis, the superior predictive performance over the traditional GARCH and stochastic volatility models. \cite{de2018machine} show how GPR can be deployed to approximate the derivative pricing function, for instance, pricing of exotic options under advanced models. They also study the application of GPR in the fitting of sophisticated Greek curves and implied volatility surfaces. Their numerical findings suggest that GPR could deliver a speed-up of a factor of several magnitudes relative to other benchmark methods for respective problems. \cite{cousin2016kriging} introduce shape-constrained Gaussian processes for yield-curve and credit default swaps (CDS) curve interpolation. \cite{filipovic2022stripping} and \cite{filipovic2022shrinking} introduce an economically motivated kernel ridge regression method for estimating the yield and return curves of treasury bonds. Our paper seems to be the first that applies GPR to predicting conditional expected stock returns.

Our paper also contributes to the literature on scalable GPR. Our ensemble learning method complements existing approaches, such as inducing point methods \cite{quinonero2005unifying} and subset of regressors (SoR) \cite{silverman1985some}, which approximate the kernel matrix using a reduced set of inducing points. Unlike inducing point methods, our ensemble learning approach does not rely on selecting a subset of points. Instead, it leverages the diversity of multiple Gaussian processes applied to different subsets of the data. Alternative approaches include Kronecker and Toeplitz methods, which exploit data structure to enable computationally efficient modeling for gridded or separable datasets (\cite{wilson2014covariance}). These methods were further advanced in \cite{wilson2015kernel}, where the authors introduce Kernel Interpolation for Scalable Structured Gaussian Processes (KISS-GP), combining the inducing point framework with structure-exploiting techniques. This approach is extended in \cite{wilson2016deep} by integrating it with neural networks. More recently, \cite{gardner2018gpytorch} utilized GPU acceleration to perform efficient matrix computations for large-scale GPR. This GPU-based acceleration can also enhance our ensemble method, where each individual Gaussian process model is implemented to take advantage of GPU computing.

The paper is organized as follows. Section~\ref{section_methodology} introduces the model framework and our ensemble learning approach. Section~\ref{empirical_study} contains our empirical analysis.  Section~\ref{section_conclusion} concludes. The appendix contains background material on Gaussian processes, on the kernel selection, and additional results.

\section{Methodology}\label{section_methodology}

Consider a financial market consisting of assets $i$ in discrete time $t=0,1,2,\ldots$, where $t$ represents the end of a month. More specifically, we denote by $\Ical_{t+1}$ the index set of assets $i$ that exist during the period $[t,t+1]$. At any time $t$, for any asset $i\in \Ical_{t+1}$, we observe the vector of predictor variables $x=x_{i,t}$ with values in a feature space $\mathcal{X}$ consisting of asset $i$-specific characteristics and common macro-economic variables. 

We denote by $r_{i,t+1}$ the excess return (henceforth simply referred to as ``return'', if there is no risk of confusion) of asset $i$ over the period $[t,t+1]$. Following \cite{gu_kel_xiu_20}, we describe it as an additive prediction error model,
\begin{equation}\label{predictive_model1}
r_{i,t+1}=E_{t}(r_{i,t+1})+\epsilon_{i,t+1}, 
\end{equation}
where $E_{t}$ denotes the conditional expectation given the information available at time $t$. We model the conditional expected return by a common function $f:\Xcal\to\R$,  
\begin{equation}\label{predictive_model2}
E_{t}(r_{i,t+1})=f(x_{i,t}), \quad \text{for all $i\in \Ical_{t+1}$ and $t$.} 
\end{equation}
The unexpected return  component, $\epsilon_{i,t+1}$, satisfies $E_{t}[\epsilon_{i,t+1}]=0$ and represents the aleatoric uncertainty arising due to idiosyncratic risk.  


At any given time $t$, our goal is to learn the function $f$ from past data $\mathcal{D}=\{(x_{i,j-1},r_{i,j}),~i\in \Ical_j,~j=1,2,\ldots,t\}$, and then predict next period conditional expected returns. We assume that the function $f$ does neither directly depend on the index $i$ nor time $t$, so that it can be learned efficiently using all instances in the panel data set $\mathcal{D}$.



\subsection{Ensemble learning method based on Gaussian process regression}

We use Gaussian process regression (GPR) to learn the process $f$ in \eqref{predictive_model2}. Thereto, in a first round, we assume that $f$ is a Gaussian process with some pre-specified prior mean function $m(\cdot)$ and covariance function, or, kernel $k(\cdot,\cdot)$. We assume that the unexpected error components $\epsilon_{i,t+1}$ are i.i.d.\ Gaussian random variables with mean zero and variance $\sigma_\epsilon^2$ and independent of $f$. This means that the joint distribution of the past returns ${\bm r}_{1:t}=\{r_{i,j},~i\in \Ical_j,~j=1,2,\ldots,t\}$ and the conditional expected returns $f({\bm x}_{t})=\{f(x_{i,t}),\ i\in \Ical_{t+1}\}$ is Gaussian of the form
\begin{equation}\label{GPdistr}
  \begin{pmatrix}
  {\bm r}_{1:t} \\  f({\bm x}_{t})
\end{pmatrix}  \sim\Ncal \left(\begin{pmatrix}
  m({\bm x}_{0:t-1}) \\ m({\bm x}_t)
\end{pmatrix} ,\begin{pmatrix}
  k({\bm x}_{0:t-1},{\bm x}_{0:t-1}) + \sigma_\epsilon^2 {I} & k({\bm x}_{0:t-1},{\bm x}_t) \\
  k({\bm x}_t,{\bm x}_{0:t-1}) & k({\bm x}_t,{\bm x}_t)
\end{pmatrix}\right),
\end{equation}
where ${I}$ is the identity matrix, and ${\bm x}_{0:t-1}=\{x_{i,j-1},~i\in \Ical_j,~j=1,2,\ldots,t\}$ and ${\bm x}_{t}=\{x_{i,t},\ i\in \Ical_{t+1}\}$ denote the arrays of features observed by time $t-1$ and at $t$, respectively. Here, for a function $g:\Xcal\to\R$ and an array $\bm x=\{x_i\}$ of points in $\Xcal$, we denote by $g(\bm x)=\{g(x_i)\}$ the corresponding array of function values. In particular, $k({\bm x}_{0:t-1},{\bm x}_{0:t-1})$ is the $N\times N$ matrix of the covariances evaluated at all pairs of the past features ${\bm x}_{0:t-1}$, where $N=\sum_{j=1}^t|\Ical_j|$ denotes the size of the sample. Similarly, $k({\bm x}_{0:t-1},{\bm x}_t)$ is the $N \times |\Ical_{t+1}|$ matrix of the covariances evaluated at all pairs of the past features ${\bm x}_{0:t-1}$ and current features ${\bm x}_{t}$.  

The predictive (posterior) distribution of $f$ given the data $\mathcal{D}$ is again Gaussian with mean and covariance functions given by
\begin{align} 
\label{mean_pred}
\hat m(x)&= m(x) + k(x,{\bm x}_{0:t-1}) \left(k({\bm x}_{0:t-1},{\bm x}_{0:t-1})+ \sigma_\epsilon^2 {I}\right)^{-1} ({\bm r}_{1:t}- m({\bm x}_{0:t-1})),\\
\label{cov_pred}\hat{k}(x,x')&=k(x,x')-k(x,{\bm x}_{0:t-1})\left(k({\bm x}_{0:t-1},{\bm x}_{0:t-1})+ \sigma_\epsilon^2 {I}\right)^{-1} k({\bm x}_{0:t-1},x').    
\end{align}
The mean vector $\hat m({{\bm x}_{t}})=E[f(\bm x_t)\mid\mathcal{D}]$ is then our prediction of the conditional expected returns $E_t[{\bm{r}}_{t+1}]$. The covariance matrix $\hat{k}({\bm x}_{t},{\bm x}_{t})=\operatorname{Var}\!\big(f(\bm x_{t})\mid \mathcal{D}\big)$ quantifies the epistemic uncertainty of our prediction. In contrast, the unexpected return components, $\bm\varepsilon_{t+1}$, represent the aleatoric uncertainty of the returns. The 
realized residual (or realized prediction error), from Equation (\ref{predictive_model1}), therefore can be decomposed as 
\begin{equation}\label{residual_realized}
r_{i,t+1}-\hat{m}(x_{i,t})
=
\underbrace{\big(f(x_{i,t})-\hat{m}(x_{i,t})\big)}_{\text{epistemic uncertainty}}
+
\underbrace{\epsilon_{i,t+1}}_{\text{aleatoric uncertainty}},    
\end{equation}
implying the total variance
\begin{equation}\label{residual_realized2}
\operatorname{Var}(\bm r_{t+1}\mid \bm x_{t},\mathcal{D})
=
\hat{k}(\bm x_{t},\bm x_{t})+ \sigma_{\epsilon}^2 I. 
\end{equation}

In our empirical analysis, we set the prior mean function $m(\cdot)$ equal to zero. This is motivated by the empirical fact that zero predictions perform better than the historical mean of excess returns, see \cite{gu_kel_xiu_20}. Moreover, it is well documented in the literature that a zero prior mean usually works well for GPR, see \cite{de2018machine,williams2006gaussian}. 


The kernel $k(\cdot,\cdot)$ of the prior distribution of $f$ in \eqref{GPdistr} depends on hyperparameters, which are estimated along with $\sigma_{\epsilon}^2$ from the training data by maximizing the marginal log-likelihood, given by
\begin{equation}\label{marginal}
\begin{aligned}
\log p({\bm r}_{1:t}\mid {\bm x}_{0:t-1})&=-\frac{N\log(2\pi)}{2}-\frac{{\bm r}_{1:t}^\top \big(k({\bm x}_{0:t-1},{\bm x}_{0:t-1})+\sigma_\epsilon^2I\big)^{-1}{\bm r}_{1:t}}{2}\\
&\quad -\frac{\log \det(k({\bm x}_{0:t-1},{\bm x}_{0:t-1})+\sigma_\epsilon^2 I)}{2}.
\end{aligned}
  \end{equation}
A known challenge in GPR is that the computation of the log-likelihood function \eqref{marginal} involves repeated inversion of the regularized kernel matrix, which takes time of the order $O(N^3)$. This is only feasible for a small $N$ (less than several thousand), which is not the case for the problem at hand for which $N$ is of the order of millions. 

We tackle the computational bottleneck of GPR and introduce an ensemble learning approach in the spirit of the mixture-of-experts method.\footnote{Alternative ensemble learning approaches proposed in the literature include the product of GP experts in \cite{ng2014hierarchical}, the generalised product of experts in \cite{cao2014generalized}, the Bayesian Committee Machine in \cite{tresp2000bayesian}, 
the robust Bayesian Committee Machine in \cite{deisenroth2015distributed}, and Distributed Kriging (DISK) in \cite{guhaniyogi2017divide}. The product of experts approach obtains the joint prediction by the product of all predictions from trained GPR models, while the generalized product of experts approach adds the flexibility by assigning weights to the contributions from independent GPR models thus increasing/reducing their importance. These approaches are further generalized in the Bayesian Committee Machine and the robust Bayesian Committee Machine, where the GP priors are explicitly incorporated when combining predictions. In contrast to these product of experts approaches, Distributed Kriging obtains the combined predictions as the Wasserstein barycenter of the subset posterior distributions.} Thereto, we partition the training data into subsets, apply a GPR on each subset in parallel, and obtain a predictive distribution conditional on the full training data by mixing the predictive distributions over the subsets. In contrast to ad hoc partitioning schemes used in the literature, such as random or clustering based partitioning, we use that our training data is naturally divided into monthly subsets. That is, we treat data from each month $j=1,2,\ldots,t$ as a training subset on which we train an individual Gaussian process $f^{(j)}$. Specifically, we estimate hyperparameters by maximizing the log-likelihood function \eqref{marginal}, and obtain the predictive Gaussian distribution of $f^{(j)}$ with mean and covariance functions $\hat m^{(j)}(\cdot)$ and $\hat{k}^{(j)}(\cdot,\cdot)$ as in \eqref{mean_pred} and \eqref{cov_pred}, with the training data $\{{\bm x}_{0:t-1},{\bm r}_{1:t}\}$ replaced by $\{{\bm x}_{j-1}, {\bm r}_{j}\}$.

Finally, we obtain the predictive distribution of $f$ given the full training data by mixing the individual predictive Gaussian distributions of $f^{(j)}({\bm x}_{t})$ using some weights $w_j\ge 0$ with $\sum_{j}w_j=1$. The mean vector and covariance matrix of this Gaussian mixture distribution are given by  
\begin{align}
  \hat{\bm{r}}_{t+1}&=\sum_{j} w_j\hat m^{(j)}({\bm x}_{t}), \label{ensmean}\\
   \hat{{\bm\Sigma}}_{t+1}&=\sum_{j} w_j\hat{{\bm M}}^{(j)}_{t+1}-\hat{{\bm r}}_{t+1}\hat{{\bm r}}_{t+1}^\top,\label{enscov}
\end{align}
where $\hat{{\bm M}}^{(j)}_{t+1}=\hat{k}^{(j)}({\bm x}_{t},{\bm x}_{t})+\hat m^{(j)}({\bm x}_{t}) \hat m^{(j)}({\bm x}_{t})^\top$ denotes the second order moment matrix of $f^{(j)}({\bm x}_{t})$. The mean $\hat{\bm{r}}_{t+1}$ is our ensemble prediction of the conditional expected return vector ${\bm{r}}_{t+1}$, and $\hat{{\bm\Sigma}}_{t+1}$ quantifies the epistemic uncertainty of our predictions. For our empirical analysis, we implement the following two mixing weight schemes:
\begin{enumerate}
\item\label{WS1} Equal weights: we select the $K_t\le t$ most recent training months (more details on the choice of $K_t$ follow in the next subsection), set $w_j=0$ for $j\le t-K_t$, and apply equal weights $w_j=1/K_t$ to the GPR models $j=t-K_t+1,\dots,\,t-1,\,t$. The sums in \eqref{ensmean} and \eqref{enscov} are effectively over $j=t-K_t+1,\dots,\,t-1,\,t$.
\item\label{WS2} MSE weights: as above we select the $K_t\le t$ most recent training months and set $w_j=0$ for $j\le t-K_t$. We also hold out the most recent training month, $w_t=0$, which we call \emph{calibration} month, and define the weights for the remaining $K_t-1$ months as proportional to the mean squared error
\begin{equation*}
MSE_j=\frac{1}{|\Ical_t|}\sum_{i\in \Ical_t}(r_{i,t}-\hat r^{(j)}_{i,t} )^2, 
\end{equation*}
of GPR model $j$ on the calibration month $t$ with corresponding predicted returns $\hat r^{(j)}_{i,t}=\hat m^{(j)}(x_{i,t-1})$. The MSE weights are thus given by
\begin{equation*}
w_j=\frac{1/MSE_j}{\sum_{s=t-K_t+1}^{t-1}1/MSE_s},\quad j=t-K_t+1, \dots,t-2,\,t-1.
\end{equation*}
That is, the smaller $MSE_j$ the larger the weight $w_j$ we give to GPR model $j$. The sums in \eqref{ensmean} and \eqref{enscov} are effectively over $j=t-K_t+1,\dots,\,t-2,\,t-1$.
\end{enumerate}

Our ensemble learning approach has several advantages. First, it offers a substantial computational speed-up, due to a straightforward parallel implementation of the individual GPR models on subsets of the training data, compared to training a GPR model on the full training data. Second, the flexibility that each GPR model can have its own set of optimal hyperparameters (the parametric kernel is the same for each model) allows to account for the non-stationarity and heteroscedasticity present in the financial data. Third, our approach scales well with sample size and provides an online learning framework. This is in contrast to other sophisticated machine learning algorithms, which are hard to train recursively every month due to their high computational costs. Specifically, we only need to train one additional GPR model on the newly observed data from month $t+1$ and combine it with the already trained GPR models on months $1,2,\ldots,t$ to obtain a mixed predictive distribution for the conditional expected returns ${\bm r}_{t+2}$.

\subsection{Sample splitting: training, validation and test samples}\label{train_valid_test}

The process of estimating hyperparameters, predicting, and evaluating the predictions requires the modeller to partition the full sample into training, validation and test samples. To achieve this goal, we conduct an empirical analysis of rolling and expanding schemes on the training and validation samples, and then use the scheme that performs best in terms of prediction accuracy on the validation sample, for our out-of-sample test data analysis.{\footnote{Another scheme, known as the fixed scheme, divides the sample into fixed training, validation, and test data, estimates the model once from the training and validation samples and makes predictions on the test sample. Although the fixed scheme is not very expensive in terms of the computation cost, it fails to capture the changes in the behaviour of data over time, thus affecting the model's performance.}}

The underlying idea of the \emph{rolling} scheme is to gradually shift the training and validation samples forward in time to include more recent data and exclude the oldest data points such that a fixed size of the rolling window is maintained. At each rolling step, one re-fits the model on the prevailing training and validation samples and obtains the predictions on the next test data, thus resulting in a sequence of performance measures, i.e., one corresponding to each window. Although this approach has the benefit that it can potentially leverage more recent data for predictions, it can significantly impact the performance of the model if the excluded data contains essential information, e.g., a financial crisis period. The \emph{expanding} scheme also gradually includes more recent data points in the training and validation samples. But in contrast to the rolling scheme it retains the entire history in the training sample. In terms of the mixing weight schemes \ref{WS1} and \ref{WS2}, we set $K_t=K$ for the rolling scheme, where $K$ is a fixed constant, and $K_t=t$ for the expanding scheme.


\subsection{Predictive performance evaluation}

We evaluate the model performance in predicting conditional expected returns, using three measures. The first is the predictive out-of-sample pooled $R$-squared, 
\begin{equation*}
R_{pool}^2=1-\frac{\sum_{t\in \Tcal_3}\sum_{i\in \Ical_t}(r_{i,t}-\hat{r}_{i,t})^2}{\sum_{t\in \Tcal_3}\sum_{i\in \Ical_t}r_{i,t}^2},
\end{equation*}
where $\mathcal{T}_3$ denotes the collection of test months. $R_{pool}^2$ provides a metric for the grand panel-level performance of the model by pooling the prediction errors across stocks and over time. Being a pooled performance measure, $R^2_{pool}$ places more weight on months with a comparatively larger cross-section of stocks. However, the size of the cross-section varies considerably across the sample, as shown in Figure \ref{sample_split_1}. A monthly rebalancing portfolio manager is more concerned with the average monthly predictive performance. 

Therefore, we also consider a second performance measure, the predictive out-of-sample average $R$-squared,
\begin{eqnarray*}
R_{avg}^2=\frac{1}{|\Tcal_3|}\sum_{t\in\Tcal_3}R_t^2,
\end{eqnarray*}
where $R_t^2$ denotes the $R$-squared for the predictions in month $t$, 
\begin{eqnarray*}
R_t^2=1-\frac{\sum_{i\in \Ical_t}(r_{i,t}-\hat{r}_{i,t})^2}{\sum_{i\in \Ical_t}r_{i,t}^2}.
\end{eqnarray*}
Both measures, $R_{pool}^2$ and $R^2_{avg}$, compare our model predictions against the naive forecast of zero excess log returns and not against the historical mean excess log returns. This is because the latter are known to predict excess log returns worse than zero by a large margin, see \cite{gu_kel_xiu_20}. 

Investors can use our model to construct portfolios based on the predicted relative performance of the stocks. For example, a long-short investor will go long in top-ranked stocks and short in bottom-ranked stocks to earn the difference between the relative returns of the two buckets of stocks. The performance metrics, $R^2_{pool}$ and $R^2_{avg}$, measure the extent to which the levels of predicted excess returns differ from realized excess returns and thus are not necessarily suitable for a long-short investor. Therefore, we consider a third performance measure, the information coefficient, defined as the average
\begin{eqnarray*}
IC&=&\frac{1}{|\Tcal_3|}\sum_{t\in\Tcal_3}\rho_t
\end{eqnarray*}
of the cross-sectional Spearman's rank correlation coefficients between the realized excess returns and predictions,
\begin{eqnarray*}
\rho_t&=&1-\frac{6\sum_{i\in \Ical_t}d_i^2}{|\mathcal{I}_t|(|\mathcal{I}_t|^2-1)},
\end{eqnarray*}
where $d_i$ is the difference in ranks between the $i$th largest elements of $\{\hat{r}_{j,t}\}_{j\in\mathcal{I}_t}$ and $\{r_{j,t}\}_{j\in\mathcal{I}_t}$. The IC, originally proposed by \cite{ambachtsheer1974profit}, is a widely used performance measure in investment management to measure predictive ability. It disregards the absolute levels, is less sensitive to outliers, and quantifies the model's ability to differentiate the relative performance among stocks.

\section{An empirical study of US equities}\label{empirical_study}

This section contains our empirical analysis. Section \ref{data} describes the data and the steps we follow to prepare it for the empirical analysis. Section \ref{model_selection} discusses the model selection where we conduct a comparison study on the validation sample to select the model parameters for out-of-sample analysis, namely the sample splitting scheme, the mixing weight scheme to create an ensemble and the kernel. Section \ref{model_performance} contains the performance results of our model using statistical and economic criteria. Section \ref{cross_insights} focuses on the cross-section insights of the model performance such as variable importance, the relationship between the features and the predicted returns, and the relationship between the features and the prediction uncertainty. Section \ref{residual_analysis_sec} discusses residual analysis.

\subsection{Data}\label{data}


We consider the monthly returns of approximately 30{,}000 individual stocks from the three major stock exchanges in the US, namely, NYSE, AMEX and NASDAQ. The data is collected from CRSP over a period spanning 55 years from February 1962 to December 2016. We use the monthly Treasury bill rate as a proxy for the risk-free rate to determine the excess simple return of a stock.

The literature on empirical asset pricing has constructed a large collection of features that help predict future stock returns. The conditioning information we use includes 94 stock-specific characteristics that is considered in \cite{gu_kel_xiu_20} and \cite{gu2021autoencoder}.\footnote{Data is available on the homepage of Dacheng Xiu (\url{https://dachxiu.chicagobooth.edu/}).} The stock-level characteristics pertain to several categories, including past returns, investment profitability, value, trading frictions etc. Thereof, 61 characteristics are updated annually, 13 are updated quarterly, and 20 are updated on a monthly basis. Since most characteristics are lagged in the sense that there is a delay in their release to the public, we follow the common conventions concerning the usage of these characteristics to avoid a forward-looking bias. More precisely, we assume that there is a lag of at most one month, four months and six months in the monthly, quarterly and annually reported characteristics respectively. Consequently, we predict the returns ${\bm r}_{t+1}$ over the period $[t,t+1]$ as a function of the most recent publicly available characteristics at $t$. That is, we use the most recent monthly, quarterly and annual characteristics at the end of the months $t-1$, $t-4$ and $t-6$, respectively.

To prepare the data for empirical analysis, we apply transformations to the stock-specific characteristics. This is a common practice in machine learning since different features have different absolute scales, and some of them are highly skewed and leptokurtic. To address the difference in scale and remove the influence of outliers, at any time $t$, we standardize the non-missing values across any specific characteristic by subtracting the cross-sectional mean and dividing by the cross-sectional standard-deviation. We then replace the missing observations by zero.

\subsection{Model selection}\label{model_selection}
\begin{figure}
        \centering
        \includegraphics[scale=0.26]{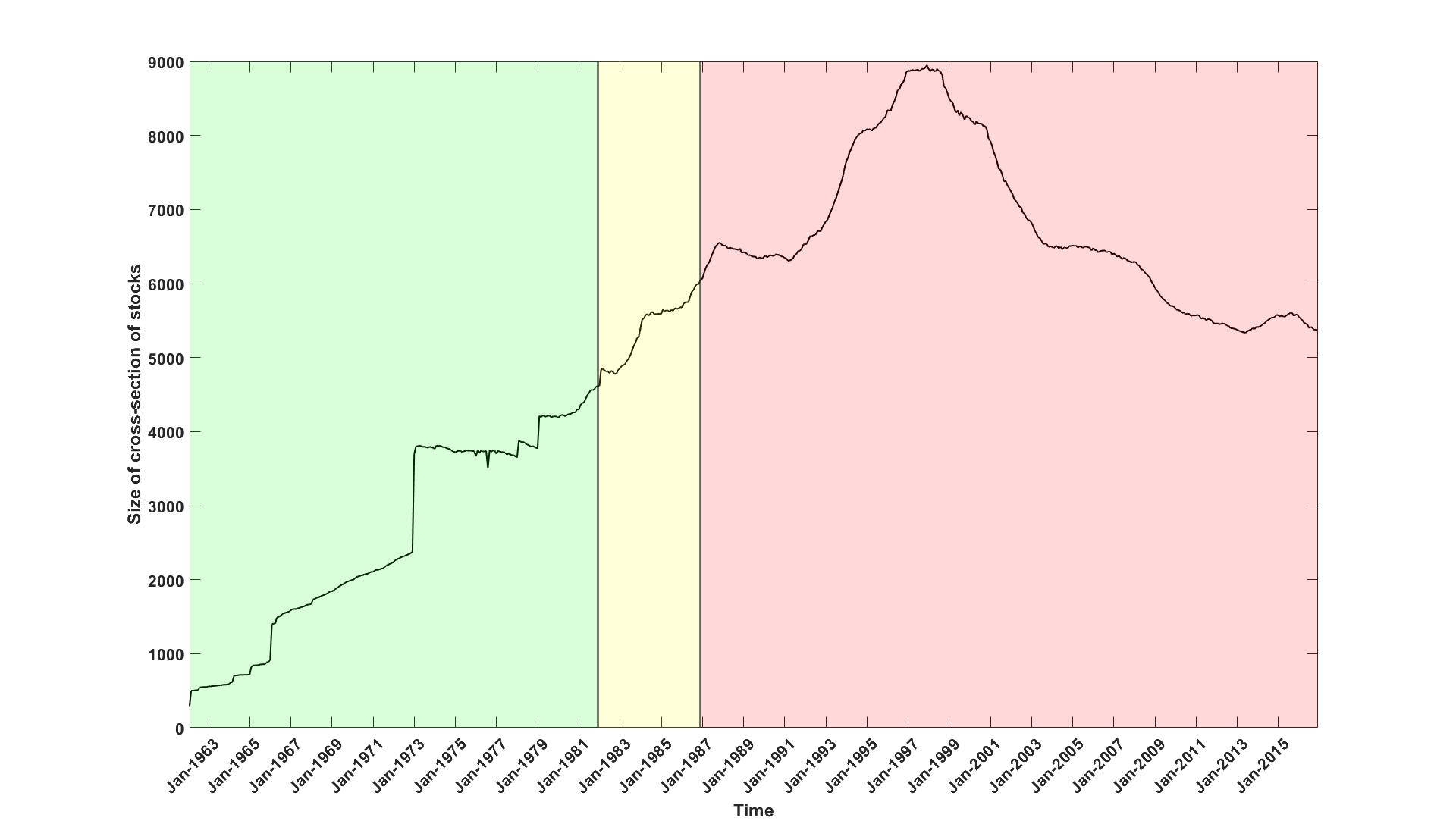}
\caption{This figure shows the size of the cross-section of stocks in each month of the sample. The full sample is split into training sample (green), Feb 1962 to Dec 1981, validation sample (yellow), Jan 1982 to Dec 1986, and test sample (red), Jan 1987 to Dec 2016.} \label{sample_split_1}
\end{figure}

We divide the data into three consecutive non-overlapping samples while maintaining the temporal ordering of the data, as shown in Figure \ref{sample_split_1}. The training sample, Feb 1962 to Dec 1981, is used to estimate the hyperparameters of the Gaussian process. The validation sample, Jan 1982 to Dec 1986, is used for model selection, that is, the kernel function, the mixing weight scheme, the sample splitting scheme (rolling or expanding), and the size of the rolling window if we adopt a rolling scheme. The test sample, Jan 1987 to Dec 2016, is then used to evaluate the performance of the selected model.

\begin{figure}
\begin{center}
\includegraphics[scale=0.62]{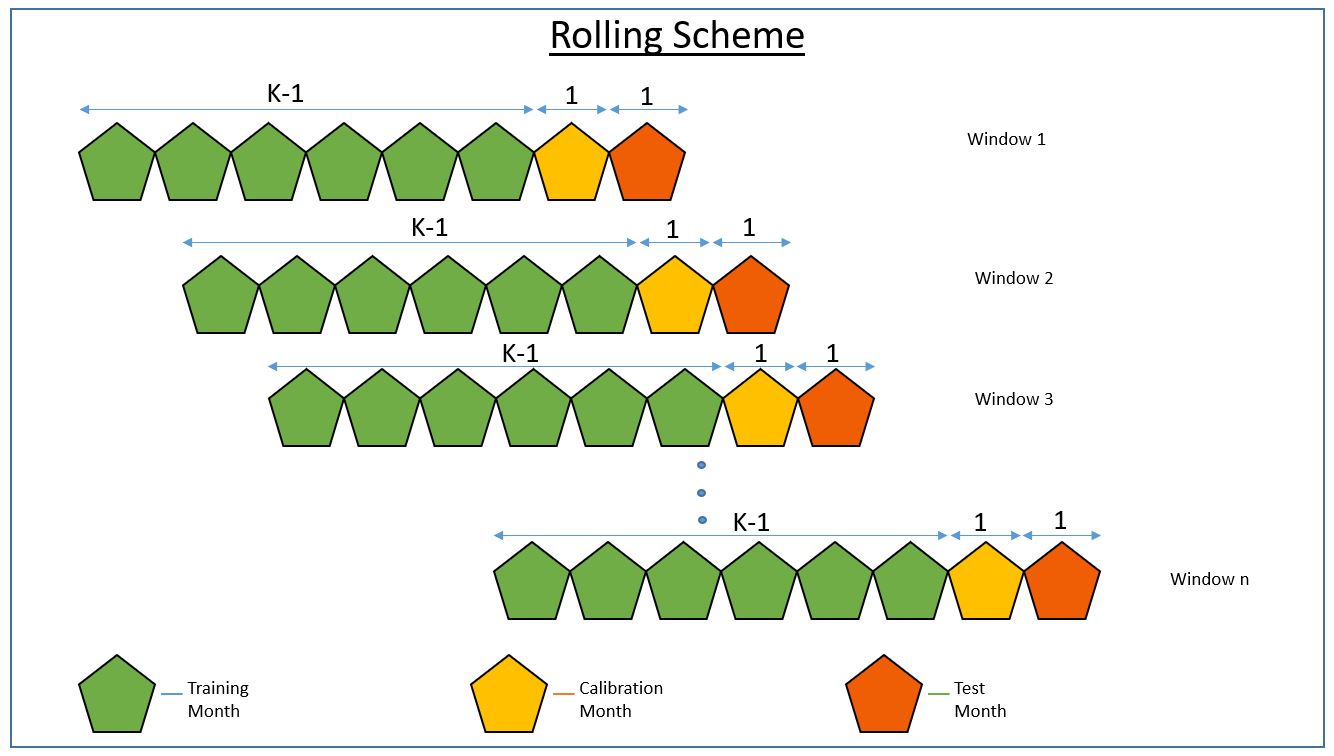}
\caption{This figure describes the mechanism of the rolling scheme with training window, including a calibration month, of length $K$.}
\label{rollingscheme}
\end{center}
\end{figure}

For model selection, we follow a data-driven approach and select the configuration, among all possible combinations, that performs best on the validation sample in terms of prediction accuracy, as measured by $R_{pool}^2$. More specifically, we implement our ensemble method for both, rolling and expanding, schemes. In more detail, we apply the rolling scheme for different possible training window lengths $K=2,\ldots,239$. That is, for the MSE weighting scheme \ref{WS2} and $K=2$, we use Nov 1981 and Dec 1981 as training and calibration months, respectively, for the first test month, Jan 1982.\footnote{For $K=2$, as there is only one training month, the role of the calibration month is effectively redundant.} Likewise, we choose Oct 1986 and Nov 1986, respectively, for the last test month, Dec 1986. Similarly, for the maximal possible $K=239$, we use Feb 1962 to Nov 1981 as training months and Dec 1981 as calibration month for the test month Jan 1982, and shift each by one month for the next test month. This is illustrated in Figure~\ref{rollingscheme}. We also apply the expanding scheme, using the full training sample available for each test month. More specifically, we use Feb 1962 to Nov 1981 as training months and Dec 1981 as a calibration month for the test month Jan 1982, while we use Feb 1962 to Oct 1986 as training months and Nov 1986 as calibration month for the last test month Dec 1986. We follow a similar procedure for the equal weighting scheme \ref{WS1} but without holding out a calibration month.

In preliminary experiments reported in Appendix~\ref{kernel_choice} with ten kernels in predicting excess log returns over the validation period, we observe that the $\gamma$-exponential kernel outperformed the others in terms of $R^2_{pool}$,
\begin{gather*}
    K_{10}(x,x')=\sigma^2\exp\left\{-\left(\frac{||x-x'||}{\ell}\right)^\gamma\right\},
\end{gather*}
where $\sigma,\ell>0$ and $\gamma\in(0,2]$ are hyperparameters. We therefore report the validation analysis and out-of-sample test on simple excess returns below only for this kernel. We note that $\sigma$ is not relevant for the predictive mean in \eqref{mean_pred} but it is significant in quantifying the epistemic uncertainty, see \eqref{cov_pred}.

\begin{figure}
\centering
\includegraphics[scale=0.38]{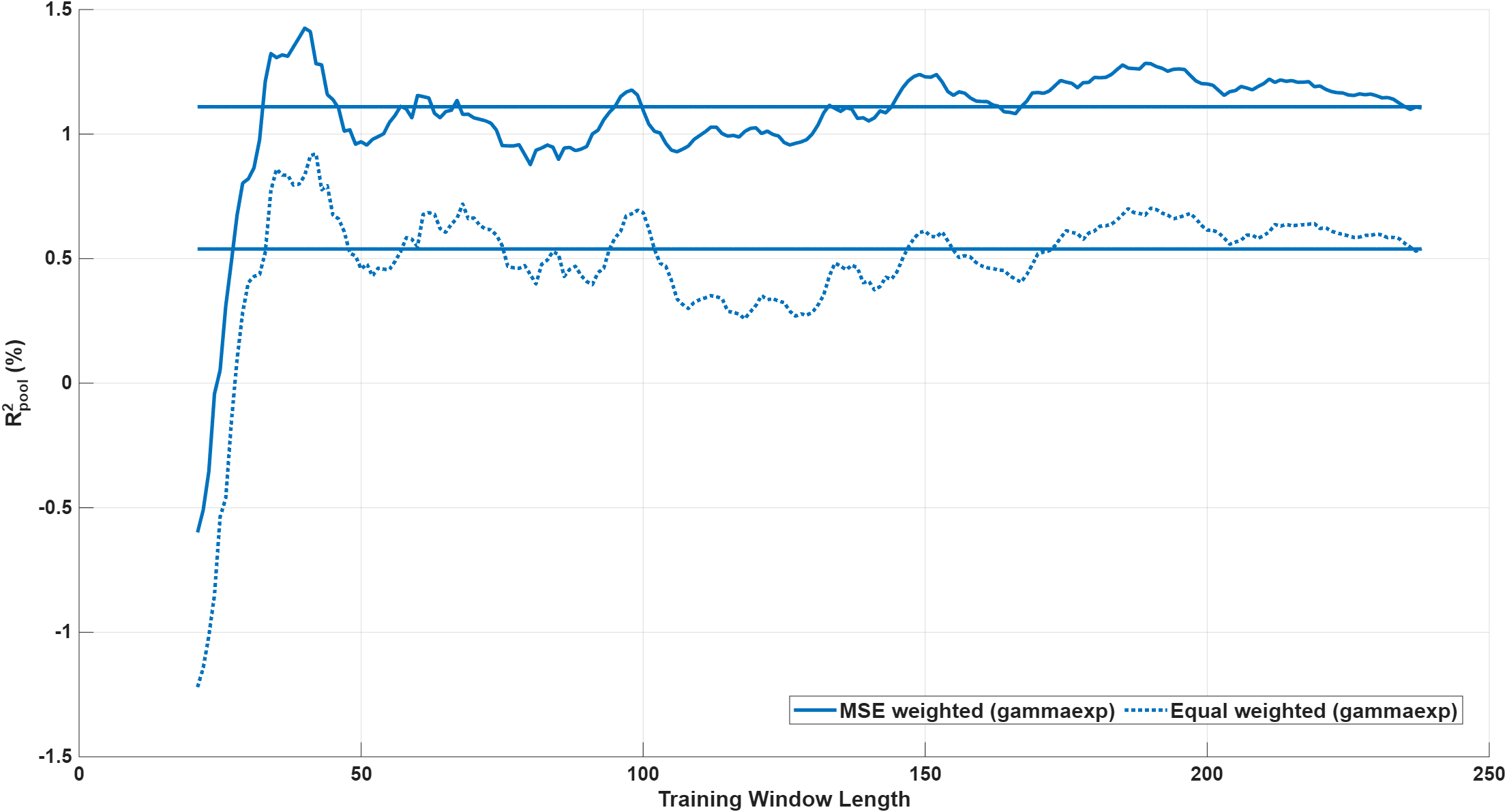}
\caption{This figure presents $R^2_{pool}$ over the validation sample, Jan 1982 to Dec 1986 for MSE- and Equal- weighting scheme against the length of the training window. To improve the visualization of the figure, we start the window length from 21 as the smaller window length exhibit unusually high volatility due to the small training data. Note that this truncation does not affect the underlying analysis.}
\label{validationRsquared1}
\end{figure}

Figure \ref{validationRsquared1} shows $R_{pool}^2$ over the validation sample for MSE- and equal-weighting schemes against varying lengths $K$ of the training window. The curved lines correspond to the rolling scheme, while the flat lines represent $R_{pool}^2$ for the expanding scheme, which does not depend on $K$. There are several findings. First, we observe that both sample splitting and mixing weight schemes generate positive $R^2_{pool}$ for $K$ large enough ($K>30$). Second, the MSE weighted ensemble yields a larger $R^2_{pool}$ than the equal weighted. A possible explanation is the temporal non-stationarity of financial data. There are non-observable changing regimes such that the regime prevailing at any given month $j$ is not explicitly captured by the observable features $\bm x_{j-1}$. But it is implicitly captured by the trained Gaussian process $f^{(j)}$ through its fitted hyperparameters and predictive distribution, given data $\{\bm x_{j-1},\,\bm r_{j}\}$. The MSE-weighting scheme in turn gives more weight to a training month $j$ the closer its regime is to the current regime prevailing in the calibration month. This results in a more accurate prediction than for equal weighting. Third, it is striking that incorporating the full available training sample, which is the expanding scheme, worsens the predictive performance; as the flat lines are below the curves for both mixing weight schemes. This finding is in line with the bias-variance trade-off in machine learning, as our ensemble model complexity grows with the size of the training window. Fourth, for the rolling scheme, $R^2_{pool}$ is relatively more volatile for small $K$ ($K<50$) and becomes stable and provides considerably more consistent performance for large $K$ ($K>50$). Further, for $K>50$, we observe a peak around $K=100$, which motivates us to choose $K=96$ as the window length for the rolling scheme.\footnote{The NBER’s business cycle dating committee maintains a chronology of US business cycles, available on \url{https://www.nber.org/research/data/us-business-cycle-expansions-and-contractions}. During our sample period, the average length of a business cycle is around eight years, which is in line with our choice of $K=96$.} To conclude, based on the performance in the validation sample, we select the rolling scheme with training window length $K=96$ and the MSE-weighting scheme for the out-of-sample test.

\subsection{Model performance}\label{model_performance}

Having selected the model configuration, we now conduct our empirical analysis over the test sample. We first train $K-1=95$ individual monthly GPR models, Jan 1979 to Nov 1986, compute MSE weights on the first calibration month, Dec 1986, and use these weights to get the predictive distribution for the returns in the first test month, Jan 1987. We proceed by induction and train one additional GPR model on Dec 1986, combine it with the already trained GPR models on the previous $94$ months, Feb 1979 to Nov 1986, compute MSE weights on the next calibration month, Jan 1987, and predict returns on Feb 1987. We repeat this online learning procedure until the full test sample is exhausted, consisting of 360 test months until Dec 2016. We report the estimated hyper-parameters of the monthly GPR models in Figure \ref{hyperparameters} in the Appendix. We evaluate the predictive performance, both statistically and economically, in the following subsections.

\subsubsection{Predictive performance across all stocks}\label{outsample_statistical1}

We first evaluate the predictive performance of our ensemble GPR model, henceforth referred to as E-GPR ($\gamma$-exp), across all stocks and benchmark it against three linear models. The first benchmark, \emph{E-GPR (affine)}, is an \emph{ensemble GPR model with an affine kernel function}, defined as  
\[
K(x, y) = c_0 + c_1 x^\top y,
\]  
where \( c_0, c_1 >0\) are hyperparameters. The second benchmark, \emph{ensemble linear regression (E-LR)}, adopts the ensemble GPR framework by fitting a separate linear regression for each training month. The third benchmark is \emph{standard linear regression (LR)}, fitted using all available training data for each test month.

\begin{figure}
    \centering
    \includegraphics[scale=0.45]{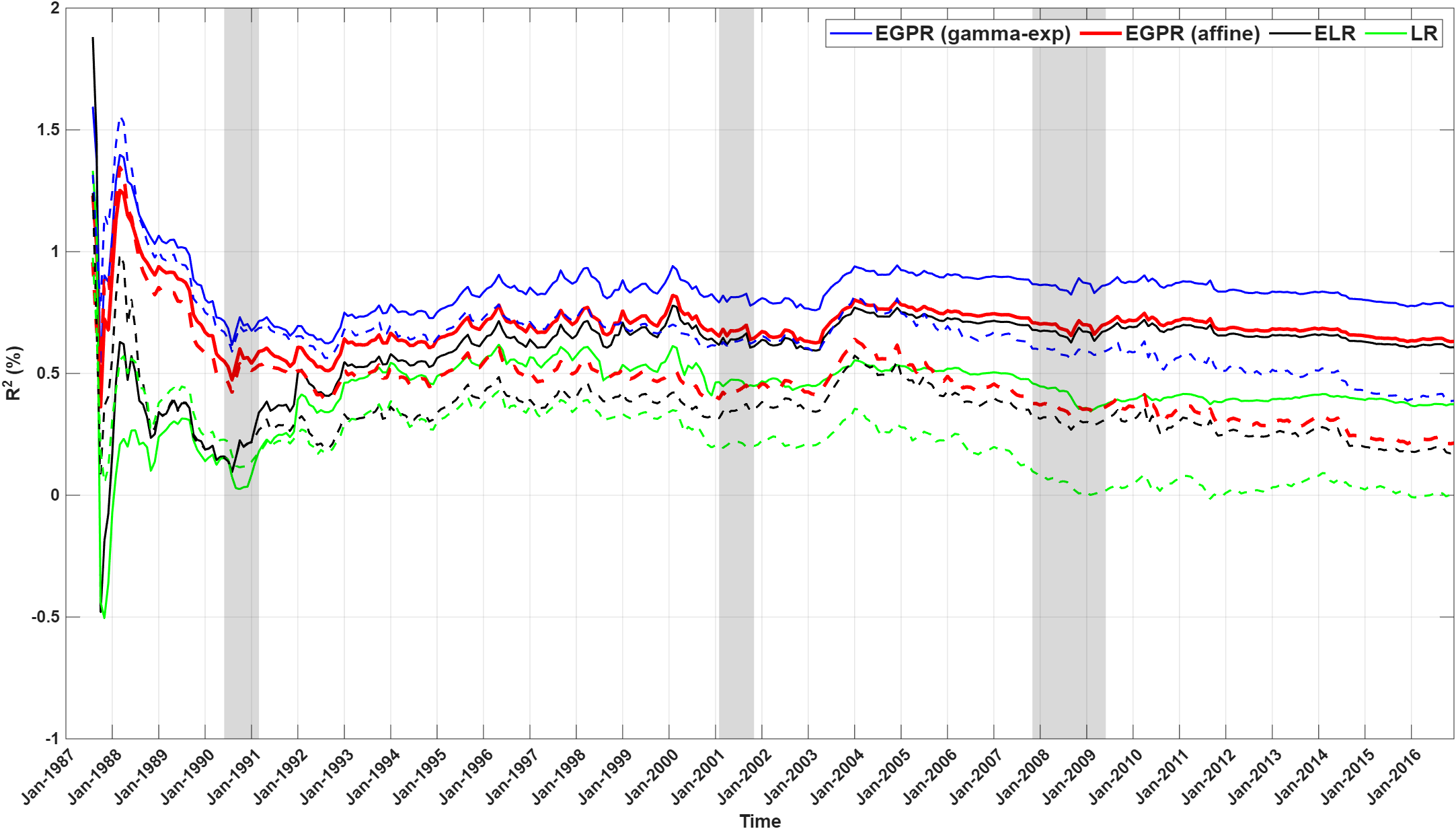}
    \caption{This figure shows the evolution of $R^2_{pool}$ (solid lines) and $R^2_{avg}$ (dotted lines) over an expanding test subsample for our model, E-GPR ($\gamma$-exp), against the linear benchmark models, E-GPR (affine), E-LR and LR. The shaded periods indicate NBER recessions. The first six months exhibit unusually high volatility due to the small size of the initial out-of-sample window, which leads to unstable estimates. To improve the visualization of the figure, we truncate the vertical axis at a value of 2. As a result, observations exceeding this threshold during the first six months are not fully displayed in the figure. Note that this truncation does not affect the underlying analysis.}
    \label{comparison_R2}
\end{figure}

Figure~\ref{comparison_R2} shows the out-of-sample performance in predicting the returns across time. It illustrates the evolution of $R_{pool}^2$ (solid lines) and $R^2_{avg}$ (dotted lines) across an expanding test sample for our model and the linear benchmark models.\footnote{We also plot the time series of monthly $R$-squared, $R_t^2$, from our model in Figure \ref{outsamplersquared2} in the appendix.} For example, $R^2_{pool}$ ($R^2_{avg}$) in Jan 2000 is the pooled $R$-squared (average $R$-squared) evaluated on the test sample, Jan 1987 to Jan 2000.

Further, Table~\ref{rsq_comparison} presents the final values of $R^2_{pool}$ and $R^2_{avg}$ for each model over the entire test sample. These final values correspond to the endpoints of the curves depicted in Figure \ref{comparison_R2}.

\begin{table}[ht!]
    \centering
    \begin{tabular}{c|c|c}
Model &  $R^2_{pool}$ (\%) & $R^2_{avg} (\%)$\\
\hline
E-GPR ($\gamma$-exp)&0.78& 0.39\\
E-GPR (affine)&0.63& 0.21\\
E-LR& 0.61 & 0.18\\
LR & 0.37 & 0.003\\
\hline
    \end{tabular}
    \caption{Comparison of predictive performance across all stocks, i.e., $R^2 (\%)$, among various models}
    \label{rsq_comparison}
\end{table}

There are several observations. The positive values of $R^2_{pool}$ and $R_{avg}^2$ over the expanding test subsample indicates that our model outperforms the zero predictions consistently over time. Moreover, our model achieves $R^2_{pool}=0.78\%$ over the full test sample.\footnote{This is substantially greater than the corresponding numbers, 0.4\% and 0.58\%, for the neural network models in \cite{gu_kel_xiu_20} (Table 1 on Page 2250) and \cite{gu2021autoencoder} (Table 2 on Page 11), respectively.} The value $R^2_{avg}=0.39\%$ over the full test sample further confirms the superior performance of our model. To assure that small stocks do not drive this unprecedented predictive performance, i.e., that our model is not simply picking up small-scale inefficiencies driven by illiquidity, we also measured $R^2_{pool}$ on two test subsamples. The first consists of the top-1{,}000 stocks and the second of the bottom-1{,}000 stocks by market capitalization each month. The values of $R^2_{pool}$ for the two subsamples are 0.625\% and 1.002\%, respectively. This is indicative that our model is capable of capturing the systematic structure in the large-cap as well as small-cap stocks, although the performance is slightly better for the small-cap stocks.

Compared to the linear benchmark models, we observe a significant improvement in prediction accuracy when employing the non-linear kernel instead of the affine kernel. Our model, E-GPR ($\gamma$-exp), consistently outperforms the linear benchmarks in terms of both $R^2_{pool}$ and $R^2_{avg}$. Figure~\ref{comparison_R2} also illustrates the temporal evolution of the contribution of non-linearity to predictive performance, highlighting that the ranking of model performance presented in Table \ref{rsq_comparison} remains consistent throughout the sample period. This consistency reinforces the robustness of our model across different time horizons. Additionally, the improved $R^2$ of ensemble linear regression (E-LR) compared to standard linear regression (LR) underscores the benefits of our proposed ensemble learning approach.

\begin{figure}
    \centering
    \includegraphics[scale=0.23]{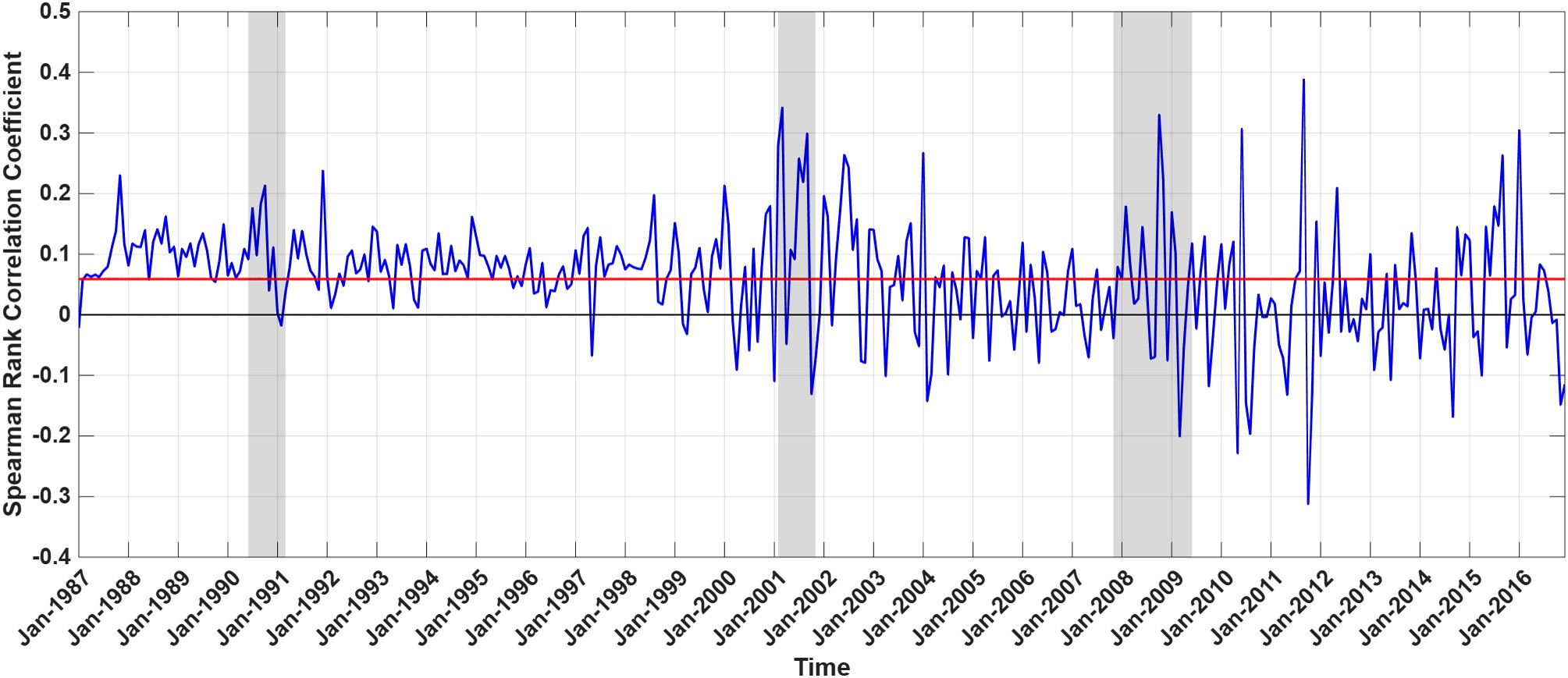}
    \caption{This figure shows the evolution of Spearman's rank correlation $\rho_t$ between the realized and predicted returns over the test sample. The flat line gives the information coefficient. The shaded periods indicate NBER recessions.}
    \label{spearman}
\end{figure}

Figure \ref{spearman} shows the time series of monthly Spearman's rank correlations $\rho_t$ between the predicted and realized returns. The flat line gives the information coefficient, $IC$, our third performance measure. It is evident that there is a substantial variation over time in the ability of our model to differentiate relative performance between stocks, as can be seen by correlation coefficients ranging from -31.2\% to 38.79\%. The information coefficient equals 5.89\% and is significantly greater than zero at the 95\% confidence level. Remarkably, we observe that our model performs equally well during the NBER recession months, as shown in the shaded periods.

\subsubsection{Predictive performance across sorted portfolios}\label{outsample_statistical2}

So far, our model's predictive performance assessment has been based on individual stock returns. Next, we analyze the predictive ability of our model at the portfolio level. Why should we assess the portfolio-level predictions when our model is optimized for predicting individual stock returns? We know that one of the main applications of predicting stock returns is to construct portfolios. A model performing better at predicting stock returns need not provide accurate predictions at the portfolio level. Therefore, assessing portfolio forecasts give an additional measure to evaluate the predictive ability of our model.

Given our predictions at the beginning of each test month, we sort stocks into deciles, which we denote by $D1, D2, \ldots, D10$, where $D1$ corresponds to the lowest predicted returns and $D10$ corresponds to the largest predicted returns. Within each decile, we then construct three different portfolios. The first is \emph{equal weighted (EW)}, and the second is \emph{value weighted (VW)} by market capitalization. These are the standard portfolios in the empirical asset pricing literature. For the third portfolio, we minimize the epistemic uncertainty of the return by solving the following optimization problem,
\begin{equation}\label{eqUW}
    \min_{{\bm w}\in\mathcal{W}}{\bm w}^\top\hat{{\bm\Sigma}}_{t+1} {\bm w},
\end{equation}
for the predictive covariance matrix $\hat{{\bm\Sigma}}_{t+1}$, and where $\mathcal{W}=\{\bm w;\ \sum_{j\in D } w_j=1,\ w_j\geq 0,\ j\in D\}$ denotes the feasible set of portfolio weights, for the respective decile $D$. We call it the \emph{uncertainty-weighted (UW)} portfolio. The goal of studying the predictive performance of the UW portfolio is to examine the role of accuracy estimates in portfolio selection and finally, study its economic contribution when we construct a risk-adjusted portfolio below.

Table \ref{statistical_across_portfolio1} shows the out-of-sample predictive performance of EW and VW portfolios from our ensemble GPR model with $\gamma$-exponential kernel. We also compare to the performance of EW and VW portfolios from the benchmark models. We report the $R^2_{pool}$ between the predicted excess returns and realized excess returns of the decile portfolios over the test sample, along with $R^2_{pool}$, pooled over all deciles, for each of the two portfolio strategies. It provides a grand portfolio-level assessment of portfolio predictions against the realized portfolio returns.

\begin{table}
    \centering
\begin{tabular}{c|cccccccccc|c} 	
    & \multicolumn{11}{c}{Equal-Weighted}\\
    \hline
&	$D1$ &$D2$&$D3$&$D4$&$D5$&	$D6$ &$D7$&$D8$&$D9$&$D10$&	$All$\\
\hline
E-GPR ($\gamma$-exp)  & 2.33	&1.33&	2.20&	3.58&	5.52&	7.29&	9.28&	11.42&	13.65&	19.83&	8.04 \\
E-GPR (affine)  & -2.00&	-0.23&	1.37&	3.43&	5.57&	7.15&	9.08&	11.06&	13.70&	20.29&	6.97 \\
E-LR   & 0.42  & 0.68  & 1.33  & 2.27  & 3.14  & 3.70  & 5.19  & 6.10   & 9.36   & 16.35  & 5.29 \\
LR & -2.08 & -1.88 & -0.52 & 0.68  & 1.60  & 2.15  & 2.47  & 2.44   & 4.41   & 10.29  & 2.93 \\
    \hline
    & \multicolumn{11}{c}{Value-Weighted}\\
    \hline
&	$D1$ &$D2$&$D3$&$D4$&$D5$&	$D6$ &$D7$&$D8$&$D9$&$D10$&	$All$\\
\hline
E-GPR ($\gamma$-exp) & -0.87&	-0.68&	0.77&	1.42&	3.23&	5.88&	6.18&	9.10&	9.45&	10.36&	3.85 \\
E-GPR (affine)  & -5.11 &	-2.44&	-0.49&	1.49&	3.39&	5.05&	6.83&	7.36&	9.39&	6.66	&1.95 \\
E-LR   & -2.33  & -1.16  & -0.41  & 0.20  & 0.86  & 1.00  & 1.80  & 0.97  & 2.39  & -0.68 & -0.18 \\
LR & -5.42  & -2.65  & -0.59  & 1.05  & 2.04  & 2.30  & 2.14  & 1.80  & 2.07  & -3.66 & -0.81 \\
\hline  
\end{tabular}
\caption{In this table, we report the out-of-sample predictive performance, measured by $R^2_{pool}$ in percentage points, across the sorted portfolios obtained using equal-weighting and value-weighting. We compare the performance of our ensemble GPR model with $\gamma$-exponential kernel to ensemble GPR model with affine kernel and different variations of linear regression model. In each panel, the first ten columns ($D1$ to $D10$) report $R^2_{pool}$ for each decile while the last column (All) reports $R^2_{pool}$ calculated over the grand panel of all deciles.}
\label{statistical_across_portfolio1}
\end{table}

Table \ref{statistical_across_portfolio2} compares the same measures for UW portfolios from our ensemble GPR model with $\gamma$-exponential kernel to that from the ensemble GPR model with affine kernel. Note that UW portfolios are a unique feature of GPR models and thus are not available for E-LR and LR. It is evident from Table \ref{statistical_across_portfolio1} that the out-of-sample predictive performance at the portfolio level aligns very closely with the results on the prediction performance at the individual stocks level reported earlier. That is, GPR based models outperform both the variants of the linear regression model (E-LR and LR) across all the portfolio strategies. Remarkably, the UW portfolios significantly outperform the EW and VW portfolios in terms of predictive performance across all the deciles. Further, the UW portfolio yields a positive grand panel $R^2_{pool}$ of 13.39\% in contrast to EW and VW portfolios that generate a grand panel $R^2_{pool}$ of 8.04\% and 3.85\%, respectively.  These findings reveal that prediction uncertainties matter.

\begin{table}
    \centering
    \resizebox{.98\textwidth}{!}{
    \begin{tabular}{c|cccccccccc|c} 	
    & \multicolumn{11}{c}{Uncertainty-Weighted Portfolios}\\
    \hline
&	$D1$ &$D2$&$D3$&$D4$&$D5$&	$D6$ &$D7$&$D8$&$D9$&$D10$&	$All$\\
\hline
E-GPR ($\gamma$-exp)  & 10.25&5.69
&5.10&7.13&9.60&9.73&14.43&17.61&20.09&
28.40&13.39\\
E-GPR (affine)  &-0.80 & 2.63  & 6.02  & 3.14  & 4.75  & 6.92  & 4.96  & 7.04  & 6.82  & -12.81 & 2.21 \\
    \hline
\end{tabular}}
\caption{In this table, we report the out-of-sample predictive performance, measured by $R^2_{pool}$ in percentage points, across the sorted portfolios obtained using uncertainty-weighted strategy. We compare the performance of our ensemble GPR model with $\gamma$-exponential kernel to ensemble GPR model with affine kernel. The first ten columns ($D1$ to $D10$) report $R^2_{pool}$ for each decile while the last column (All) reports $R^2_{pool}$ calculated over the grand panel of all deciles.}
\label{statistical_across_portfolio2}
\end{table}

\subsubsection{Economic performance of sorted portfolios}\label{outsample_economic}

Next, we assess whether the improved statistical performance of our prediction-sorted portfolios translates into better economic performance. On top of the EW and VW portfolios discussed in the previous subsection, we introduce two additional portfolios.

First, for the \emph{prediction-weighted (PW)} portfolio we assign weights, within each decile, to the stocks based on their predicted returns. The goal is to take advantage of the relative strength of the prediction signal in addition to the rankings. Specifically, for each of the top five deciles ($D6$ to $D10$), we subtract the smallest predicted return within the decile to aim at maximal predicted return. Similarly, for each of the bottom five deciles ($D1$ to $D5$), we subtract each predicted return from the largest predicted return within that decile to aim at the minimal predicted return. We then normalized the level-adjusted predicted returns such that they sum up to 1 within each decile. More specifically, for any stock $i$ we define the level-adjusted predicted return
\begin{equation}\label{eqPW}
     \hat{s}_{i,t+1}=\begin{cases}
     \hat{r}_{i,t+1}-\min_{j\in D_i}\hat{r}_{j,t+1}, &\text{if $i$ lies in the top 5 deciles, $D6$ to $D10$,}\\
     \max_{j\in D_i}\hat{r}_{j,t+1}-\hat{r}_{i,t+1}, &\text{if $i$ lies in the bottom 5 deciles, $D1$ to $D5$,}
     \end{cases}
\end{equation}
where $\hat{r}_{j,t+1}$ are the predicted returns in decile $D_i$ corresponding to stock $i$. These level-adjusted predicted returns are then normalized to obtain portfolio weights, 
\begin{equation*}
w_i=\frac{\hat{s}_{i,t+1}}{\sum_{j\in D_i}\hat{s}_{j,t+1}}.    
\end{equation*}

Second, the \emph{prediction-uncertainty weighted (PUW)} portfolio is motivated by the findings of the previous subsection that the UW portfolio yields a substantially higher $R^2_{pool}$ (for each decile) than the EW and VW portfolios. We utilize the predictive covariance matrix and the predicted returns to construct a mean-variance type portfolio for a uncertainty averse investor. We aim at maximizing the portfolio return by investing in stocks with high predicted returns with high accuracy at the same time. More precisely, we combine \eqref{eqUW} and \eqref{eqPW} by solving the following optimization problem,
\begin{equation}\label{eqPUW}
    \max_{{\bm w}\in\mathcal{W}}    
	{\bm w}^\top\hat{{\bm s}}_{t+1}-\frac{\zeta}{2}{\bm w}^\top\hat{{\bm\Sigma}}_{t+1} {\bm w},
\end{equation}
where $\hat{{\bm s}}_{t+1}=\{ \hat{s}_{j,t+1},\ j\in D\}$ is the vector of level-adjusted predicted returns in the respective decile $D$, and $\zeta$ is the uncertainty-aversion parameter.

\begin{figure}
\centering
\includegraphics[scale=0.23]{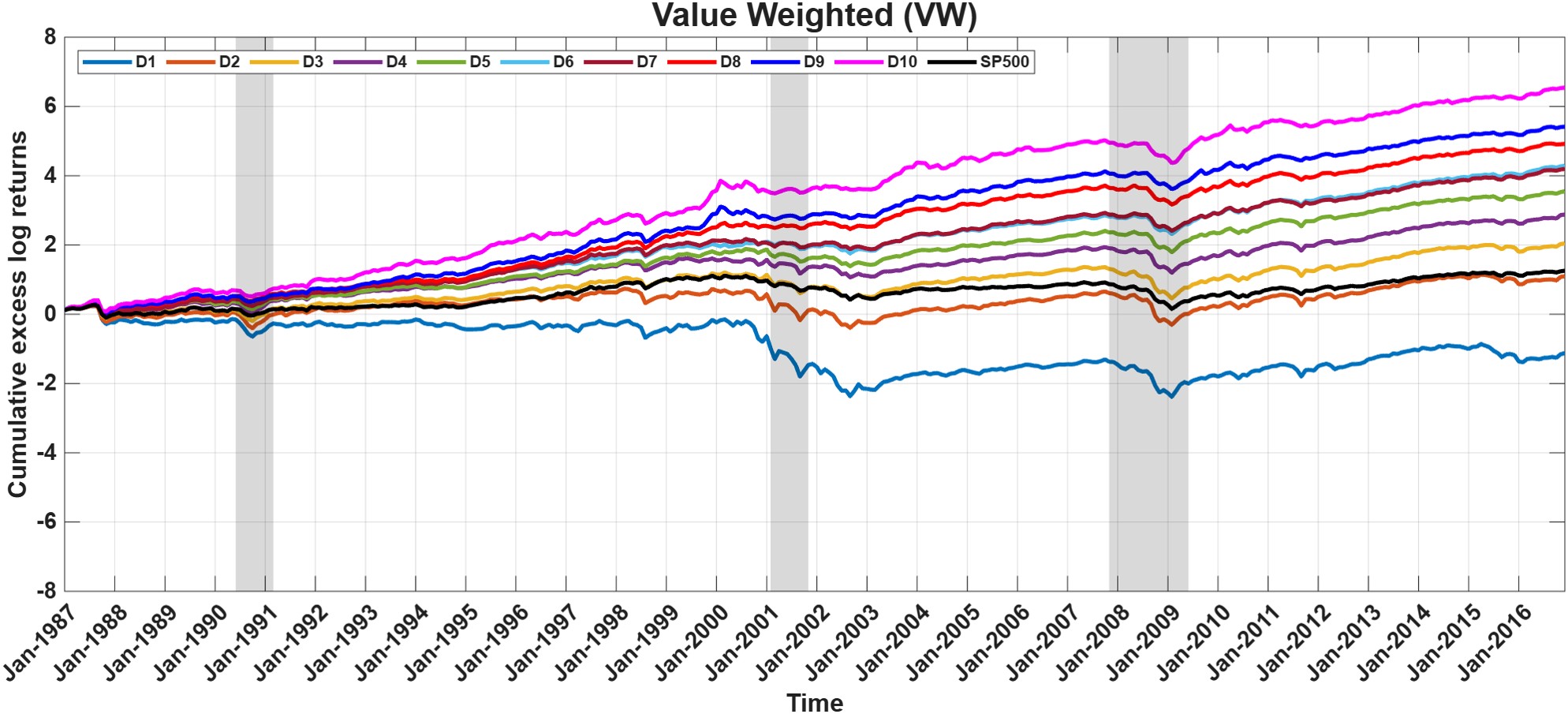}
\caption{This figure shows the out-of-sample cumulative excess log returns of the VW decile portfolios sorted based on our predicted returns. It also shows the S\&P 500. The shaded periods indicate NBER recessions.}
\label{valueweighted}
\end{figure}

Figure \ref{valueweighted} shows the cumulative excess log returns of the VW decile portfolios $D1$ to $D10$. It also shows the cumulative excess log returns of the S\&P 500 for comparison. A clear pattern emerges. First, realized returns of the decile portfolios essentially increase monotonically from $D1$ to $D10$. Portfolios based on higher predictions have higher subsequent returns. In particular, the $D1$ portfolio and $D10$ portfolio clearly separate. Second, the $D10$ portfolio outperforms the S\&P 500 by a large margin. Similar patterns emerge for the EW, PW, and PUW20 decile portfolios, as shown in Figures \ref{equalweighted} to \ref{predictionuncertaintyweighted} in the appendix. These findings indicate that our sorted portfolio strategies can consistently dissect the market into high- and low-performing stocks.

\begin{figure}
\centering
\includegraphics[scale=0.23]{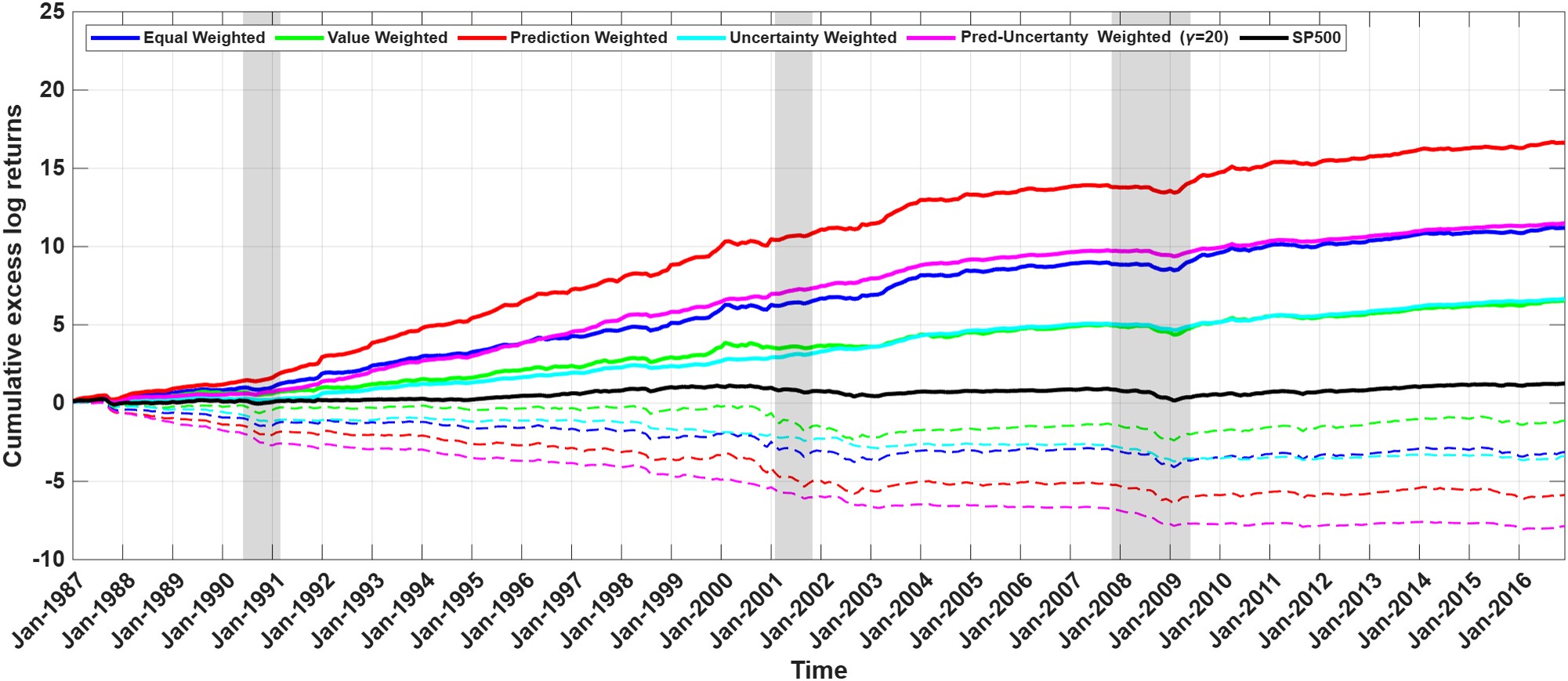}
\caption{This figure shows the out-of-sample cumulative excess log returns of the $D10$ (solid lines) and $D1$ (dashed lined) portfolios for EW, VW, PW, UW and PUW ($\gamma=20$), sorted based on our predicted returns. It also shows the S\&P 500. The shaded periods indicate NBER recessions.}
\label{D10D1}
\end{figure}

In Figure \ref{D10D1}, we plot the cumulative excess log returns of the top ($D10$) and bottom ($D1$) decile portfolios for EW, VW, UW, PW, PUW20, and the S\&P 500. The PUW $D10$ portfolio yields a cumulative excess log return of 11.49. The PUW $D1$ portfolio, on other side, yields a cumulative excess log return of -7.86. The corresponding numbers for PW portfolio are 16.64 and -5.86, and for EW portfolio are 11.20 and -3.12 respectively. The large difference in cumulative returns from PW portfolios shows that our model is not only good at dissecting the stock market into deciles but also at predicting the relative return levels within a decile. Further, the uncertainty averse PUW portfolio dominates the performance of the VW by a large margin in both directions, top and bottom and compares favorably to that of EW. These findings show that exploiting the Bayesian nature of GPR by incorporating uncertainty in the predictions can significantly improve portfolio performance. Interestingly, the bottom decile portfolio performance is essentially flat in the post-2000 sample, except for PUW. This observation has also been reported in \cite{gu_kel_xiu_20}. 

\begin{figure}
\centering
\includegraphics[scale=0.23]{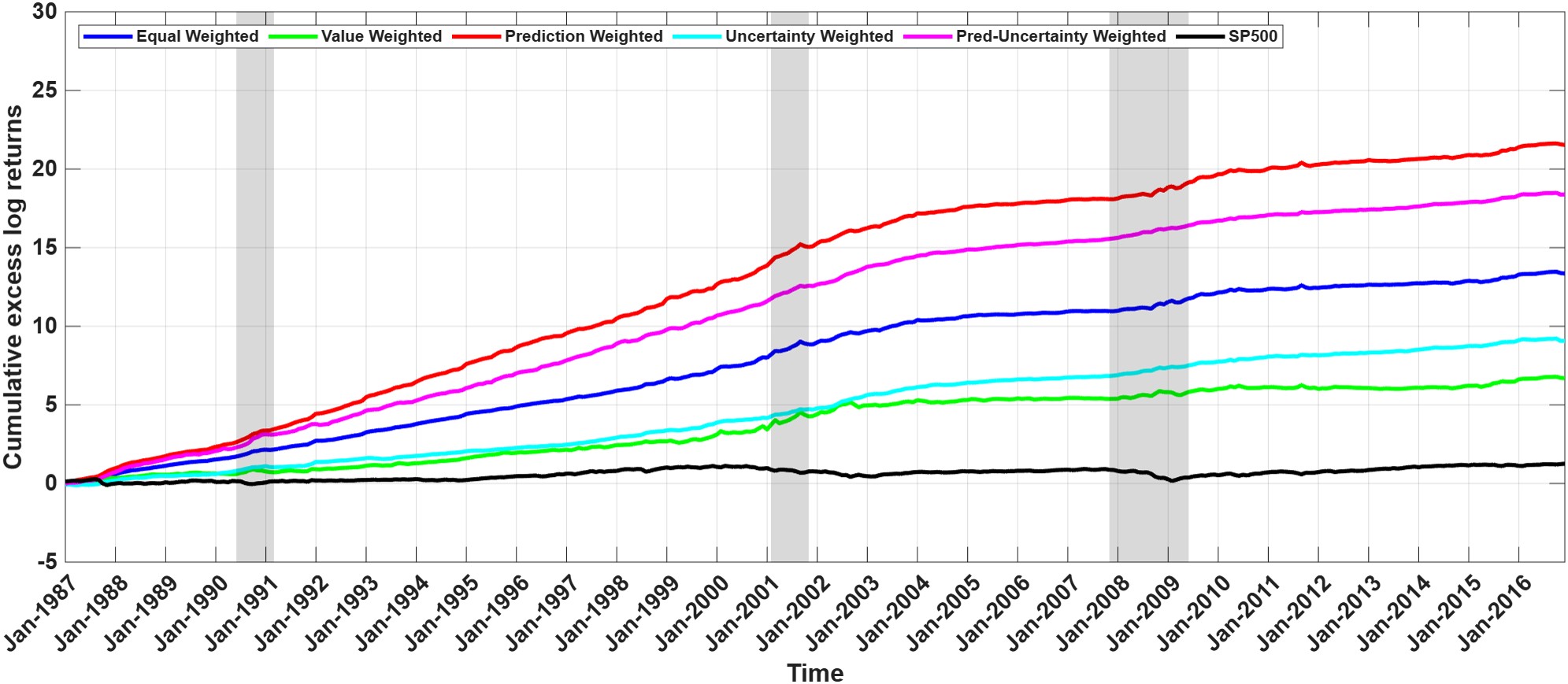}
\caption{This figure shows the cumulative excess log returns of the long-short portfolios for EW, VW, PW and PUW, obtained by taking a long position in the $D10$ and short position in $D1$ portfolios. It also shows the S\&P 500. The shaded periods indicate NBER recessions.}
\label{LS_}
\end{figure}

We also construct a zero-net-investment portfolio that is long $D10$ and short $D1$ for all the five portfolio strategies. Figure \ref{LS_} shows their cumulative excess log returns. The qualitative conclusions are identical to what we observed in Figure~\ref{D10D1}.

We evaluate the performance of decile portfolios, $D1$ to $D10$, and the long-short portfolio $LS$ from our five portfolio strategies, namely EW, VW, PW, UW and PUW, in terms of their predicted monthly returns, the average realized monthly returns, their standard deviations, and Sharpe ratios. These metrics are annualized and are calculated based on simple excess returns, and are collected in Table~\ref{performance_table1}. We observe that realized portfolio returns generally increase monotonically with our model's predictions. The average realized monthly returns, their standard deviations, and Sharpe ratios for S\&P 500 are 0.054, 0.150 and 0.360 respectively. It is evident that each of our portfolios generate a higher Sharpe ratio in comparison to the S\&P 500 index. The Sharpe ratio of 2.44 of the $LS$ portfolio from EW outperforms Sharpe ratio of 0.91 of the corresponding portfolio from the VW strategy, an observation coinciding with the findings of \cite{gu_kel_xiu_20}. Further, taking into account the levels of predictions, the PW portfolio outperforms the EW portfolio in terms of Sharpe ratio, but it exhibits a higher standard deviation of 24\% compared to 17\% from the EW. This reconfirms that our model is good at predicting the ranks of stock returns as well as their levels relative to other stocks within the same decile. The best long-short strategy comes from the PUW20 portfolio, which gives on average 59\% annualized  return with annualized volatility of 17\%, amounting to an annualized Sharpe ratio of 3.44. The annualized Sharpe ratio of the LS portfolio from the PW strategy is 2.98, slightly higher than that from PUW20 but it comes at the expense of a high standard deviation of 24\%. Nonetheless, PW could potentially be a suitable choice for a risk-seeking investor. The UW portfolio also generates a high Sharpe ratio of 2.40, with an annualized standard deviation of just 12\%. The UW portfolio is an ideal choice for a risk-averse investor who by taking least risk can still beat the market. As we increase $\zeta$ in the PUW portfolios, we approach the performance of the UW portfolios, which is as expected. These finding suggests that taking into account the uncertainty estimates of the predictions significantly improves the performance of the EW and VW portfolios, typically studied in the literature.\footnote{Our performance compares favorably to \cite{gu_kel_xiu_20}, who report Sharpe ratios of 1.0 and 0.8 for EW and VW $D10$ portfolios, -0.4 and -0.19 for EW and VW $D1$ portfolios, and 2.36 (2.63) and 1.2 (1.53) for EW and VW long-short portfolios (in \cite{gu2021autoencoder}).}

\begin{table}
  \centering
    \resizebox{.9\textwidth}{!}{    \begin{tabular}{l|rrrr|rrrr|rrrr}
          & \multicolumn{4}{c}{EW}        & \multicolumn{4}{c}{VW}        & \multicolumn{4}{c}{PW} \\
          \hline
          & \multicolumn{1}{|l}{Pred} & \multicolumn{1}{l}{Avg} & \multicolumn{1}{l}{Std} & \multicolumn{1}{l|}{SR} & \multicolumn{1}{l}{Pred} & \multicolumn{1}{l}{Avg} & \multicolumn{1}{l}{Std} & \multicolumn{1}{l|}{SR} & \multicolumn{1}{l}{Pred} & \multicolumn{1}{l}{Avg} & \multicolumn{1}{l}{Std} & \multicolumn{1}{l}{SR} \\
          \hline
$D1$    & -0.17 & -0.12 & 0.27  & -0.45 & -0.15 & -0.05 & 0.27  & -0.2  & -0.24 & -0.21 & 0.29  & -0.7 \\
    $D2$    & -0.07 & -0.01 & 0.23  & -0.03 & -0.07 & 0.01  & 0.22  & 0.03  & -0.08 & -0.02 & 0.23  & -0.08 \\
    $D3$    & -0.03 & 0.04  & 0.2   & 0.2   & -0.03 & 0.03  & 0.18  & 0.17  & -0.04 & 0.03  & 0.2   & 0.16 \\
    $D4$    & -0.01 & 0.06  & 0.18  & 0.34  & -0.01 & 0.06  & 0.17  & 0.34  & -0.01 & 0.06  & 0.18  & 0.32 \\
    $D5$    & 0.02  & 0.09  & 0.17  & 0.51  & 0.02  & 0.08  & 0.16  & 0.49  & 0.01  & 0.08  & 0.17  & 0.48 \\
    $D6$    & 0.04  & 0.1   & 0.16  & 0.6   & 0.04  & 0.1   & 0.15  & 0.66  & 0.04  & 0.1   & 0.17  & 0.62 \\
    $D7$    & 0.06  & 0.12  & 0.16  & 0.71  & 0.06  & 0.1   & 0.15  & 0.65  & 0.07  & 0.12  & 0.16  & 0.72 \\
    $D8$    & 0.09  & 0.15  & 0.17  & 0.87  & 0.09  & 0.12  & 0.15  & 0.8   & 0.1   & 0.15  & 0.17  & 0.88 \\
    $D9$    & 0.13  & 0.18  & 0.18  & 1     & 0.12  & 0.14  & 0.18  & 0.82  & 0.13  & 0.2   & 0.19  & 1.04 \\
    $D10$   & 0.21  & 0.35  & 0.25  & 1.44  & 0.19  & 0.19  & 0.21  & 0.89  & 0.28  & 0.56  & 0.31  & 1.78 \\
    $LS$    & 0.33  & 0.42  & 0.17  & 2.44  & 0.28  & 0.19  & 0.21  & 0.91  & 0.47  & 0.71  & 0.24  & 2.98 \\
          \hline
          & \multicolumn{4}{c}{UW}        & \multicolumn{4}{c}{PUW1}      & \multicolumn{4}{c}{PUW10} \\
          \hline
          & \multicolumn{1}{l}{Pred} & \multicolumn{1}{l}{Avg} & \multicolumn{1}{l}{Std} & \multicolumn{1}{l|}{SR} & \multicolumn{1}{l}{Pred} & \multicolumn{1}{l}{Avg} & \multicolumn{1}{l}{Std} & \multicolumn{1}{l|}{SR} & \multicolumn{1}{l}{Pred} & \multicolumn{1}{l}{Avg} & \multicolumn{1}{l}{Std} & \multicolumn{1}{l}{SR} \\
          \hline
$D1$    & -0.14 & -0.15 & 0.16  & -0.97 & -0.48 & -0.68 & 0.53  & -1.27 & -0.31 & -0.41 & 0.23  & -1.73 \\
    $D2$    & -0.07 & -0.07 & 0.13  & -0.52 & -0.10 & -0.08 & 0.17  & -0.46 & -0.08 & -0.08 & 0.13  & -0.60 \\
    $D3$    & -0.03 & -0.03 & 0.11  & -0.29 & -0.05 & -0.06 & 0.13  & -0.46 & -0.04 & -0.04 & 0.11  & -0.35 \\
    $D4$    & -0.01 & -0.01 & 0.10  & -0.12 & -0.02 & -0.01 & 0.12  & -0.07 & -0.01 & -0.02 & 0.10  & -0.15 \\
    $D5$    & 0.02  & 0.01  & 0.09  & 0.12  & 0.01  & 0.00  & 0.11  & -0.01 & 0.01  & 0.01  & 0.10  & 0.06 \\
    $D6$    & 0.04  & 0.03  & 0.10  & 0.33  & 0.05  & 0.04  & 0.11  & 0.34  & 0.04  & 0.04  & 0.10  & 0.36 \\
    $D7$    & 0.06  & 0.05  & 0.09  & 0.57  & 0.07  & 0.07  & 0.11  & 0.57  & 0.07  & 0.06  & 0.09  & 0.60 \\
    $D8$    & 0.09  & 0.07  & 0.09  & 0.79  & 0.10  & 0.09  & 0.12  & 0.73  & 0.09  & 0.08  & 0.10  & 0.80 \\
    $D9$    & 0.12  & 0.11  & 0.11  & 1.00  & 0.15  & 0.15  & 0.16  & 1.00  & 0.13  & 0.13  & 0.11  & 1.13 \\
    $D10$   & 0.18  & 0.18  & 0.13  & 1.41  & 0.56  & 1.76  & 1.20  & 1.47  & 0.31  & 0.56  & 0.26  & 2.14 \\
    $LS$    & 0.27  & 0.28  & 0.12  & 2.40  & 0.98  & 2.38  & 1.27  & 1.88  & 0.56  & 0.91  & 0.28  & 3.29 \\
              \hline
          & \multicolumn{4}{c}{PUW20}     & \multicolumn{4}{c}{PUW100}    & \multicolumn{4}{c}{PUW250} \\
                    \hline
          & \multicolumn{1}{l}{Pred} & \multicolumn{1}{l}{Avg} & \multicolumn{1}{l}{Std} & \multicolumn{1}{l|}{SR} & \multicolumn{1}{l}{Pred} & \multicolumn{1}{l}{Avg} & \multicolumn{1}{l}{Std} & \multicolumn{1}{l|}{SR} & \multicolumn{1}{l}{Pred} & \multicolumn{1}{l}{Avg} & \multicolumn{1}{l}{Std} & \multicolumn{1}{l}{SR} \\
          \hline
$D1$    & -0.23 & -0.30 & 0.19  & -1.56 & -0.15 & -0.17 & 0.16  & -1.08 & -0.15 & -0.16 & 0.16  & -1.01 \\
    $D2$    & -0.08 & -0.08 & 0.13  & -0.57 & -0.07 & -0.07 & 0.13  & -0.53 & -0.07 & -0.07 & 0.13  & -0.53 \\
    $D3$    & -0.04 & -0.04 & 0.11  & -0.32 & -0.03 & -0.03 & 0.11  & -0.30 & -0.03 & -0.03 & 0.11  & -0.29 \\
    $D4$    & -0.01 & -0.01 & 0.10  & -0.14 & -0.01 & -0.01 & 0.10  & -0.12 & -0.01 & -0.01 & 0.10  & -0.12 \\
    $D5$    & 0.02  & 0.01  & 0.09  & 0.08  & 0.02  & 0.01  & 0.09  & 0.11  & 0.02  & 0.01  & 0.09  & 0.12 \\
    $D6$    & 0.04  & 0.03  & 0.10  & 0.35  & 0.04  & 0.03  & 0.10  & 0.33  & 0.04  & 0.03  & 0.10  & 0.33 \\
    $D7$    & 0.06  & 0.05  & 0.09  & 0.59  & 0.06  & 0.05  & 0.09  & 0.57  & 0.06  & 0.05  & 0.09  & 0.57 \\
    $D8$    & 0.09  & 0.07  & 0.09  & 0.80  & 0.09  & 0.07  & 0.09  & 0.79  & 0.09  & 0.07  & 0.09  & 0.79 \\
    $D9$    & 0.13  & 0.12  & 0.11  & 1.08  & 0.12  & 0.11  & 0.11  & 1.02  & 0.12  & 0.11  & 0.11  & 1.00 \\
    $D10$   & 0.25  & 0.35  & 0.17  & 2.08  & 0.19  & 0.19  & 0.13  & 1.51  & 0.18  & 0.18  & 0.13  & 1.44 \\
    $LS$    & 0.42  & 0.59  & 0.17  & 3.44  & 0.29  & 0.31  & 0.12  & 2.65  & 0.27  & 0.29  & 0.12  & 2.49 \\
              \hline
    \end{tabular}}%
      \caption{In this table, we compare the economic performance of prediction sorted portfolios over the 30-year out-of-sample testing period for the ensemble GPR model with $\gamma$-exponential kernel. We compare the performance of the decile portfolios corresponding to EW, VW, PW, UW and PUW strategies. We report the performance of PUW portfolios for different values, $\{1, 10, 20, 100, 250\}$ of $\zeta$, the uncertainty-aversion parameter. We also compare the long-short portfolios. For each portfolio, we report the predicted monthly returns (``Pred"), the average realized monthly returns (``Avg"), their standard deviations (``Std"), and Sharpe ratios (``SR"). We calculate these measures using realized simple excess returns of the portfolios over the test sample. The values of ``Avg", ``Std" and ``SR" for the S\&P 500 are 0.054, 0.150 and 0.360 respectively. All measures are annualized.}
\label{performance_table1}%
\end{table}%

Next, we investigate the factors contributing to the improved performance of our portfolios, aiming to identify whether this enhancement stems from the non-linearity of the Gaussian process regression or from the ensemble learning approach. For that, we compare the performance of the portfolios from our ensemble GPR model with $\gamma$-exponential kernel to the corresponding portfolios from the ensemble GPR model with affine kernel, the ensemble linear regression model, and the standard linear regression model. The results for the benchmark models are reported in Tables~\ref{performance_table2} and \ref{performance_table3}. An important implication emerges from a comparison of the portfolio performance in Table \ref{performance_table2}, with that in Table \ref{performance_table1}. It demonstrates that the non-linear kernel outperforms the affine kernel in several aspects: dissecting the cross-section (EW and VW portfolios), accurately predicting stock return levels (PW portfolios), and achieving high precision in return predictions (UW and PUW portfolios). While the performance of the PW portfolio with affine kernel is comparable to that with $\gamma$-exponential kernel, the uncertainty-based portfolios, UW and PUW, exhibit significant improvements when using the non-linear kernel over the affine kernel. Thus, it suggests that the affine kernel is not able to deliver good estimates of the prediction uncertainty, as is also evident from the performance of the UW portfolio in Table \ref{performance_table2} where the annualized standard deviation of the $LS$ portfolio is 18\%, much higher than 11\% of the corresponding portfolio resulting from the $\gamma$-exponential kernel. Furthermore, the performance of the EW, VW and PW portfolios of the GPR with affine kernel (Table \ref{performance_table2}) is similar to the corresponding portfolios from linear regression (Table \ref{performance_table3}). The performance of the ensemble linear regression is broadly comparable to that of the standard linear regression model: EW and VW portfolios perform essentially identically, while the top decile PW portfolios of E-LR slightly outperform those of LR. In summary, this observations suggest that the improved portfolio performance of our ensemble GPR model with $\gamma$-exponential kernel stems from several aspects: the non-linearity, the ensemble learning approach, and the prediction uncertainty estimates.

\begin{table}
  \centering
      \resizebox{.9\textwidth}{!}{
    \begin{tabular}{l|rrrr|rrrr|rrrr}
          & \multicolumn{4}{c}{EW}        & \multicolumn{4}{c}{VW}        & \multicolumn{4}{c}{PW} \\
          \hline
          & \multicolumn{1}{|l}{Pred} & \multicolumn{1}{l}{Avg} & \multicolumn{1}{l}{Std} & \multicolumn{1}{l|}{SR} & \multicolumn{1}{l}{Pred} & \multicolumn{1}{l}{Avg} & \multicolumn{1}{l}{Std} & \multicolumn{1}{l|}{SR} & \multicolumn{1}{l}{Pred} & \multicolumn{1}{l}{Avg} & \multicolumn{1}{l}{Std} & \multicolumn{1}{l}{SR} \\
          \hline
$D1$    & -0.25 & -0.09 & 0.27  & -0.31 & -0.23 & -0.03 & 0.26  & -0.13 & -0.36 & -0.16 & 0.31  & -0.53 \\
    $D2$    & -0.11 & 0.01  & 0.22  & 0.07  & -0.10 & 0.02  & 0.21  & 0.08  & -0.12 & 0.00  & 0.22  & 0.02 \\
    $D3$    & -0.05 & 0.05  & 0.20  & 0.23  & -0.05 & 0.05  & 0.18  & 0.26  & -0.06 & 0.04  & 0.20  & 0.21 \\
    $D4$    & -0.01 & 0.06  & 0.18  & 0.36  & -0.01 & 0.07  & 0.16  & 0.44  & -0.02 & 0.06  & 0.18  & 0.34 \\
    $D5$    & 0.02  & 0.08  & 0.17  & 0.49  & 0.02  & 0.07  & 0.15  & 0.48  & 0.01  & 0.09  & 0.17  & 0.50 \\
    $D6$    & 0.05  & 0.09  & 0.16  & 0.57  & 0.05  & 0.09  & 0.15  & 0.60  & 0.05  & 0.10  & 0.16  & 0.60 \\
    $D7$    & 0.08  & 0.11  & 0.16  & 0.69  & 0.08  & 0.10  & 0.15  & 0.66  & 0.08  & 0.11  & 0.16  & 0.70 \\
    $D8$    & 0.11  & 0.13  & 0.17  & 0.79  & 0.11  & 0.10  & 0.15  & 0.69  & 0.12  & 0.14  & 0.17  & 0.80 \\
    $D9$    & 0.16  & 0.18  & 0.19  & 0.95  & 0.16  & 0.13  & 0.16  & 0.80  & 0.17  & 0.19  & 0.20  & 0.97 \\
    $D10$   & 0.29  & 0.32  & 0.24  & 1.30  & 0.25  & 0.14  & 0.19  & 0.76  & 0.40  & 0.46  & 0.29  & 1.58 \\
    $LS$    & 0.49  & 0.35  & 0.17  & 2.08  & 0.42  & 0.12  & 0.19  & 0.65  & 0.70  & 0.56  & 0.22  & 2.60 \\
    
          \hline
          & \multicolumn{4}{c}{UW}        & \multicolumn{4}{c}{PUW1}      & \multicolumn{4}{c}{PUW10} \\
          \hline
          & \multicolumn{1}{l}{Pred} & \multicolumn{1}{l}{Avg} & \multicolumn{1}{l}{Std} & \multicolumn{1}{l|}{SR} & \multicolumn{1}{l}{Pred} & \multicolumn{1}{l}{Avg} & \multicolumn{1}{l}{Std} & \multicolumn{1}{l|}{SR} & \multicolumn{1}{l}{Pred} & \multicolumn{1}{l}{Avg} & \multicolumn{1}{l}{Std} & \multicolumn{1}{l}{SR} \\
          \hline
D1    & -0.19 & -0.07 & 0.17  & -0.42 & -0.80 & -0.74 & 0.89  & -0.84 & -0.61 & -0.43 & 0.42  & -1.02 \\
    D2    & -0.10 & -0.01 & 0.15  & -0.05 & -0.14 & 0.02  & 0.31  & 0.07  & -0.13 & -0.01 & 0.21  & -0.06 \\
    D3    & -0.05 & 0.00  & 0.13  & 0.01  & -0.07 & 0.00  & 0.18  & 0.00  & -0.06 & 0.02  & 0.16  & 0.11 \\
    D4    & -0.01 & 0.01  & 0.13  & 0.05  & -0.02 & -0.02 & 0.14  & -0.15 & -0.02 & -0.01 & 0.14  & -0.04 \\
    D5    & 0.02  & 0.02  & 0.12  & 0.14  & 0.01  & 0.05  & 0.15  & 0.37  & 0.01  & 0.04  & 0.13  & 0.30 \\
    D6    & 0.05  & 0.01  & 0.12  & 0.08  & 0.06  & 0.04  & 0.13  & 0.32  & 0.06  & 0.03  & 0.12  & 0.28 \\
    D7    & 0.08  & 0.03  & 0.12  & 0.27  & 0.09  & 0.06  & 0.17  & 0.35  & 0.09  & 0.03  & 0.13  & 0.22 \\
    D8    & 0.11  & 0.06  & 0.12  & 0.48  & 0.13  & 0.03  & 0.15  & 0.21  & 0.13  & 0.06  & 0.13  & 0.44 \\
    D9    & 0.16  & 0.08  & 0.16  & 0.47  & 0.20  & 0.07  & 0.21  & 0.33  & 0.19  & 0.07  & 0.17  & 0.42 \\
    D10   & 0.24  & 0.04  & 0.15  & 0.24  & 0.85  & 1.10  & 1.14  & 0.97  & 0.66  & 0.28  & 0.44  & 0.64 \\
    LS    & 0.38  & 0.05  & 0.18  & 0.29  & 1.60  & 1.79  & 1.34  & 1.33  & 1.22  & 0.65  & 0.58  & 1.13 \\
              \hline
          & \multicolumn{4}{c}{PUW20}        & \multicolumn{4}{c}{PUW100}      & \multicolumn{4}{c}{PUW250} \\
          \hline
          & \multicolumn{1}{l}{Pred} & \multicolumn{1}{l}{Avg} & \multicolumn{1}{l}{Std} & \multicolumn{1}{l|}{SR} & \multicolumn{1}{l}{Pred} & \multicolumn{1}{l}{Avg} & \multicolumn{1}{l}{Std} & \multicolumn{1}{l|}{SR} & \multicolumn{1}{l}{Pred} & \multicolumn{1}{l}{Avg} & \multicolumn{1}{l}{Std} & \multicolumn{1}{l}{SR} \\
          \hline
$D1$    & -0.49 & -0.27 & 0.33  & -0.82 & -0.27 & -0.11 & 0.18  & -0.62 & -0.22 & -0.08 & 0.17  & -0.50 \\
    $D2$    & -0.13 & -0.01 & 0.19  & -0.05 & -0.11 & 0.00  & 0.15  & 0.00  & -0.10 & 0.00  & 0.15  & -0.03 \\
    $D3$    & -0.06 & 0.02  & 0.14  & 0.16  & -0.05 & 0.01  & 0.13  & 0.06  & -0.05 & 0.00  & 0.13  & 0.03 \\
    $D4$    & -0.02 & 0.00  & 0.13  & -0.01 & -0.01 & 0.01  & 0.13  & 0.08  & -0.01 & 0.01  & 0.13  & 0.08 \\
    $D5$    & 0.01  & 0.03  & 0.12  & 0.21  & 0.02  & 0.02  & 0.12  & 0.17  & 0.02  & 0.02  & 0.12  & 0.16 \\
    $D6$    & 0.06  & 0.03  & 0.12  & 0.24  & 0.05  & 0.01  & 0.11  & 0.09  & 0.05  & 0.01  & 0.12  & 0.09 \\
    $D7$    & 0.09  & 0.03  & 0.12  & 0.24  & 0.08  & 0.03  & 0.12  & 0.25  & 0.08  & 0.03  & 0.12  & 0.28 \\
    $D8$    & 0.12  & 0.06  & 0.13  & 0.46  & 0.12  & 0.07  & 0.13  & 0.53  & 0.11  & 0.06  & 0.12  & 0.51 \\
    $D9$    & 0.18  & 0.09  & 0.19  & 0.49  & 0.17  & 0.07  & 0.15  & 0.49  & 0.16  & 0.07  & 0.16  & 0.47 \\
    $D10$   & 0.56  & 0.16  & 0.26  & 0.62  & 0.31  & 0.08  & 0.16  & 0.52  & 0.27  & 0.05  & 0.15  & 0.33 \\
    $LS$    & 0.99  & 0.38  & 0.41  & 0.93  & 0.53  & 0.14  & 0.20  & 0.69  & 0.43  & 0.08  & 0.18  & 0.43 \\              \hline
    \end{tabular}}%
      \caption{In this table, we compare the economic performance of prediction sorted portfolios over the 30-year out-of-sample testing period for the ensemble GPR model with affine kernel. We compare the performance of the decile portfolios corresponding to EW, VW, PW, UW and PUW strategies. We report the performance of PUW portfolios for different values, $\{1, 10, 20, 100, 250\}$ of $\zeta$, the uncertainty-aversion parameter. We also compare the long-short portfolios. For each portfolio, we report the predicted monthly returns (``Pred"), the average realized monthly returns (``Avg"), their standard deviations (``Std"), and Sharpe ratios (``SR"). We calculate these measures using realized simple excess returns of the portfolios over the test sample. The values of ``Avg", ``Std" and ``SR" for the S\&P 500 are 0.054, 0.150 and 0.360 respectively. All measures are annualized.}
\label{performance_table2}%
\end{table}%

\begin{table}
  \centering
    \resizebox{.9\textwidth}{!}{    \begin{tabular}{l|rrrr|rrrr|rrrr}
          & \multicolumn{12}{c}{Linear Regression (LR)}\\
          \hline
          & \multicolumn{4}{c}{EW}        & \multicolumn{4}{c}{VW}        & \multicolumn{4}{c}{PW} \\
          \hline
          & \multicolumn{1}{|l}{Pred} & \multicolumn{1}{l}{Avg} & \multicolumn{1}{l}{Std} & \multicolumn{1}{l|}{SR} & \multicolumn{1}{l}{Pred} & \multicolumn{1}{l}{Avg} & \multicolumn{1}{l}{Std} & \multicolumn{1}{l|}{SR} & \multicolumn{1}{l}{Pred} & \multicolumn{1}{l}{Avg} & \multicolumn{1}{l}{Std} & \multicolumn{1}{l}{SR} \\
          \hline
$D1$ & -0.19 & -0.09 & 0.25 & -0.36 & -0.17 & -0.04 & 0.24 & -0.17 & -0.27 & -0.13 & 0.26 & -0.52 \\
$D2$ & -0.04 & 0.01 & 0.20 & 0.06 & -0.04 & 0.02 & 0.19 & 0.10 & -0.06 & 0.02 & 0.20 & 0.09 \\
$D3$ & 0.01 & 0.04 & 0.18 & 0.20 & 0.01 & 0.05 & 0.17 & 0.28 & -0.01 & 0.05 & 0.18 & 0.31 \\
$D4$ & 0.05 & 0.06 & 0.17 & 0.36 & 0.05 & 0.07 & 0.15 & 0.48 & 0.03 & 0.07 & 0.17 & 0.40 \\
$D5$ & 0.08 & 0.08 & 0.16 & 0.46 & 0.08 & 0.07 & 0.15 & 0.47 & 0.06 & 0.07 & 0.16 & 0.44 \\
$D6$ & 0.11 & 0.09 & 0.17 & 0.55 & 0.11 & 0.09 & 0.15 & 0.63 & 0.10 & 0.09 & 0.17 & 0.52 \\
$D7$ & 0.14 & 0.11 & 0.17 & 0.66 & 0.14 & 0.10 & 0.15 & 0.68 & 0.13 & 0.10 & 0.18 & 0.59 \\
$D8$ & 0.17 & 0.13 & 0.18 & 0.73 & 0.17 & 0.11 & 0.16 & 0.71 & 0.17 & 0.12 & 0.20 & 0.64 \\
$D9$ & 0.23 & 0.18 & 0.21 & 0.88 & 0.23 & 0.14 & 0.18 & 0.81 & 0.23 & 0.19 & 0.23 & 0.81 \\
$D10$ & 0.38 & 0.34 & 0.27 & 1.30 & 0.33 & 0.16 & 0.21 & 0.72 & 0.49 & 0.49 & 0.32 & 1.53 \\
$LS$ & 0.51 & 0.38 & 0.16 & 2.36 & 0.44 & 0.14 & 0.18 & 0.80 & 0.70 & 0.57 & 0.22 & 2.63 \\
              \hline
          & \multicolumn{12}{c}{Ensemble Linear Regression (E-LR)}\\
          \hline
          & \multicolumn{4}{c}{EW}        & \multicolumn{4}{c}{VW}      & \multicolumn{4}{c}{PW} \\
          \hline
          & \multicolumn{1}{l}{Pred} & \multicolumn{1}{l}{Avg} & \multicolumn{1}{l}{Std} & \multicolumn{1}{l|}{SR} & \multicolumn{1}{l}{Pred} & \multicolumn{1}{l}{Avg} & \multicolumn{1}{l}{Std} & \multicolumn{1}{l|}{SR} & \multicolumn{1}{l}{Pred} & \multicolumn{1}{l}{Avg} & \multicolumn{1}{l}{Std} & \multicolumn{1}{l}{SR} \\
          \hline
$D1$ & -0.19 & -0.09 & 0.25 & -0.36 & -0.17 & -0.04 & 0.24 & -0.17 & -0.33 & -0.16 & 0.31 & -0.51 \\
$D2$ & -0.04 & 0.01 & 0.20 & 0.06 & -0.04 & 0.02 & 0.19 & 0.10 & -0.06 & 0.01 & 0.22 & 0.04 \\
$D3$ & 0.01 & 0.04 & 0.18 & 0.20 & 0.01 & 0.05 & 0.17 & 0.28 & 0.00 & 0.04 & 0.20 & 0.21 \\
$D4$ & 0.05 & 0.06 & 0.17 & 0.36 & 0.05 & 0.07 & 0.15 & 0.48 & 0.04 & 0.06 & 0.18 & 0.33 \\
$D5$ & 0.08 & 0.08 & 0.16 & 0.46 & 0.08 & 0.07 & 0.15 & 0.47 & 0.07 & 0.08 & 0.17 & 0.47 \\
$D6$ & 0.11 & 0.09 & 0.17 & 0.55 & 0.11 & 0.09 & 0.15 & 0.63 & 0.11 & 0.09 & 0.16 & 0.57 \\
$D7$ & 0.14 & 0.11 & 0.17 & 0.66 & 0.14 & 0.10 & 0.15 & 0.68 & 0.14 & 0.11 & 0.17 & 0.69 \\
$D8$ & 0.17 & 0.13 & 0.18 & 0.73 & 0.17 & 0.11 & 0.16 & 0.71 & 0.18 & 0.14 & 0.17 & 0.79 \\
$D9$ & 0.23 & 0.18 & 0.21 & 0.88 & 0.23 & 0.14 & 0.18 & 0.81 & 0.23 & 0.19 & 0.20 & 0.97 \\
$D10$ & 0.38 & 0.34 & 0.27 & 1.30 & 0.33 & 0.16 & 0.21 & 0.72 & 0.49 & 0.45 & 0.29 & 1.56 \\
$LS$ & 0.51 & 0.38 & 0.16 & 2.36 & 0.44 & 0.14 & 0.18 & 0.80 & 0.76 & 0.56 & 0.22 & 2.57 \\
              \hline
    \end{tabular}}%
      \caption{In this table, we compare the economic performance of prediction sorted portfolios over the 30-year out-of-sample testing period for the standard linear regression model (LR) and ensemble linear regression model (E-LR). We compare the performance of the decile portfolios corresponding to EW, VW, and PW strategies. We also compare the long-short portfolios. For each portfolio, we report the predicted monthly returns (``Pred"), the average realized monthly returns (``Avg"), their standard deviations (``Std"), and Sharpe ratios (``SR"). We calculate these measures using realized simple excess returns of the portfolios over the test sample. The values of ``Avg", ``Std" and ``SR" for the S\&P 500 are 0.054, 0.150 and 0.360 respectively. All measures are annualized.}
\label{performance_table3}%
\end{table}

We also examine how the contributions from these factors, the non-linearity, the ensemble learning approach, and the prediction uncertainty estimates, change over time. To this end, we plot the cumulative excess log-return of long-short portfolios from these models in Figures \ref{ew_comparison} to \ref{pw_puw_comparison} in the appendix. In Figure \ref{ew_comparison} for EW long-short portfolios, E-GPR ($\gamma$-exp) consistently outperforms the linear benchmark models, demonstrating the significant contribution of non-linearity in improved portfolio performance. Additionally, 
E-GPR (affine) and E-LR perform comparably, with both achieving slightly better results than LR. A similar pattern is observed for VW long-short portfolios underscoring the robustness of non-linear approaches. Moreover, ensemble models 
exhibit considerably better performance than LR, further confirming the importance of adopting an ensemble learning approach. When incorporating uncertainty, PUW from our model surpasses PUW from E-GPR (affine) for both levels of uncertainty averse-ness ($\zeta=1$ and $\zeta=20$) and the PW portfolios from linear benchmark models, in Figure \ref{pw_puw_comparison}.\footnote{These levels of uncertainty averse-ness were chosen based on their superior Sharpe ratios.}

We end this section by a disclaimer noting that the performance of the above portfolios does not take into account any transaction costs, which, when considered, could offset the gains.\footnote{\cite{avramov2020machine} pointed out that transaction costs could significantly deteriorate the performance of machine learning portfolios due to high turnover or extreme positions.} Importantly, our dataset contains highly illiquid stocks with extremely small market capitalization. These stocks are thus unlikely to be accessible to investors and may incur significant transaction costs due to high bid-ask spreads and low liquidity. Hence, an investor could not potentially exploit the higher gains by shorting the $D1$ portfolio. But this applies to the benchmarks in the literature as well.

\subsection{Cross-sectional insights}\label{cross_insights}

This section focuses on cross-sectional insights of the model performance. Section \ref{varimp} analyzes the contribution of individual features to the model's performance. Section \ref{feat_ret} examines the relationship between the features and the return predictions, while Section \ref{feat_unc} examines how the features relate to the prediction uncertainty.

\subsubsection{Variable importance}\label{varimp}

\begin{figure}
    \centering
    \includegraphics[width=1.05\linewidth]{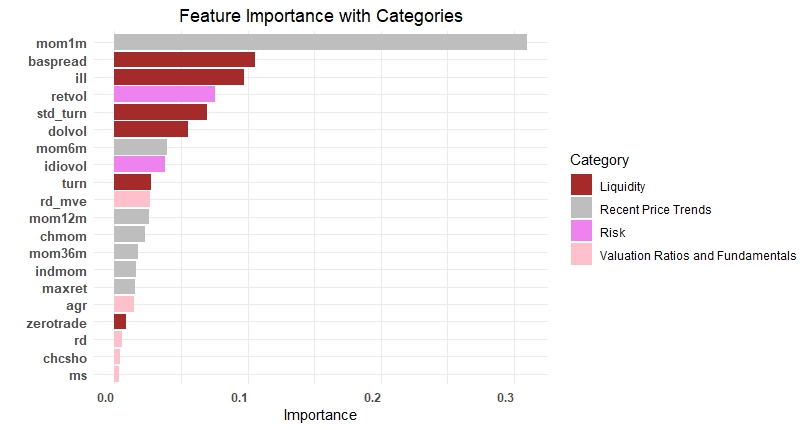}
    \caption{Top twenty features by variable importance, averaged over all training samples. The variable importance of these twenty features is normalized to sum to one.}
    \label{var_importance}
\end{figure}

We analyze the relative importance of individual features based on their contribution to the performance of our model. Following the approach in \cite{gu_kel_xiu_20}, we rank the features based on a variable importance metric, denoted as $VI_j$ for any feature $j$. $VI_j$ is defined as the reduction in pooled $R$-squared resulting from assigning all values of feature $j$ to zero, while keeping the estimates of the remaining model parameters unchanged. Specifically, we re-obtain the predictions for each test month using the pre-trained GPR models from the training period, utilizing all but the $j$th feature, which is set to zero. Figure \ref{var_importance} reports the resultant importance of the top twenty features, normalized to sum to one.\footnote{There were features for which the reduction in $R^2_{pool}(\%)$ was negative but negligibly close to zero. We replaced the variable importance for such features with zero.} 
Beyond these, variable importance hovers near zero. The total contribution by the top twenty features is 97.96\%.

Following \cite{gu_kel_xiu_20}, we have grouped the features into four categories. The first category is based on recent price trends, which includes variables such as short-term reversal (mom1m and mom6m), stock momentum (mom12m), momentum change (chmom), long-term reversal (mom36m), recent maximum return (maxret) and industry momentum (indmom). The second category consists of liquidity variables, which include bid-ask spread (baspread), dollar volume (dolvol), turnover volatility (SD$\_$turn), Amihud illiquidity (ill) and turnover (turn). The third category consists of risk measures, which include total and idiosyncratic return volatility (retvol, idiovol). The fourth category includes valuation ratios and fundamental signals consisting of features such as R\&D increase (rd), R\&D to market capitalization (rd\_mev), asset growth (agr), change in shares outstanding (chcsho) and financial statement score (ms). We also report the category-wise contributions of these features in Table \ref{varImp}, which shows that our model is inclusive and extracts predictive information from a wide range of features in the sense that it assigns significant importance to features across all the four categories, rather than being dependent on any single category.

\begin{table}
    \centering
    \begin{tabular}{c|c}
   Category & Importance (\%)\\
    \hline
   Recent Price Trends  & 45.71\\ 
   Liquidity Variables & 37.15\\
   Risk Measures & 11.56\\
   Valuation Ratios and Fundamental Signals& 5.58 \\
\hline
    \end{tabular}
    \caption{Variable Importance by Categories}
    \label{varImp}
\end{table}

\subsubsection{Association between features and predicted returns}\label{feat_ret}

Next, we investigate the cross-sectional heterogeneity in predicted returns and prediction uncertainty, and relate them to the above top twenty features.

To analyze the relation between predicted returns and the features, in each of the test months, we divide the cross-section into deciles based on the predicted returns. That is, given our predictions in each of the test month, we sort stocks into deciles, which we denote by $D1, D2, \ldots, D10$, where $D1$ corresponds to the stocks with lowest predicted returns and $D10$ corresponds to the stocks with highest predicted returns. Within each decile, we find the mean of each of the top twenty features resulting into a vector of length 360 (number of test months) for each feature. Since the cross-sectional mean of each feature is standardized to zero, we also conduct $t$-tests to determine whether the mean feature values within each decile are significantly different from zero. We report the results of the $t$-test in Table \ref{predbased_source} in the appendix. 

Figure \ref{predbased} presents the mean values of the top twenty features, categorized into four groups, averaged across the testing period for each decile. In the panel on liquidity variables, we observe that stocks with the highest predicted returns tend to be less liquid, as indicated by higher values of the bid-ask spread (baspread) and illiquidity (ill). Additionally, higher predicted returns are associated with lower values of turnover (turn) and dollar volume (dolvol), further reinforcing the observation that less liquid stocks are linked to higher predicted returns. For momentum variables, we observe an increasing trend in 6-month (mom6m) and 12-month (mom12m) momentum, coupled with a declining trend in 1-month (mom1m) momentum. This pattern suggests a positive momentum effect over longer horizons and a reversal effect over shorter horizons. Further, the stocks in deciles $D1$ and $D10$ tend to carry higher levels of risk as depicted by a U-shaped pattern for risk measures, retvol and idivol, a finding that aligns with the risk-return trade-off: the stocks with higher expected returns, positive or negative, have high values of risk levels. The bottom right panel, valuation ratios and fundamentals, shows that our models predict higher returns for growth-oriented and R\&D-intensive firms as depicted by rising trend in rd\_mve and agr. In contrast, chcsho exhibits a declining trend, suggesting higher predicted returns for firms with less equity dilution.

\begin{figure}
    \centering
    \includegraphics[scale=0.67]{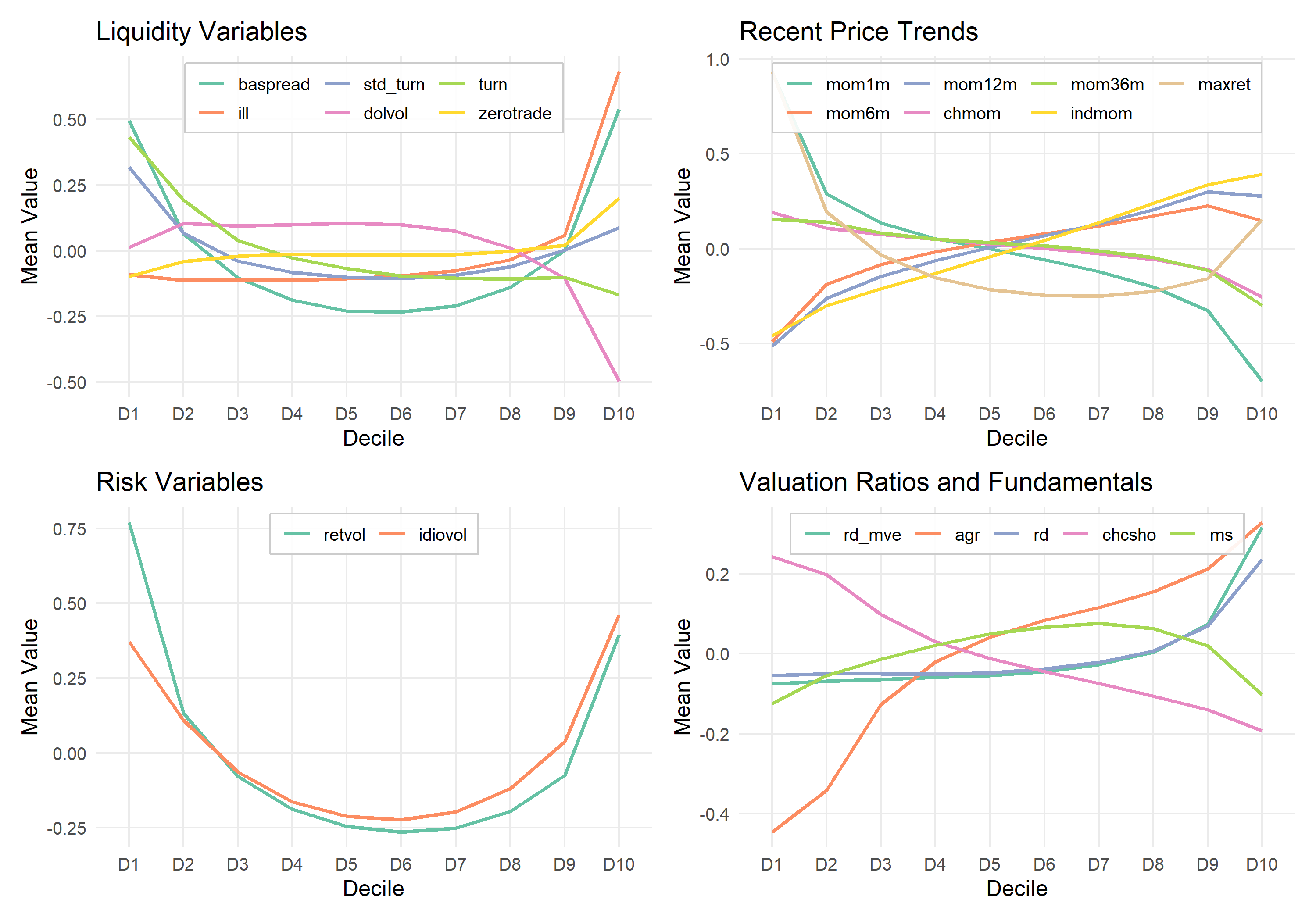}
    \caption{This figure shows the relationship between the features and predicted returns. Horizontal axes show cross-sectional deciles of predicted returns ($D1$ is lowest, $D10$ is highest). Vertical axes show time averages of conditional means of feature values given deciles.}
    \label{predbased}
\end{figure}

In essence, the above findings also give insights into the shape of the predictive function as they reveal how our model translates features to predictions. That is, analyzing the behavior of features across deciles based on predicted returns offers an understanding of the structural relationship and feature interactions that govern the model’s predictive behavior. In conclusion, we observe significant heterogeneity in the features across deciles, suggesting that predicted returns are influenced by a diverse set of factors.

\subsubsection{Association between features and prediction uncertainty}\label{feat_unc}

To analyze the relation between prediction uncertainty and the features, we repeat the above exercise. Given our prediction uncertainty in each of the test month, we sort stocks into deciles, which we denote by $D1, D2, \ldots, D10$, where $D1$ corresponds to the stocks with lowest prediction uncertainty and $D10$ corresponds to the stocks with highest prediction uncertainty.

\begin{figure}
    \centering
    \includegraphics[scale=0.67]{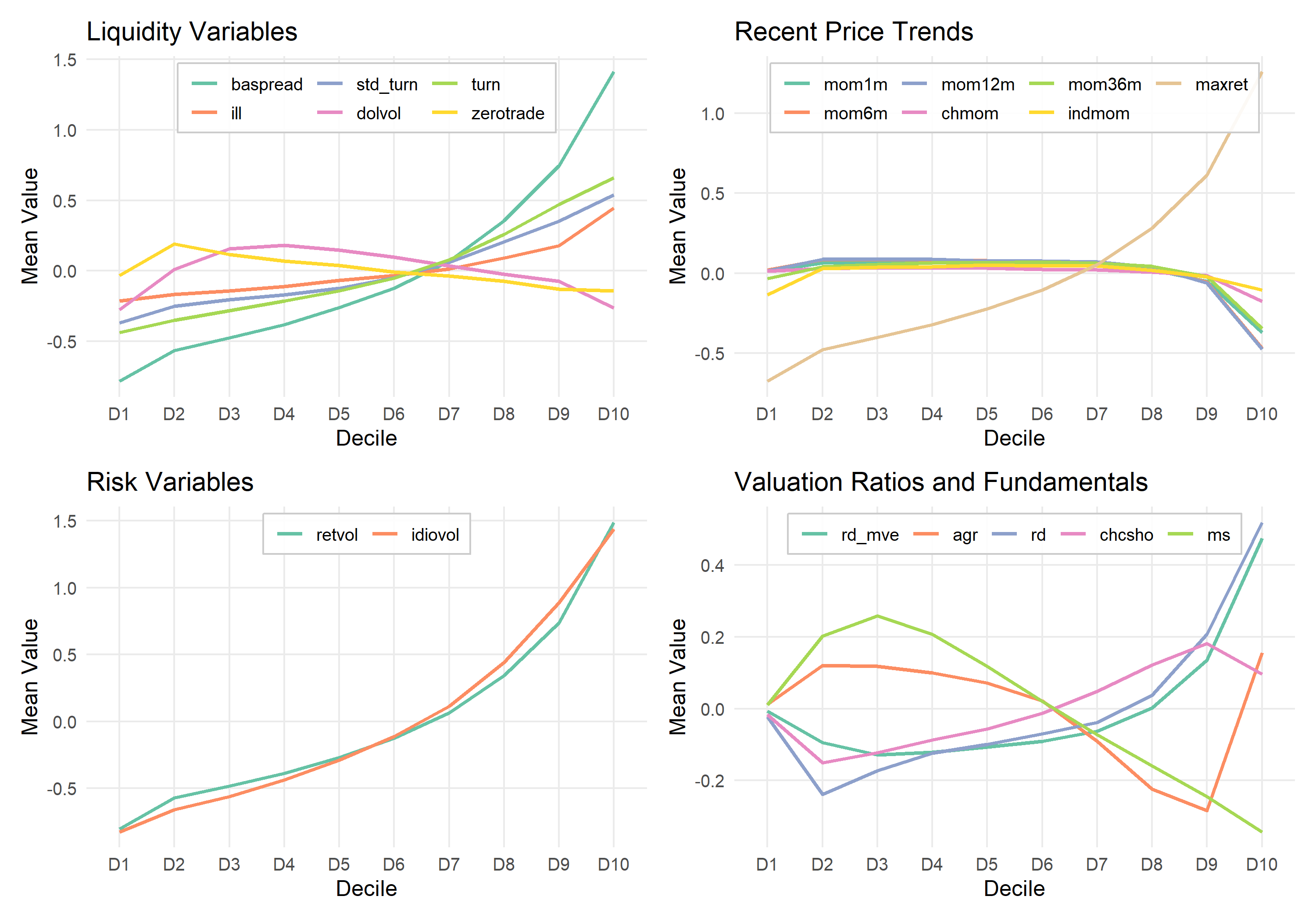}
    \caption{This figure shows the relationship between the features and prediction uncertainty. Horizontal axes show cross-sectional deciles of prediction uncertainty ($D1$ is lowest, $D10$ is highest). Vertical axes show time averages  of conditional means of feature values given deciles.}
    \label{uncbased}
\end{figure}

Figure \ref{uncbased} shows the mean value of each of the top twenty features, categorized into four groups, averaged over the testing period for each decile. We report the results of the $t$-test in Table \ref{uncbased_source} in the appendix.  There are several interesting observations. We observe that the stocks with high prediction uncertainty are indeed the ones with limits to arbitrage frictions and that exhibit extreme illiquidity as depicted by high value of bid-ask spread, a high value of illiquidity (ill) and a low value of dollar volume (dolvol). Further, the stocks with high uncertainty have high values of risk measures, namely return volatility (retvol) and idiosyncratic volatility (idiovol), low value of asset growth (agr), a relatively higher leverage and high value of maximum return (maxret). In contrast, other recent price trends exhibit no clear association with prediction uncertainty.

\subsection{Residual diagnostics}\label{residual_analysis_sec}

Leveraging the inherent ability of the GPR model to quantify prediction uncertainty, we present another use case. Given the estimates of $\hat{\sigma}_{\epsilon,(j)}^2$ from $j$th training month, similar to Equation (\ref{ensmean}), we obtain the estimate of the aleatoric uncertainty, $\hat{\sigma}_{\epsilon,t+1}^2$, as weighted average,
\begin{equation}\label{residual_estimate_}
\hat{\sigma}_{\epsilon,t+1}^2=\sum_{j} w_j\hat {\sigma}_{\epsilon,(j)}^2.
\end{equation}

\begin{figure}[ht!]
    \centering
    \includegraphics[scale=0.55]{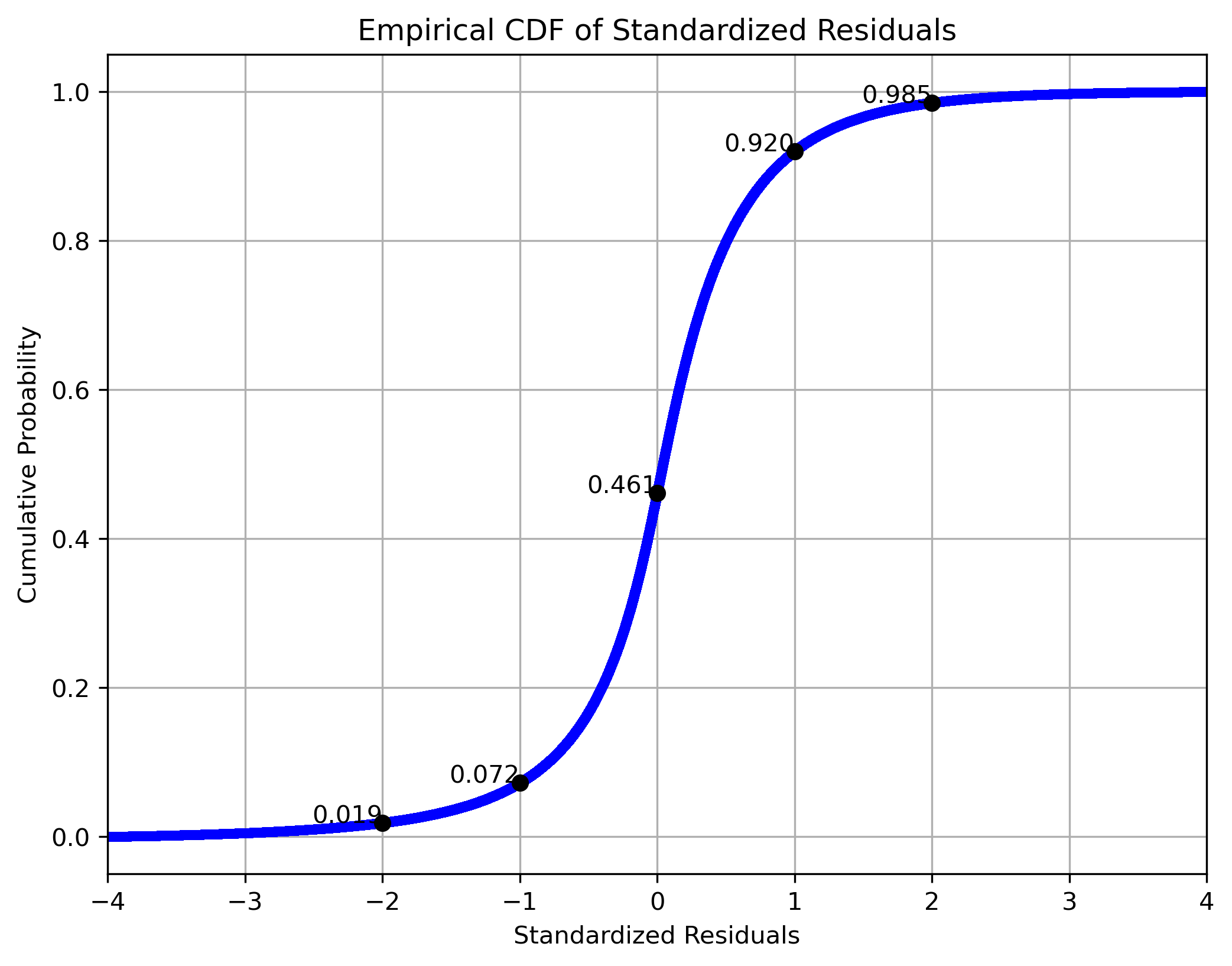}
    \caption{This figure shows empirical cumulative distribution function of the standardized residuals.}
    \label{ecdf_usecase}
\end{figure}

Given the predictions at the beginning of each test month, we can obtain the standardized realized residuals using Equations (\ref{residual_realized}) and (\ref{residual_realized2}) as
\begin{equation*}
{\bm\eta}_{t+1}=\left(\hat{\bm\Sigma}_{t+1}+\hat\sigma_{\varepsilon,t+1}^2 I\right)^{-1/2}
\big({\bm r}_{t+1}-\hat{\bm{r}}_{t+1}\big),
\end{equation*}
where ${\bm r}_{t+1}$ are the realized returns. Under standard GPR model assumptions, ${\bm\eta}_{t+1}$ would be i.i.d.\ standard normal across the test months. But a normality test leads to rejection of this hypothesis. However, this is due to the heavy tails of the residuals. In fact, Figure \ref{ecdf_usecase} presents the empirical cumulative distribution function (CDF) of the grand panel (across all test months) of standardized
realized residuals.\footnote{We have removed the standardized residuals outside $(-4,4)$, which account for 0.623\% of the test sample.} We observe that the distribution is tightly concentrated around zero but exhibits a slight right shift: the CDF at zero is 0.461, implying that 53.9\% of residuals are positive and that the empirical median is slightly above zero. Moreover, 84.8\% of the residuals lie within one standard deviation, and 96.6\% lie within two standard deviations, which broadly aligns with the two-sigma rule.\footnote{If a variable follows a normal distribution, then about 95\% of observations lie within two standard deviations of the mean.} 

We also conduct statistical tests to determine whether the mean and variance of the standardized residuals are 0 and 1, respectively. To this end, Panel A in Table \ref{table_residuals} reports summary statistics of the standardized residuals. The grand panel mean and variance are calculated by pooling residuals across all stocks and test months, providing an unconditional measure of overall bias and dispersion. We observe that mean is close to 0 and variance is 0.9896, suggesting our model is closely unbiased and residual dispersion is also stable in aggregate. Further, to assess the temporal behavior of cross-sectional residuals, we also report mean and variances of the cross-sectional mean and variance of residuals.\footnote{We present the time-series of means and variances of residuals in each month in Figure \ref{mean_var_residuals} in Appendix \ref{other_plots}.} The mean, 0.00168, and the variance, 0.00024, of the mean of cross-sectional means shows that residuals are close to zero on average and the extent of variation is also small indicating a little temporal variation in systematic prediction errors. In contrast, the variance of cross-section variance of residuals is high (0.42). 

\begin{table}[ht]
\centering
\caption{Panel-Level Statistics and Hypothesis Tests}
\label{table_residuals}
\begin{tabular}{lcc}
\toprule
 & Mean & Variance \\
\midrule
\multicolumn{3}{l}{\textbf{Panel A: Summary Statistics}} \\
Grand panel & 0.00193 & 0.98957 \\
Cross-Sectional Means& 0.00168 & 0.00024 \\
Cross-Sectional Variances & 0.98089 & 0.42304 \\
\midrule
\multicolumn{3}{l}{\textbf{Panel B: p-values}} \\
Grand panel & 0.00251 & $\approx 0$ \\
Cross-Sectional Means & 0.04191 & -- \\
Cross-Sectional Variances& 0.57754 & -- \\
\midrule
\multicolumn{3}{l}{\textbf{Panel C: Number of Rejections}} \\
90\% & 76 & 345 \\
95\% & 48 & 340 \\
99\% & 17 & 331 \\
\bottomrule
\end{tabular}
\end{table}
We also perform $t$-tests to assess whether the above numbers are statistically significant or not. That is, whether the mean and variances of residuals significantly deviate from 0 and 1, respectively. This approach allows us to evaluate the consistency of the residuals with the expected properties under the normal assumption. Panel B in Table \ref{table_residuals} reports the $p$-values of the corresponding tests at grand panel level and cross-sectional level. For instance, at the grand panel level, we test whether the pooled residual mean equals zero and the pooled residual variance equals one. The null hypothesis of zero mean is rejected, although the magnitude of the estimated mean is economically small. Similarly, the null hypothesis of unit variance is strongly rejected ($p\approx 0$), reflecting the large effective sample size in the pooled panel. We also test the performance at the cross-sectional level, that is, whether the mean of monthly cross-sectional mean of residuals equal zero, and whether the mean of monthly cross-sectional variance of residuals equals one. The $p$-values suggest that while we observe small statistically significant systematic deviations from perfect centering, still our model is well calibrated in terms of average cross-sectional dispersion ($p$ value of 0.58). Finally, Panel C in Table \ref{table_residuals} present the number of rejections, out of 360 months, of the null hypothesis of mean zero and variance one when tests are conducted (each month) on monthly cross-sectional residuals.

\begin{figure}[ht!]
    \centering
    \includegraphics[scale=0.40]{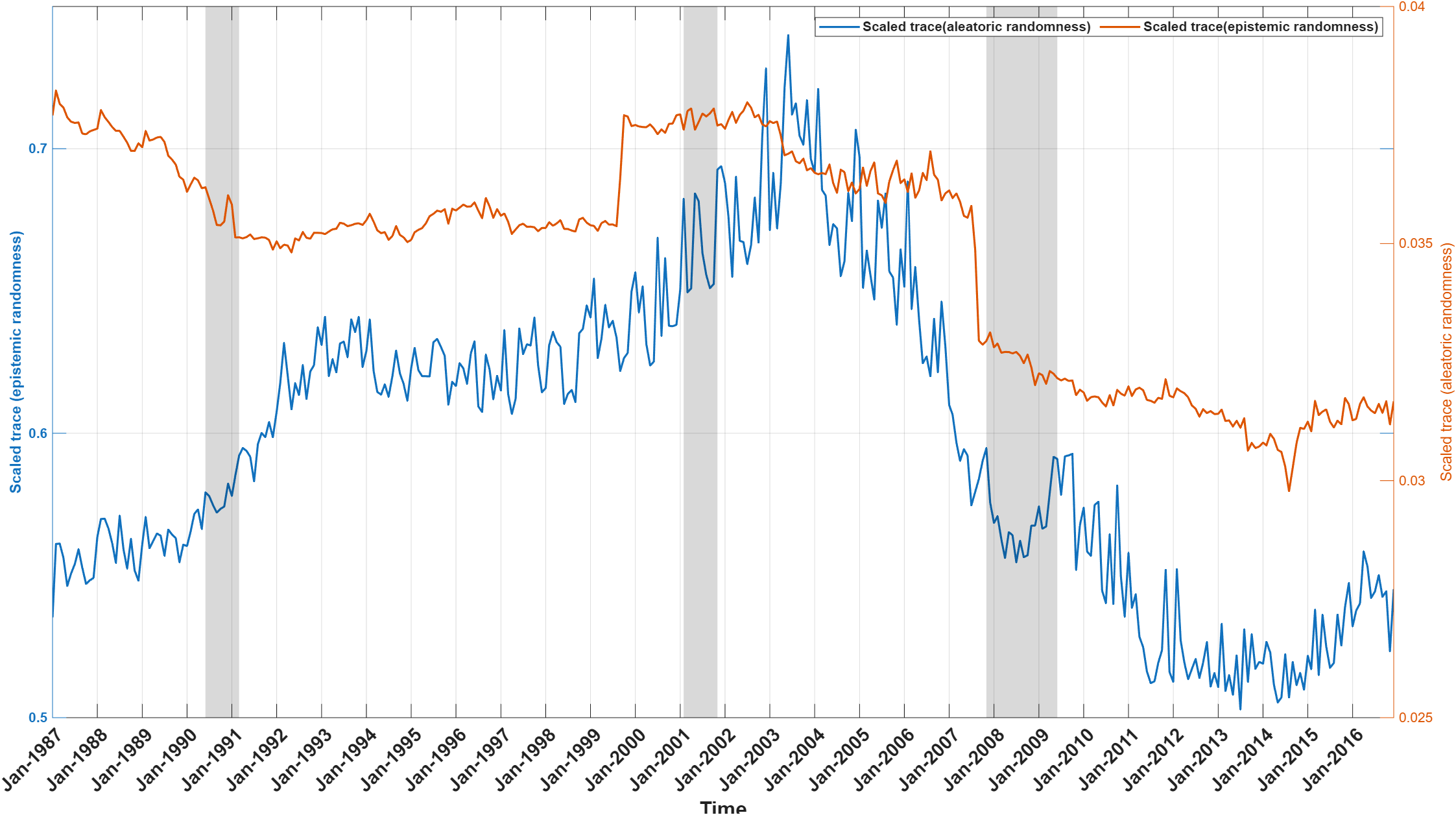}
    \caption{This figures shows the decomposition of total predictive variance. We plot the time-series of epistemic component (on left y-axis) against aleatoric component (on right y-axis).}
    \label{decomposition_posterior_variance}
\end{figure}

In addition, we decompose the estimated total predictive variance into its model-driven (epistemic) and idiosyncratic (aleatoric) components using the following trace-based measures, expressed in annualized volatility terms:
\begin{equation*}
\text{Epistemic component}
= \sqrt{\frac{12\,\operatorname{tr}(\hat{\boldsymbol{\Sigma}}_{t+1})}{n_t}},
\end{equation*}
and
\begin{equation*}
\text{Aleatoric component}
= \sqrt{\frac{12\,\operatorname{tr}(\hat{\sigma}^2_{\varepsilon,t+1} I)}{n_t}}
= \sqrt{12}\,\hat{\sigma}_{\varepsilon,t+1}.
\end{equation*}
The time series of this decomposition is displayed in Figure~\ref{decomposition_posterior_variance}.

The results show that the trace-based total predictive variance is overwhelmingly driven by the ensemble covariance component, while the idiosyncratic noise contribution is comparatively small. Importantly, the aleatoric component exhibits pronounced time variation. It declines in the early-to-mid 2000s and rises sharply around the subprime crisis, indicating clear regime dependence. The persistence of these shifts points to structural changes in the noise process rather than sampling variability. For example, we observe a marked increase around June 1999, preceding the dot-com crisis. Our ensemble GPR framework dynamically adjusts to fluctuations in the signal-to-noise ratio over time. Quantitatively, the average epistemic and aleatoric components amount to 59.88 and 3.48 percentage points in annualized volatility terms, respectively. This suggests that a sizable fraction of total predicted return variance is attributable to epistemic (model) uncertainty.


\section{Conclusion}\label{section_conclusion}

The out-of-sample prediction of conditional expected stock returns remains a central challenge in empirical asset pricing. In this paper, we introduce a novel ensemble GPR method to predict conditional expected returns. While we do not claim that our simple method is the best approach for all situations, we find that it outperforms the benchmarks by significant margins in terms of $R$-squared. Exploiting the Bayesian nature of GPR, we also model and quantify the epistemic (model) uncertainty, which leads to significant economic gains in terms of the performance of uncertainty-weighted prediction-sorted portfolios. Our ensemble learning approach reduces the computational complexity inherent in GPR and addresses the non-stationarity and heteroscedasticity in financial data. As such, it lends itself to a variety of online learning tasks to be explored in future research. Another direction of future research consists of exploiting kernel methods beyond Gaussian processes for the modeling of statistical financial risk, such as in \cite{filipovic2022copula}. 

\singlespacing
\bibliographystyle{abbrvnat}
\bibliography{ap_gpr.bib}

@article{welch2008comprehensive,
  title={A comprehensive look at the empirical performance of equity premium prediction},
  author={Welch, Ivo and Goyal, Amit},
  journal={The Review of Financial Studies},
  volume={21},
  number={4},
  pages={1455--1508},
  year={2008},
  publisher={Society for Financial Studies}
}

@article{gu_kel_xiu_20,
    author = {Gu, Shihao and Kelly, Bryan and Xiu, Dacheng},
    title = "{Empirical asset pricing via machine learning}",
    journal = {The Review of Financial Studies},
    volume = {33},
    number = {5},
    pages = {2223-2273},
    year = {2020},
}

@article{chen2019deep,
  title={Deep learning in asset pricing},
  author={Chen, Luyang and Pelger, Markus and Zhu, Jason},
  journal={Management Science, Forthcoming},
  year={2022}
}

@article{koijen2018carry,
  title={Carry},
  author={Koijen, Ralph SJ and Moskowitz, Tobias J and Pedersen, Lasse Heje and Vrugt, Evert B},
  journal={Journal of Financial Economics},
  volume={127},
  number={2},
  pages={197--225},
  year={2018},
  publisher={Elsevier}
}

@article{gu2021autoencoder,
  title={Autoencoder asset pricing models},
  author={Gu, Shihao and Kelly, Bryan and Xiu, Dacheng},
  journal={Journal of Econometrics},
  volume={222},
  number={1},
  pages={429--450},
  year={2021},
  publisher={Elsevier}
}

@book{williams2006gaussian,
  title={Gaussian processes for machine learning},
  author={Williams, Christopher K and Rasmussen, Carl Edward},
  year={2006},
  publisher={MIT press Cambridge, MA}
}

@incollection{giordano2002standard,
  title={Standard Error Estimation in Neural Network Regression Models: the {AR}-{S}ieve Bootstrap Approach},
  author={Giordano, Francesco and Rocca, Michele La and Perna, Cira},
  booktitle={Neural Nets WIRN Vietri-01},
  pages={201--206},
  year={2002},
  publisher={Springer}
}

@article{farrell2021deep,
  title={Deep neural networks for estimation and inference},
  author={Farrell, Max H and Liang, Tengyuan and Misra, Sanjog},
  journal={Econometrica},
  volume={89},
  number={1},
  pages={181--213},
  year={2021},
  publisher={Wiley Online Library}
}

@article{wager2014confidence,
  title={Confidence intervals for random forests: The jackknife and the infinitesimal jackknife},
  author={Wager, Stefan and Hastie, Trevor and Efron, Bradley},
  journal={The Journal of Machine Learning Research},
  volume={15},
  number={1},
  pages={1625--1651},
  year={2014},
  publisher={JMLR. org}
}

@article{casella2010penalized,
  title={Penalized regression, standard errors, and {B}ayesian lassos},
  author={Casella, George and Ghosh, Malay and Gill, Jeff and Kyung, Minjung},
  journal={Bayesian Analysis},
  volume={5},
  number={2},
  pages={369--411},
  year={2010},
  publisher={International Society for Bayesian Analysis}
}

@article{fama2008dissecting,
  title={Dissecting anomalies},
  author={Fama, Eugene F and French, Kenneth R},
  journal={The Journal of Finance},
  volume={63},
  number={4},
  pages={1653--1678},
  year={2008},
  publisher={Wiley Online Library}
}

@article{drobetz2021empirical,
  title={Empirical asset pricing via machine learning: evidence from the {E}uropean stock market},
  author={Drobetz, Wolfgang and Otto, Tizian},
  journal={Journal of Asset Management},
  volume={22},
  number={7},
  pages={507--538},
  year={2021},
  publisher={Springer}
}

@article{bianchi2021bond,
  title={Bond risk premiums with machine learning},
  author={Bianchi, Daniele and B{\"u}chner, Matthias and Tamoni, Andrea},
  journal={The Review of Financial Studies},
  volume={34},
  number={2},
  pages={1046--1089},
  year={2021},
  publisher={Oxford University Press}
}

@article{ambachtsheer1974profit,
  title={Profit potential in an “almost efficient” market},
  author={Ambachtsheer, Keith P},
  journal={The Journal of Portfolio Management},
  volume={1},
  number={1},
  pages={84--87},
  year={1974},
  publisher={Institutional Investor Journals Umbrella}
}

@article{han2016gaussian,
  title={Gaussian process regression stochastic volatility model for financial time series},
  author={Han, Jianan and Zhang, Xiao-Ping and Wang, Fang},
  journal={IEEE Journal of Selected Topics in Signal Processing},
  volume={10},
  number={6},
  pages={1015--1028},
  year={2016},
  publisher={IEEE}
}

@article{de2018machine,
  title={Machine learning for quantitative finance: fast derivative pricing, hedging and fitting},
  author={De Spiegeleer, Jan and Madan, Dilip B and Reyners, Sofie and Schoutens, Wim},
  journal={Quantitative Finance},
  volume={18},
  number={10},
  pages={1635--1643},
  year={2018},
  publisher={Taylor \& Francis}
}

@article{cousin2016kriging,
  title={Kriging of financial term-structures},
  author={Cousin, Areski and Maatouk, Hassan and Rulli{\`e}re, Didier},
  journal={European Journal of Operational Research},
  volume={255},
  number={2},
  pages={631--648},
  year={2016},
  publisher={Elsevier}
}

@inproceedings{deisenroth2015distributed,
  title={Distributed {G}aussian processes},
  author={Deisenroth, Marc and Ng, Jun Wei},
  booktitle={International Conference on Machine Learning},
  pages={1481--1490},
  year={2015},
  organization={PMLR}
}

@article{guhaniyogi2017divide,
  title={A divide-and-conquer {B}ayesian approach to large-scale kriging},
  author={Guhaniyogi, Rajarshi and Li, Cheng and Savitsky, Terrance D and Srivastava, Sanvesh},
  journal={arXiv preprint arXiv:1712.09767},
  year={2017}
}

@article{ng2014hierarchical,
  title={Hierarchical mixture-of-experts model for large-scale {G}aussian process regression},
  author={Ng, Jun Wei and Deisenroth, Marc Peter},
  journal={arXiv preprint arXiv:1412.3078},
  year={2014}
}

@article{cao2014generalized,
  title={Generalized product of experts for automatic and principled fusion of {G}aussian process predictions},
  author={Cao, Yanshuai and Fleet, David J},
  journal={arXiv preprint arXiv:1410.7827},
  year={2014}
}

@article{tresp2000bayesian,
  title={A {B}ayesian committee machine},
  author={Tresp, Volker},
  journal={Neural Computation},
  volume={12},
  number={11},
  pages={2719--2741},
  year={2000},
  publisher={MIT Press One Rogers Street, Cambridge, MA 02142-1209, USA journals-info~…}
}

@misc{avramov2020machine,
  title={Machine learning versus economic restrictions: Evidence from stock return predictability. Available at SSRN 3450322},
  author={Avramov, Doron and Cheng, Si and Metzker, Lior},
  year={2020}
}

@article{filipovic2022stripping,
  title={Stripping the Discount Curve—a Robust Machine Learning Approach},
  author={Filipovi{\'c}, Damir and Pelger, Markus and Ye, Ye},
  journal={Management Science},
  year={2025},
  publisher={INFORMS}
}

@article{filipovic2022shrinking,
  title={Shrinking the Term Structure},
  author={Filipovi{\'c}, Damir and Pelger, Markus and Ye, Ye},
  journal={Swiss Finance Institute Research Paper},
  number={22-61},
  year={2022}
}

@article{filipovic2022copula,
  title={Copula Process Models for Financial Risk Management},
  author={Filipovi{\'c}, Damir and Pasricha, Puneet},
  journal={Swiss Finance Institute Working Paper},
  year={2022}
}

@techreport{kaniel2022machine,
  title={Machine-learning the skill of mutual fund managers},
  author={Kaniel, Ron and Lin, Zihan and Pelger, Markus and Van Nieuwerburgh, Stijn},
  year={2022},
  institution={National Bureau of Economic Research}
}

@article{quinonero2005unifying,
  title={A unifying view of sparse approximate {G}aussian process regression},
  author={Quinonero-Candela, Joaquin and Rasmussen, Carl Edward},
  journal={The Journal of Machine Learning Research},
  volume={6},
  pages={1939--1959},
  year={2005},
  publisher={JMLR. org}
}

@article{silverman1985some,
  title={Some aspects of the spline smoothing approach to non-parametric regression curve fitting},
  author={Silverman, Bernhard W},
  journal={Journal of the Royal Statistical Society: Series B (Methodological)},
  volume={47},
  number={1},
  pages={1--21},
  year={1985},
  publisher={Wiley Online Library}
}

@phdthesis{wilson2014covariance,
  title={Covariance kernels for fast automatic pattern discovery and extrapolation with Gaussian processes},
  author={Wilson, Andrew Gordon},
  year={2014},
  school={University of Cambridge Cambridge, UK}
}

@inproceedings{wilson2015kernel,
  title={Kernel interpolation for scalable structured {G}aussian processes ({KISS-GP})},
  author={Wilson, Andrew and Nickisch, Hannes},
  booktitle={International conference on machine learning},
  pages={1775--1784},
  year={2015},
  organization={PMLR}
}

@inproceedings{wilson2016deep,
  title={Deep kernel learning},
  author={Wilson, Andrew Gordon and Hu, Zhiting and Salakhutdinov, Ruslan and Xing, Eric P},
  booktitle={Artificial intelligence and statistics},
  pages={370--378},
  year={2016},
  organization={PMLR}
}

@article{gardner2018gpytorch,
  title={Gpytorch: Blackbox matrix-matrix {G}aussian process inference with {GPU} acceleration},
  author={Gardner, Jacob and Pleiss, Geoff and Weinberger, Kilian Q and Bindel, David and Wilson, Andrew G},
  journal={Advances in Neural Information Processing Systems},
  volume={31},
  year={2018}
}

\onehalfspacing

\appendix

\renewcommand{\thesubsection}{\Alph{section}.\arabic{subsection}}
\setcounter{table}{0}
\setcounter{figure}{0}
\renewcommand{\thetable}{A.\arabic{table}}
\renewcommand{\thefigure}{A.\arabic{figure}}

\newtheorem{thm}{Theorem}[section]
\newtheorem{theorem_app}[thm]{Theorem}

\section{Gaussian Process Regression}\label{sec_app}

In this appendix we give an introduction to Gaussian process regression. For more background and theory we refer the reader to \cite{williams2006gaussian}.

\subsection{Gaussian Processes}
A Gaussian process is a collection of random variables, any finite number of which have a joint Gaussian distribution. More specifically, let $\mathcal{X}$ be a non-empty set. A random function $f:\mathcal{X}\rightarrow \R$ is a \emph{Gaussian process (GP)} with mean function $m(\cdot)$ and covariance function, or, kernel $k(\cdot,\cdot)$, if for any finite set $\bm x=(x_1,x_2,\ldots,x_n)\subset \mathcal{X}$ the random vector 
\begin{equation*}
f(\bm x)=(f(x_1),f(x_2),\ldots,f(x_n))^\top  
\end{equation*}
follows multivariate normal distribution $\mathcal{N}(m(\bm x),k(\bm x,\bm x^\top))$ with mean vector 
\[ m(\bm x)=(m(x_1),m(x_2),\ldots,m(x_n))^\top\] 
and covariance matrix 
\[k(\bm x,\bm x^\top)=(k(x_i,x_j))^n_{i,j=1}.\]

\subsection{Training GPR}
To fit the hyperparameters by maximum likelihood, we compute the partial derivatives of the marginal likelihood function \eqref{marginal} w.r.t.\ the hyperparameters,
\begin{equation}\label{gradient}
\begin{aligned}
\frac{\partial }{\partial \theta_j}\log p(\bm y\mid {\bm X},\theta)&=-\frac{{\bm y}^\top K_y^{-1}\frac{\partial K_y}{\partial \theta_j}K_y^{-1}{\bm y}}{2}-\frac{1}{2}tr\bigg(K_y^{-1}\frac{\partial K_y}{\partial \theta_j}\bigg) \\
&= \frac{1}{2}tr\bigg((K_y^{-1}{\bm y}{\bm y}^\top K_y^{-1} -K_y^{-1})\frac{\partial K_y}{\partial \theta_j}\bigg)
\end{aligned}
\end{equation}
where we write $K_y={\bm K}+\sigma_\epsilon^2I$. We then find the optimal hyperparameters $\hat{\theta}$ by any gradient based optimizer and gradient given in \eqref{gradient}.

\section{Choice of kernel}\label{kernel_choice}

A critical ingredient in the success of Gaussian process regression is the choice of kernel as it defines our prior assumptions about the function being modeled. It also drives the model's potential to capture the relationship between input features and the output variable by governing the flexibility, smoothness and the generalization ability of the model. In preliminary experiments, we tried with various kernels in predicting excess log returns in the validation period for a guided choice on the optimal kernel. We experimented with the following ten kernels,
\begin{enumerate}
\item $K_1(x,x')=\sigma^2(1+\alpha||x||)(1+\alpha||x'||)\left(1+\frac{||x-x'||^2}{2\alpha\ell^2}\right)^{-\alpha}$
\item $K_2(x,x')=\sigma^2\sqrt{(1+\alpha||x||^2)}\sqrt{(1+\alpha||x'||^2)}\left(1+\frac{||x-x'||^2}{2\alpha\ell^2}\right)^{-\alpha}$
\item $K_3(x,x')=\sigma^2\sqrt{(1+\alpha||x||^2)}\sqrt{(1+\alpha||x'||^2)}\exp\left\{-\frac{||x-x'||^2}{2\ell^2}\right\}$
\item $K_4(x,x')=\sigma^2(1+\alpha||x||)(1+\alpha||x'||)\exp\left\{-\frac{||x-x'||}{\ell}\right\}$
\item $K_5(x,x')=\sigma^2\left(1+\frac{||x-x'||}{2\alpha\ell^2}\right)^{-\alpha}$
\item $K_6(x,x')=\sigma^2(1+\alpha||x||)(1+\alpha||x'||)\left(1+\frac{||x-x'||}{\beta}\right)^{-1}$
\item $K_7(x,x')=\sigma^2\exp\left\{-\frac{||x-x'||^2}{2\ell^2}\right\}$
\item $K_8(x,x')=\sigma^2\left(1+\frac{||x-x'||^2}{2\alpha\ell^2}\right)^{-\alpha}$
\item $K_9(x,x')=\sigma^2\exp\left\{-\frac{||x-x'||}{\ell}\right\}$
\item $K_{10}(x,x')=\sigma^2\exp\left\{-\left(\frac{||x-x'||}{\ell}\right)^{\gamma}\right\}$
\end{enumerate}
where $\sigma,\alpha,\ell>0,\gamma\in(0,2]$. Here, the kernels $K_7$ to $K_{10}$ are the standard kernels: squared exponential kernel, rational quadratic kernel, exponential kernel and the $\gamma$-exponential kernel respectively. The kernels $K_1$ to $K_6$ are variations of the four standard kernels defined with an aim to introduce feature dependent variance function and utilizing the fact that any non-degenerate covariance kernel can be factorized as $k(x,x') = \sqrt{v(x)v(x')}\rho(x,x')$ where $v:\mathcal{X} \to (0,\infty)$ is a function and $\rho$ a kernel on $\mathcal{X}$ with $\rho(x,x)=1$. This factorization can always be achieved by setting $v(x)= k(x,x)$ and $\rho(x,x')=k(x,x')/\sqrt{v(x)v(x')}$, so that $v(x)$ is the variance of $f(x)$, and $\rho(x,x')$ the linear correlation of $f(x)$ and $f(x')$.

\begin{figure}
\centering
\includegraphics[scale=0.18]{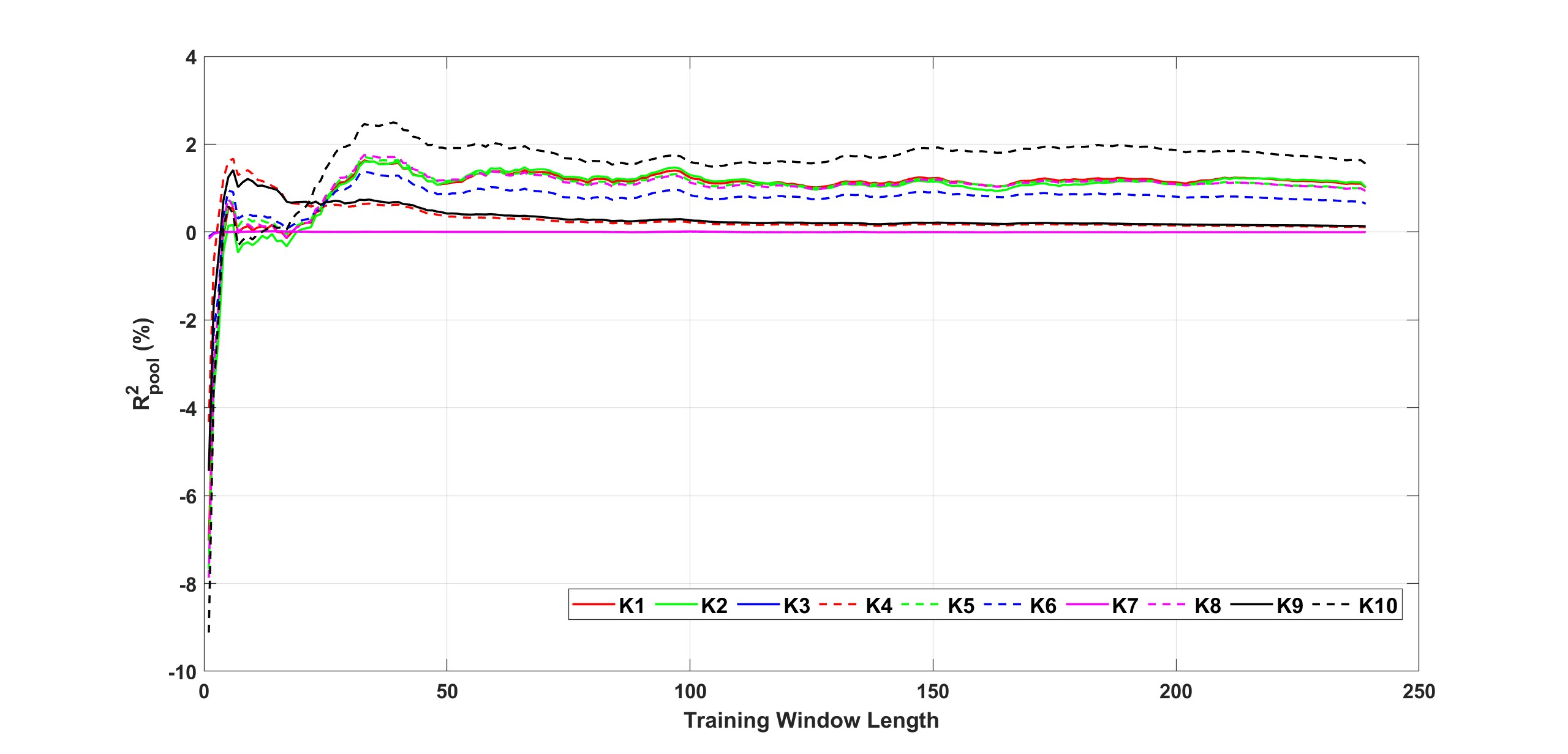}
\caption{This figure presents $R^2_{pool}$ over the validation sample, Jan 1982 to Dec 1986 for MSE-weighting scheme against the length of the training window for ten kernels under consideration.}
\label{validation_log_MSE}
\end{figure}

\begin{figure}
\centering
\includegraphics[scale=0.18]{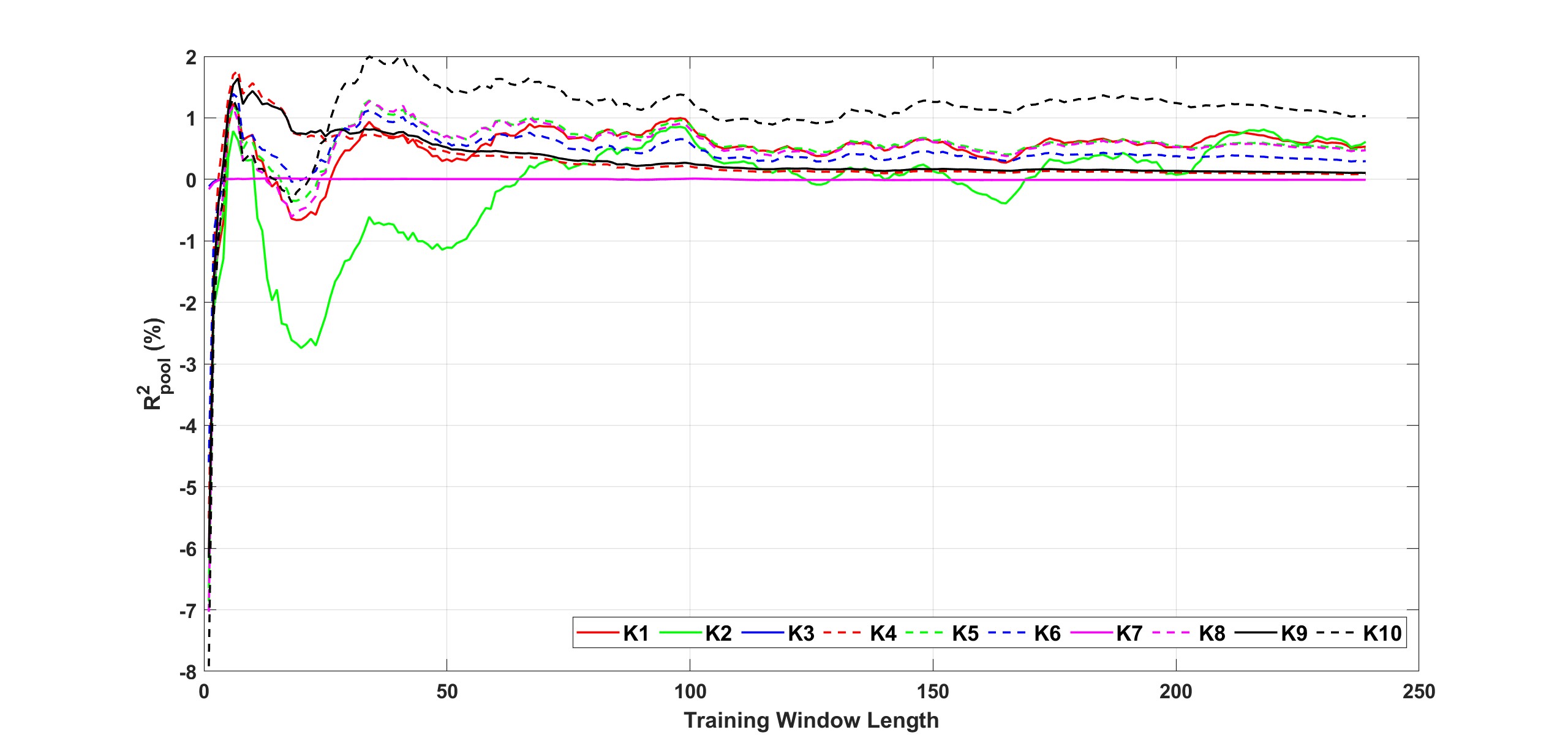}
\caption{This figure presents $R^2_{pool}$ over the validation sample, Jan 1982 to Dec 1986 for equal-weighting scheme against the length of the training window for ten kernels under consideration.}
\label{validation_log_equal}
\end{figure}

Figures \ref{validation_log_MSE} and \ref{validation_log_equal} show $R_{pool}^2$ over the validation sample for MSE- and equal-weighting schemes against varying lengths $K$ of the training window in predicting log-excess returns. It is evident from both the plots that both the weighting schemes generate positive $R^2_{pool}$ for $K$ large enough ($K>50$) with $K_2$ being an exception in equal-weighting scheme. Further, the best performing kernels are $K_5$, $K_8$ and $K_{10}$. To conclude, based on the performance in validation sample on log-excess returns, we select these three kernels for further analysis.

\section{Additional results}\label{other_plots}

This appendix includes supplementary figures and tables referenced in the main text.

\begin{figure}
    \centering
    \includegraphics[scale=0.23]{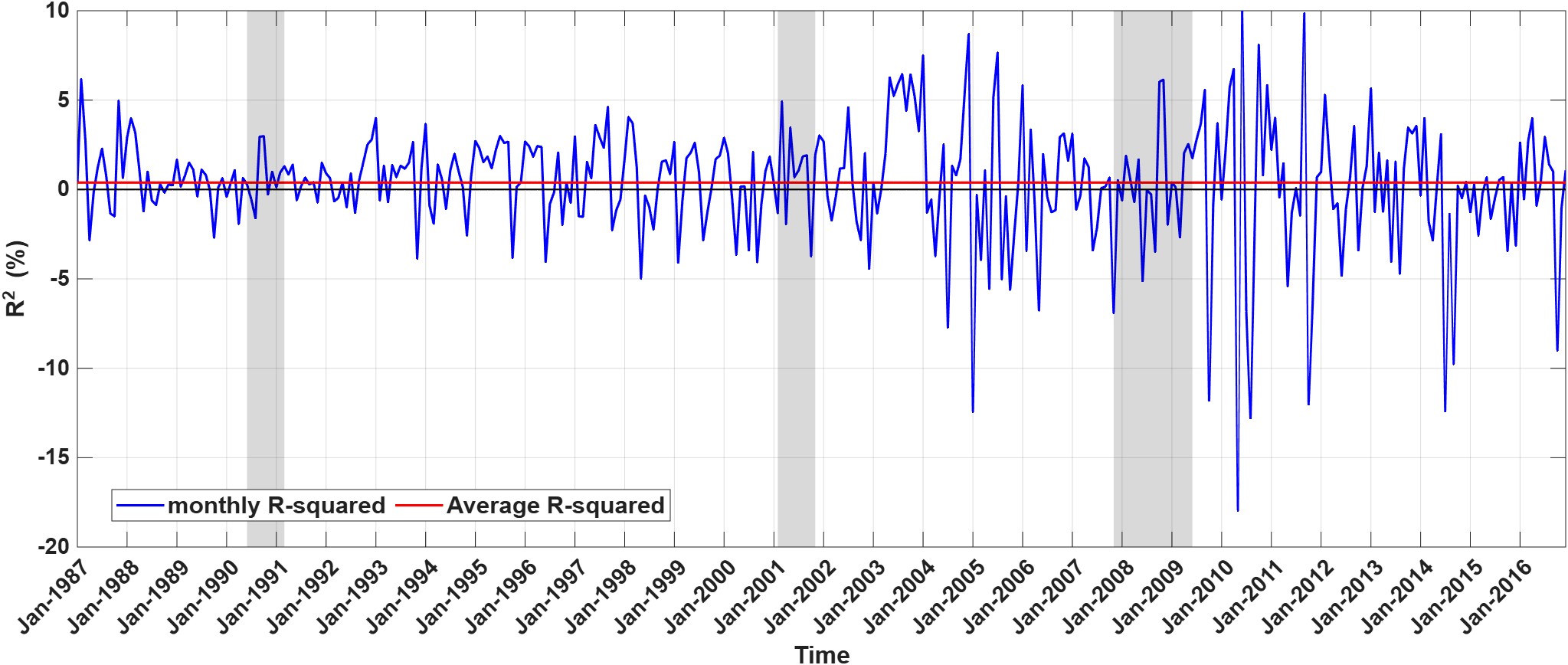}
    \caption{This figure shows the evolution of $R^2_t$ over the test sample. The shaded periods indicate NBER recessions.}
    \label{outsamplersquared2}
\end{figure}

\begin{figure}
\centering
\includegraphics[scale=0.23]{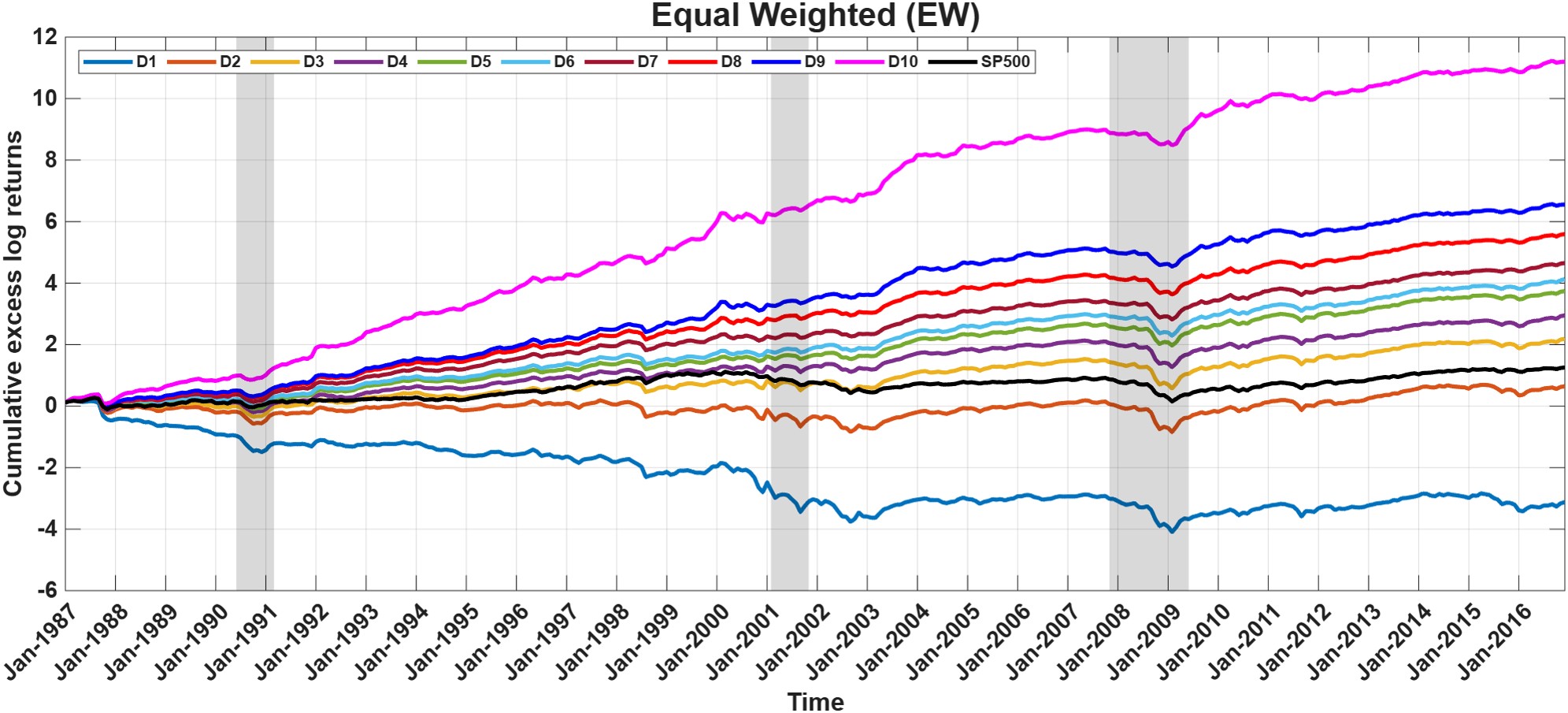}
\caption{This figure shows the cumulative excess log returns of equal weighted (EW) decile portfolios sorted based on our predicted returns. It also shows the S\&P 500. The shaded periods indicate NBER recessions.}
\label{equalweighted}
\end{figure}

\begin{figure}
\centering
\includegraphics[scale=0.23]{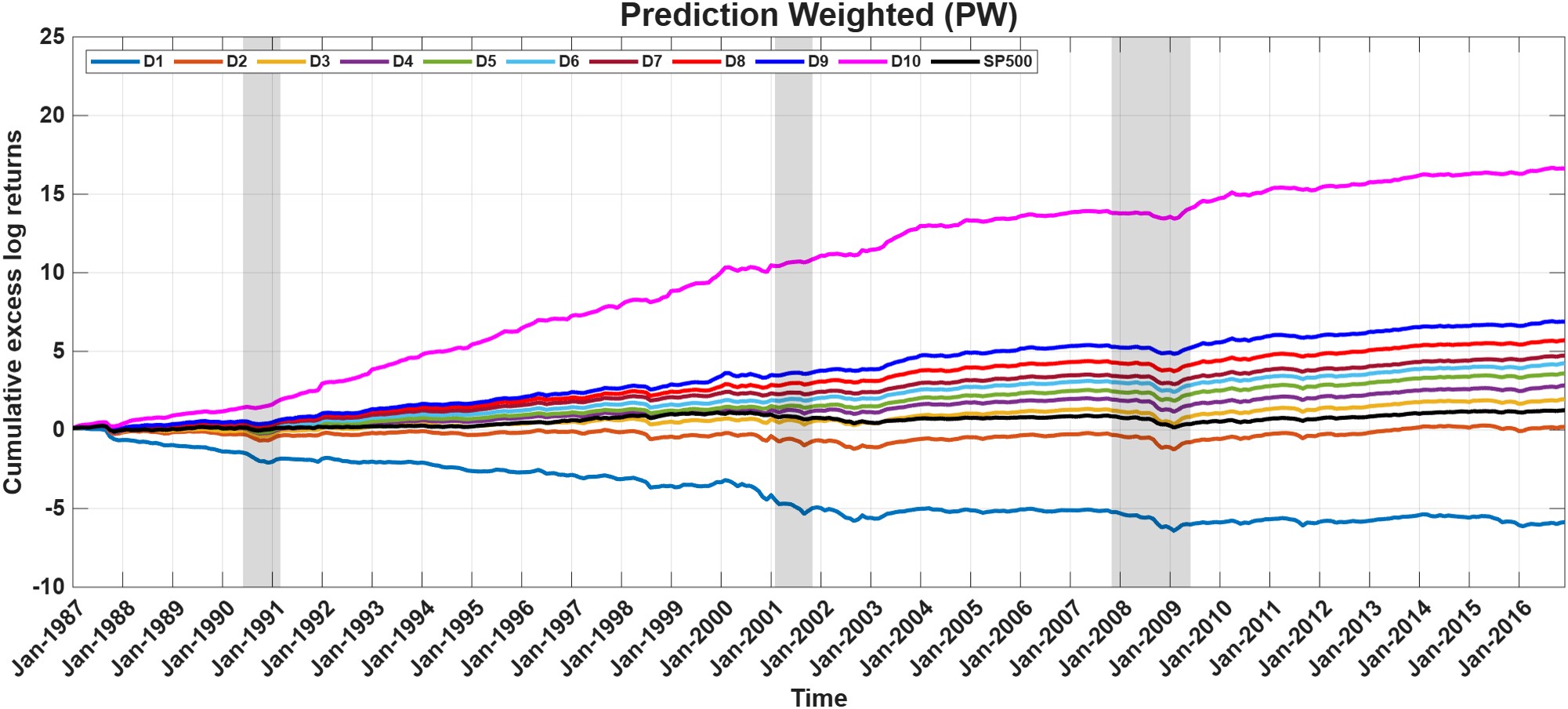}
\caption{This figure shows the cumulative excess log returns of prediction weighted (PW) decile portfolios sorted based on our predicted returns. It also shows the S\&P 500. The shaded periods indicate NBER recessions.}
\label{predictionweighted}
\end{figure}

\begin{figure}
\centering
\includegraphics[scale=0.23]{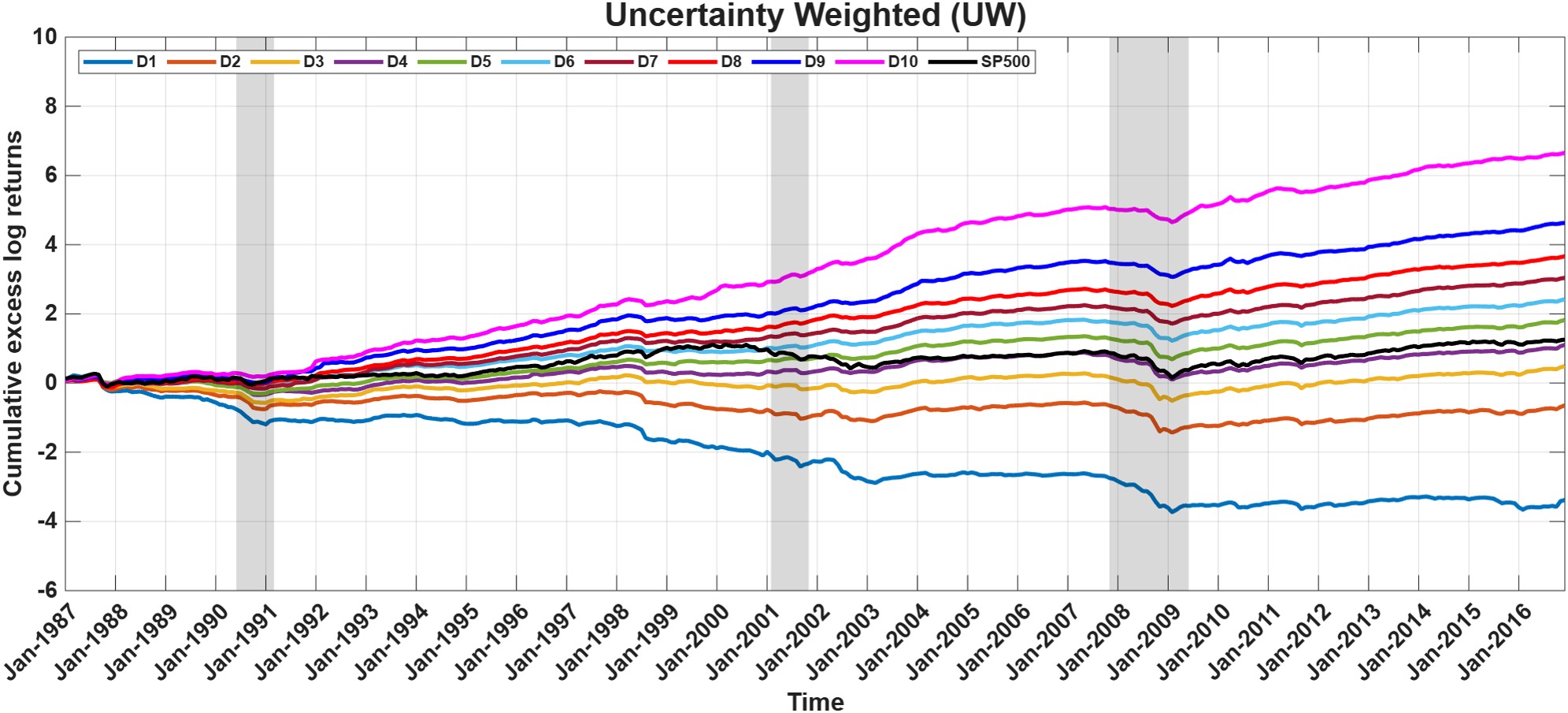}
\caption{This figure shows the cumulative excess log returns of uncertainty-weighted (UW) decile portfolios sorted based on our predicted returns. It also shows the S\&P 500. The shaded periods indicate NBER recessions.}
\label{uncertaintyweighted}
\end{figure}

\begin{figure}
\centering
\includegraphics[scale=0.23]{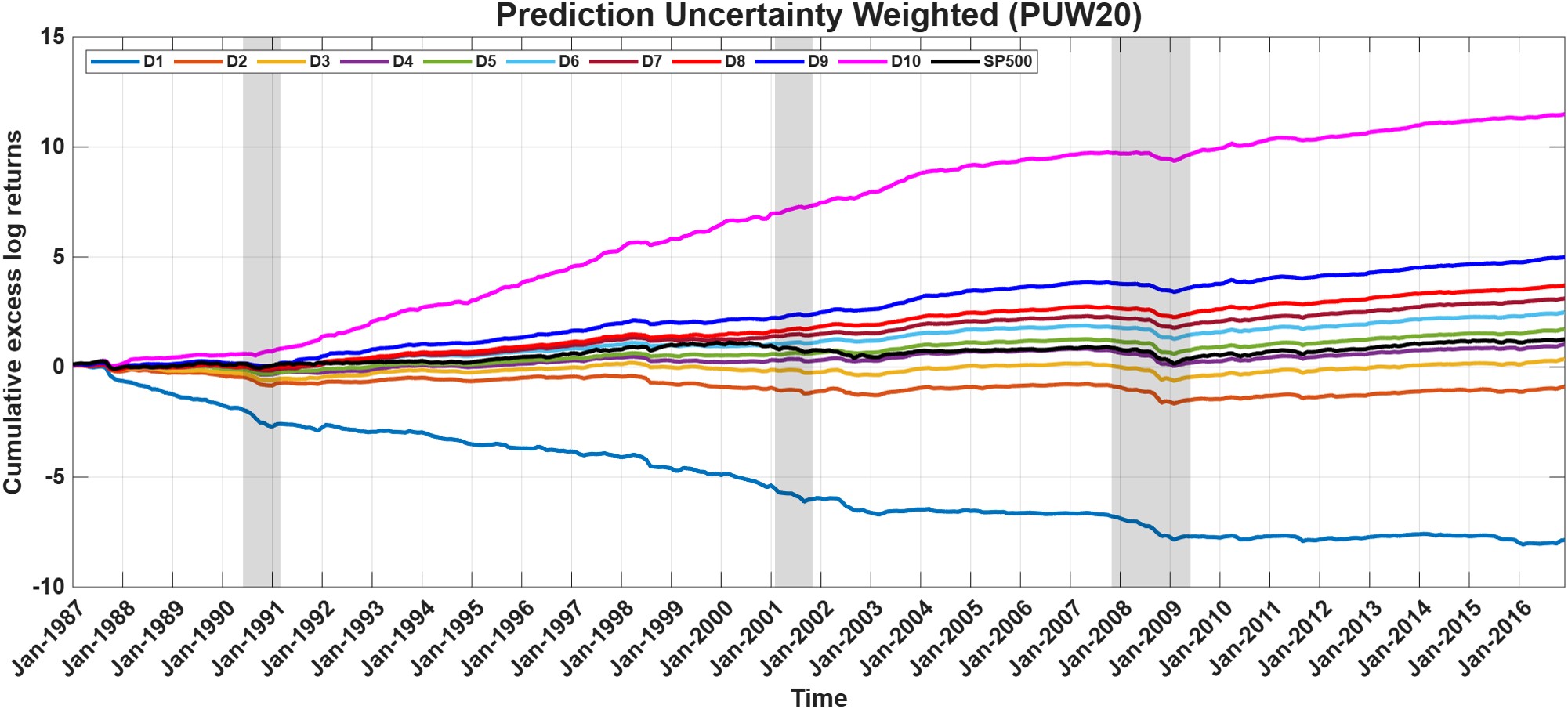}
\caption{This figure shows the cumulative excess log returns of prediction-uncertainty-weighted (PUW), with $\gamma=20$, decile portfolios sorted based on our predicted returns. It also shows the S\&P 500. The shaded periods indicate NBER recessions.}
\label{predictionuncertaintyweighted}
\end{figure}

\begin{figure}
\centering
\includegraphics[scale=0.40]{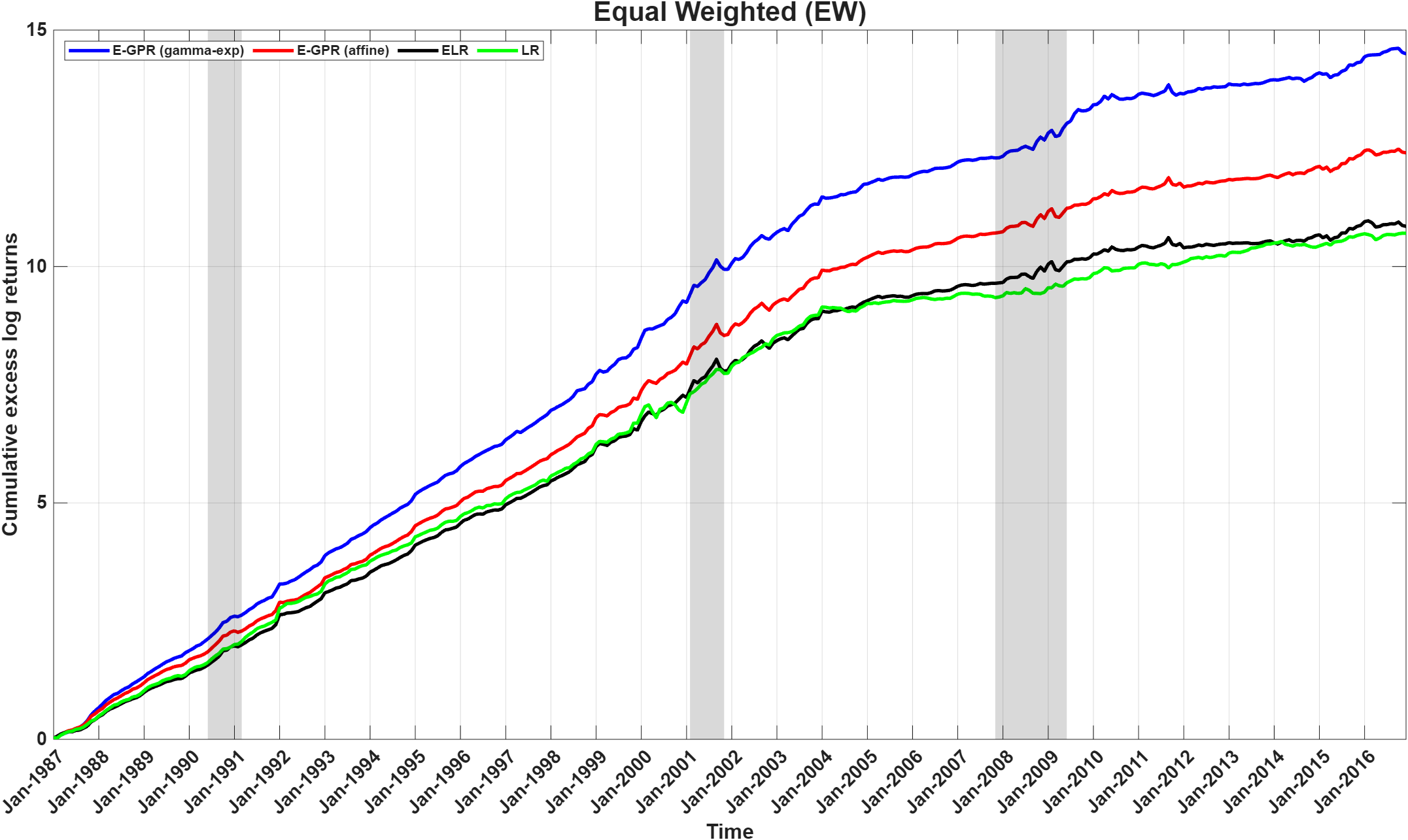}
\caption{This figure shows the cumulative excess log returns of equal-weighted (EW) long-short portfolio sorted based on the predicted returns from our GPR model (E-GPR ($\gamma$-exp)) and linear benchmark models (E-GPR (affine), E-LR and LR). The shaded periods indicate NBER recessions.}
\label{ew_comparison}
\end{figure}

\begin{figure}
\centering
\includegraphics[scale=0.40]{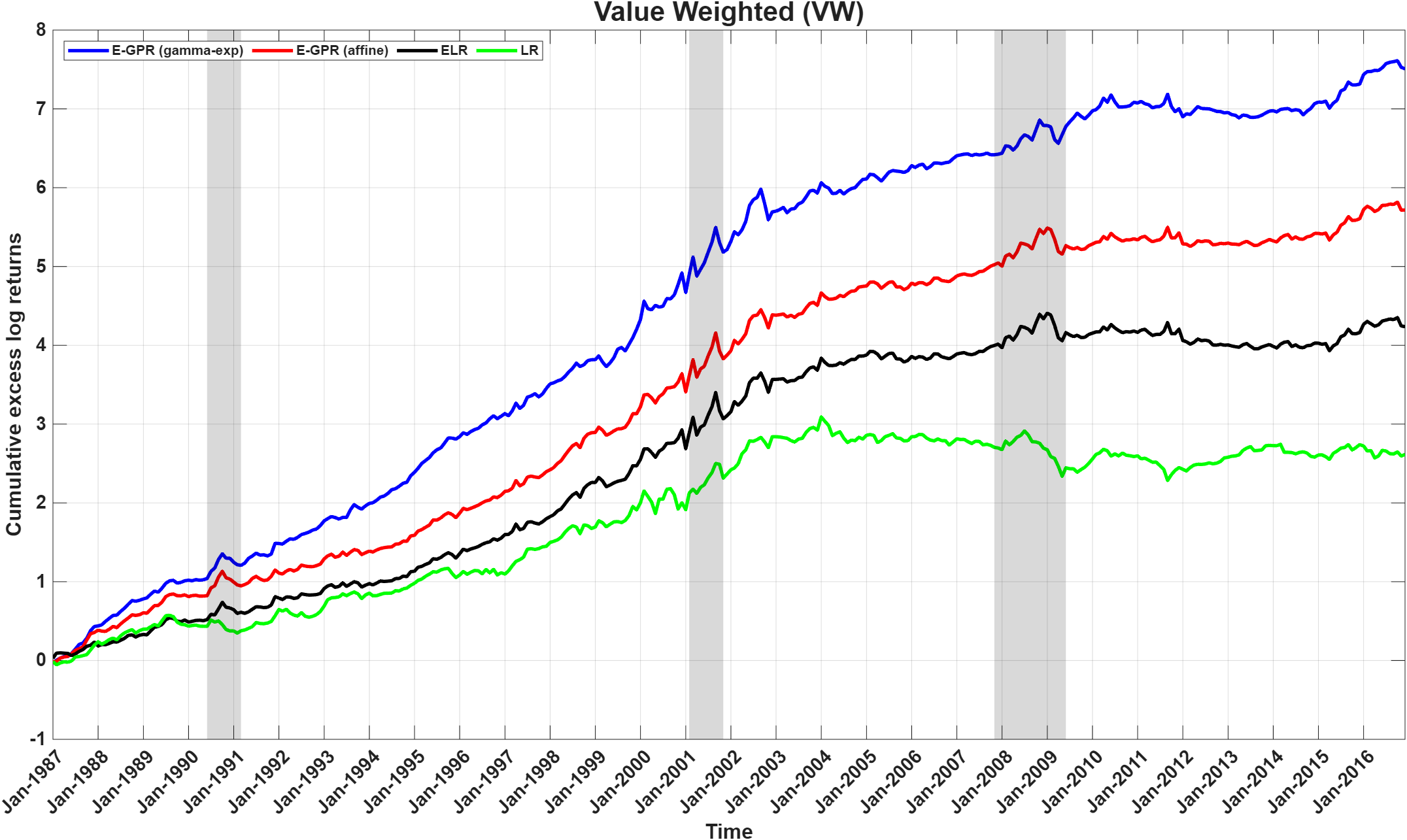}
\caption{This figure shows the cumulative excess log returns of value-weighted (VW) long-short portfolio sorted based on the predicted returns from our GPR model (E-GPR ($\gamma$-exp)) and linear benchmark models (E-GPR (affine), E-LR and LR). The shaded periods indicate NBER recessions.}
\label{vw_comparison}
\end{figure}

\begin{figure}
\centering
\includegraphics[scale=0.27]{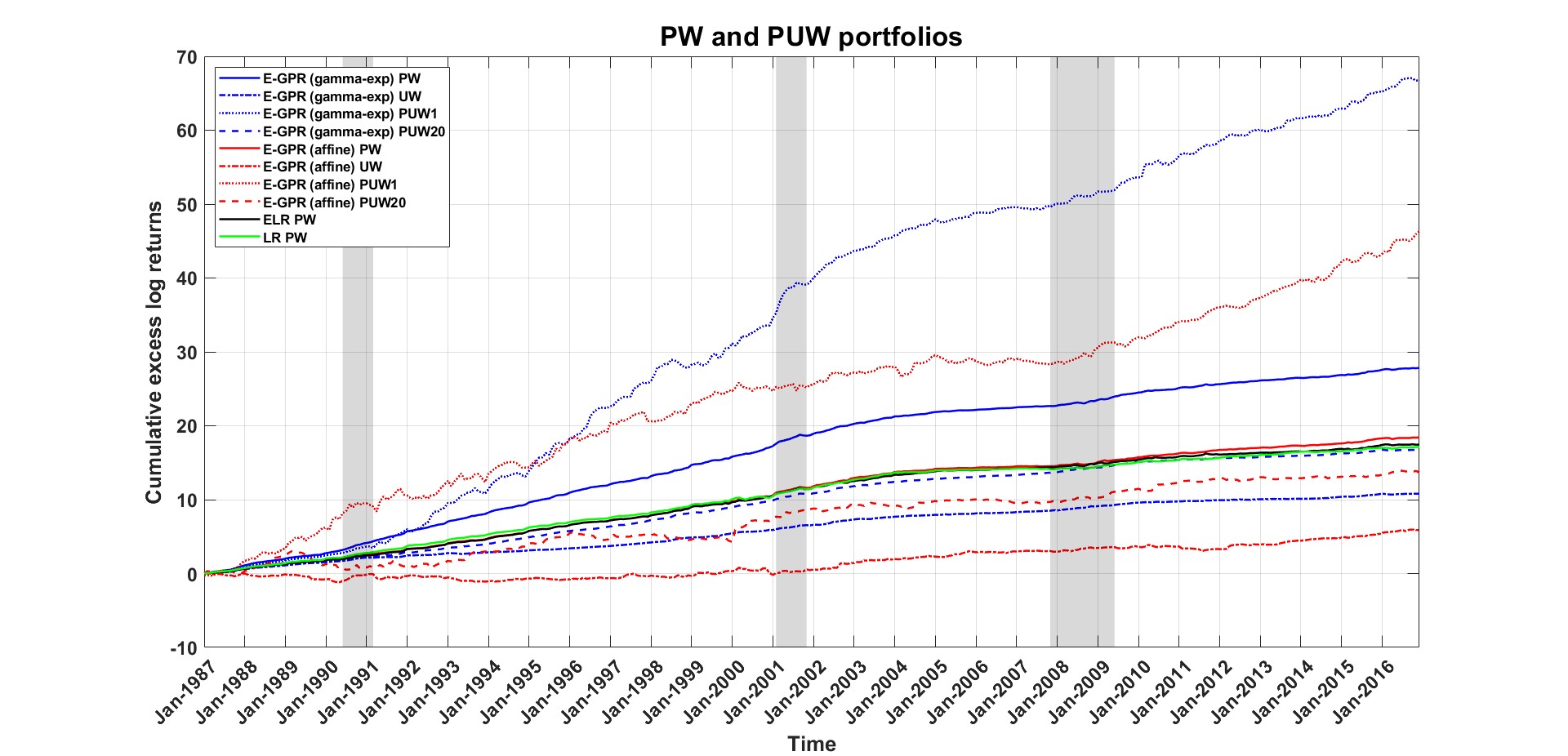}
\caption{This figure shows the cumulative excess log returns of prediction-weighted (PW) and prediction-uncertainty-weighted (PUW) long-short portfolio sorted based on the predicted returns from our GPR model (E-GPR ($\gamma$-exp)) and linear benchmark models (E-GPR (affine), E-LR and LR). The shaded periods indicate NBER recessions.}
\label{pw_puw_comparison}
\end{figure}

\begin{table}
    \centering
        \resizebox{.98\textwidth}{!}{\begin{tabular}{c|c|c|c|c|c|c|c|c|c|c}
&	$D1$ &	$D2$ &	$D3$ &	$D4$ &	$D5$ &	$D6$ &	$D7$ &	$D8$ &	$D9$ &	$D10$ \\
\hline
mom1m      &  0.929 &  0.288 &  0.136 &  0.053 & \textbf{-0.001} & -0.059 & -0.120 & -0.201 & -0.326 & -0.699 \\
baspread   &  0.495 &  0.064 & -0.102 & -0.188 & -0.229 & -0.232 & -0.209 & -0.140 &  \textbf{0.001} &  0.539 \\
ill        & -0.090 & -0.113 & -0.113 & -0.113 & -0.106 & -0.096 & -0.076 & -0.034 &  0.059 &  0.681 \\
retvol     &  0.771 &  0.133 & -0.079 & -0.189 & -0.245 & -0.264 & -0.252 & -0.196 & -0.076 &  0.395 \\
std\_turn  &  0.318 &  0.067 & -0.039 & -0.082 & -0.100 & -0.106 & -0.093 & -0.061 &  \textbf{0.003} &  0.088 \\
dolvol     &  \textbf{0.013} &  0.104 &  0.095 &  0.099 &  0.105 &  0.099 &  0.075 &  \textbf{0.011} & -0.105 & -0.497 \\
mom6m      & -0.489 & -0.188 & -0.084 & -0.016 &  0.035 &  0.078 &  0.120 &  0.172 &  0.226 &  0.146 \\
idiovol    &  0.372 &  0.109 & -0.064 & -0.164 & -0.212 & -0.224 & -0.197 & -0.120 &  0.037 &  0.461 \\
turn       &  0.433 &  0.193 &  0.040 & -0.027 & -0.067 & -0.095 & -0.105 & -0.108 & -0.101 & -0.168 \\
rd\_mve    & -0.075 & -0.069 & -0.064 & -0.059 & -0.055 & -0.045 & -0.027 &  0.004 &  0.074 &  0.316 \\
mom12m     & -0.513 & -0.261 & -0.147 & -0.064 &  \textbf{0.007} &  0.068 &  0.131 &  0.205 &  0.300 &  0.276 \\
chmom      &  0.191 &  0.108 &  0.075 &  0.049 &  0.024 &  \textbf{0.001} & -0.027 & -0.057 & -0.110 & -0.255 \\
mom36m     &  0.153 &  0.141 &  0.083 &  0.050 &  0.032 &  0.015 & -0.012 & -0.048 & -0.114 & -0.299 \\
indmom  & -0.458 & -0.301 & -0.212 & -0.131 & -0.043 &  0.042 &  0.138 &  0.239 &  0.337 &  0.391 \\
maxret     &  0.931 &  0.193 & -0.034 & -0.152 & -0.215 & -0.246 & -0.250 & -0.225 & -0.157 &  0.153 \\
agr        & -0.446 & -0.342 & -0.127 & -0.020 &  0.041 &  0.084 &  0.116 &  0.155 &  0.212 &  0.329 \\
zerotrade  & -0.097 & -0.040 & -0.021 & -0.013 & -0.017 & -0.015 & -0.014 & \textbf{-0.003} &  0.021 &  0.200 \\
rd         & -0.055 & -0.050 & -0.050 & -0.051 & -0.048 & -0.038 & -0.022 &  0.007 &  0.070 &  0.236 \\
chcsho     &  0.243 &  0.198 &  0.098 &  0.030 & -0.012 & -0.045 & -0.074 & -0.106 & -0.140 & -0.193 \\
ms         & -0.125 & -0.055 & -0.014 &  0.021 &  0.050 &  0.066 &  0.076 &  0.063 &  0.020 & -0.103 \\
    \end{tabular}}
    \caption{Source of heterogeneity (based on predicted returns). Bold means not rejected at 1\% otherwise rejected at 1\%.}
    \label{predbased_source}
\end{table}

\begin{table}
    \centering
    \resizebox{.98\textwidth}{!}{\begin{tabular}{c|c|c|c|c|c|c|c|c|c|c}
&	$D1$ &	$D2$ &	$D3$ &	$D4$ &	$D5$ &	$D6$ &	$D7$ &	$D8$ &	$D9$ &	$D10$ \\
\hline
mom1m      &  \textbf{0.014} &  0.066 &  0.071 &  0.067 &  0.068 &  0.064 &  0.052 &  0.020 & -0.051 & -0.371 \\
baspread   & -0.783 & -0.563 & -0.475 & -0.381 & -0.260 & -0.122 &  0.071 &  0.356 &  0.745 &  1.411 \\
ill        & -0.214 & -0.165 & -0.143 & -0.109 & -0.068 & -0.032 &  0.014 &  0.091 &  0.179 &  0.447 \\
retvol     & -0.803 & -0.570 & -0.482 & -0.386 & -0.269 & -0.125 &  0.065 &  0.343 &  0.737 &  1.488 \\
std\_turn  & -0.367 & -0.251 & -0.205 & -0.170 & -0.123 & -0.048 &  0.062 &  0.206 &  0.352 &  0.540 \\
dolvol     & -0.275 &  \textbf{0.010} &  0.156 &  0.183 &  0.149 &  0.099 &  0.036 & -0.023 & -0.074 & -0.262 \\
mom6m      &  \textbf{0.022} &  0.085 &  0.088 &  0.085 &  0.078 &  0.071 &  0.065 &  0.031 & -0.057 & -0.469 \\
idiovol    & -0.825 & -0.660 & -0.562 & -0.437 & -0.289 & -0.112 &  0.115 &  0.443 &  0.886 &  1.439 \\
turn       & -0.437 & -0.349 & -0.282 & -0.212 & -0.142 & -0.051 &  0.078 &  0.259 &  0.472 &  0.659 \\
rd\_mve    & -0.007 & -0.094 & -0.128 & -0.121 & -0.107 & -0.091 & -0.063 &  \textbf{0.002} &  0.135 &  0.475 \\
mom12m     &  \textbf{0.015} &  0.088 &  0.089 &  0.088 &  0.076 &  0.075 &  0.069 &  0.038 & -0.061 & -0.476 \\
chmom      &  \textbf{0.013} &  0.029 &  0.036 &  0.031 &  0.031 &  0.024 &  0.021 &  \textbf{0.007} & \textbf{-0.015} & -0.176 \\
mom36m     & -0.033 &  0.041 &  0.053 &  0.064 &  0.064 &  0.071 &  0.062 &  0.042 & \textbf{-0.019} & -0.345 \\
indmom     & -0.136 &  \textbf{0.030} &  0.038 &  0.040 &  0.050 &  0.046 &  0.045 &  \textbf{0.017} & \textbf{-0.024} & -0.105 \\
maxret     & -0.676 & -0.477 & -0.402 & -0.321 & -0.224 & -0.104 &  0.054 &  0.280 &  0.612 &  1.258 \\
agr        &  0.012 &  0.120 &  0.118 &  0.100 &  0.072 &  0.021 & -0.091 & -0.224 & -0.284 &  0.156 \\
zerotrade  & -0.033 &  0.190 &  0.115 &  0.071 &  0.040 & -0.007 & -0.035 & -0.072 & -0.130 & -0.140 \\
rd         & -0.022 & -0.239 & -0.173 & -0.124 & -0.099 & -0.070 & -0.038 &  0.038 &  0.207 &  0.519 \\
chcsho     & -0.016 & -0.151 & -0.123 & -0.087 & -0.056 & -0.013 &  0.048 &  0.122 &  0.181 &  0.096 \\
ms         &  0.010 &  0.202 &  0.258 &  0.207 &  0.118 &  0.020 & -0.073 & -0.159 & -0.244 & -0.343 \\
    \end{tabular}}
    \caption{Source of heterogeneity (based on prediction uncertainty). Bold means not rejected at 1\%  otherwise rejected at 1\%.}
    \label{uncbased_source}
\end{table}

\begin{figure}[!ht]
\begin{center}
\subfigure[Output Scale 
$\sigma^2$]{%
\includegraphics[height=4.5cm,width=12cm]{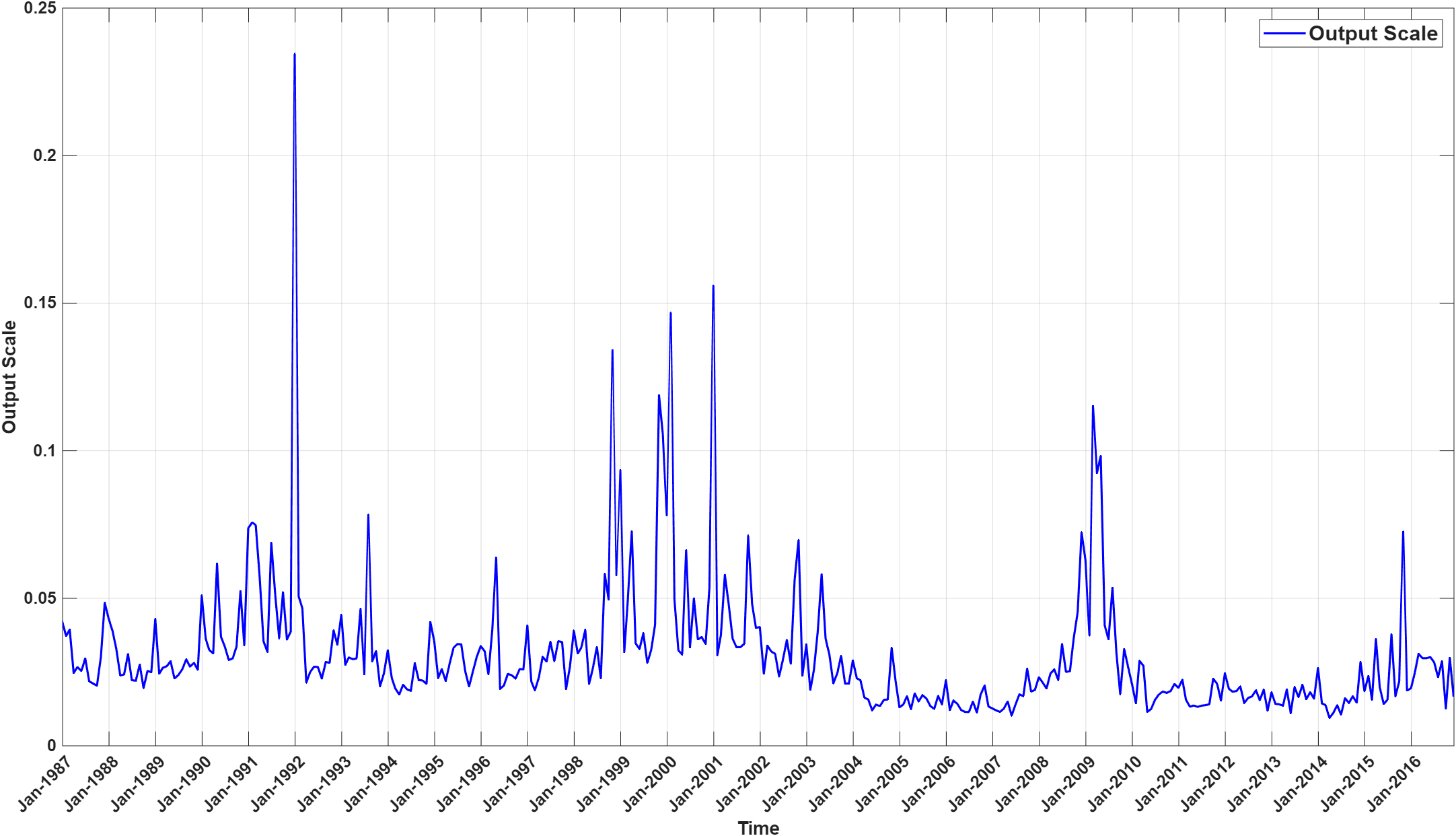}
\label{outputscale}}
\quad
\subfigure[Noise Variance $\sigma_{\epsilon}^2$]{%
\includegraphics[height=4.5cm,width=12cm]{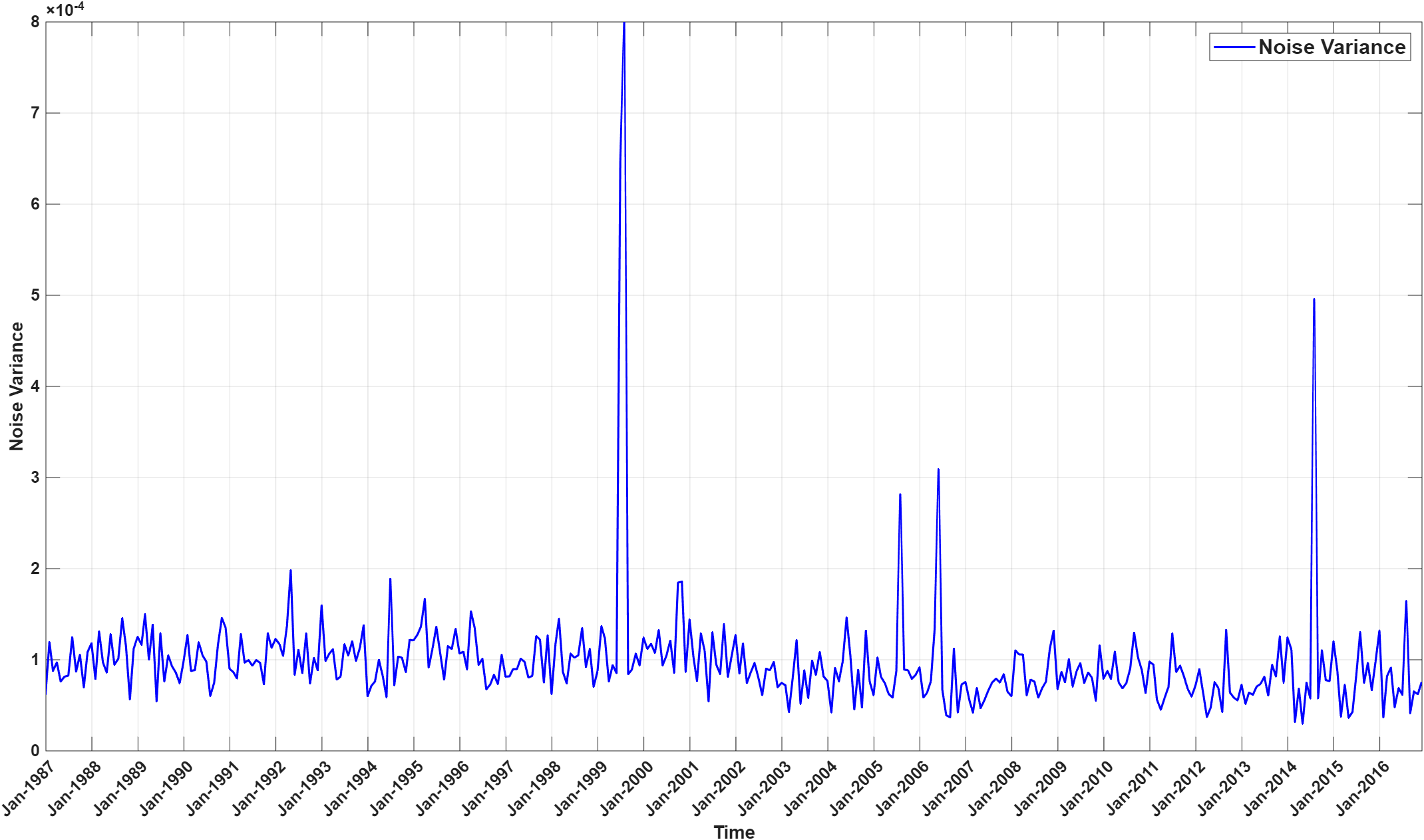}
\label{noisevariance}}
\quad
\subfigure[Length Scale $\ell$]{%
\includegraphics[height=4.5cm,width=12cm]{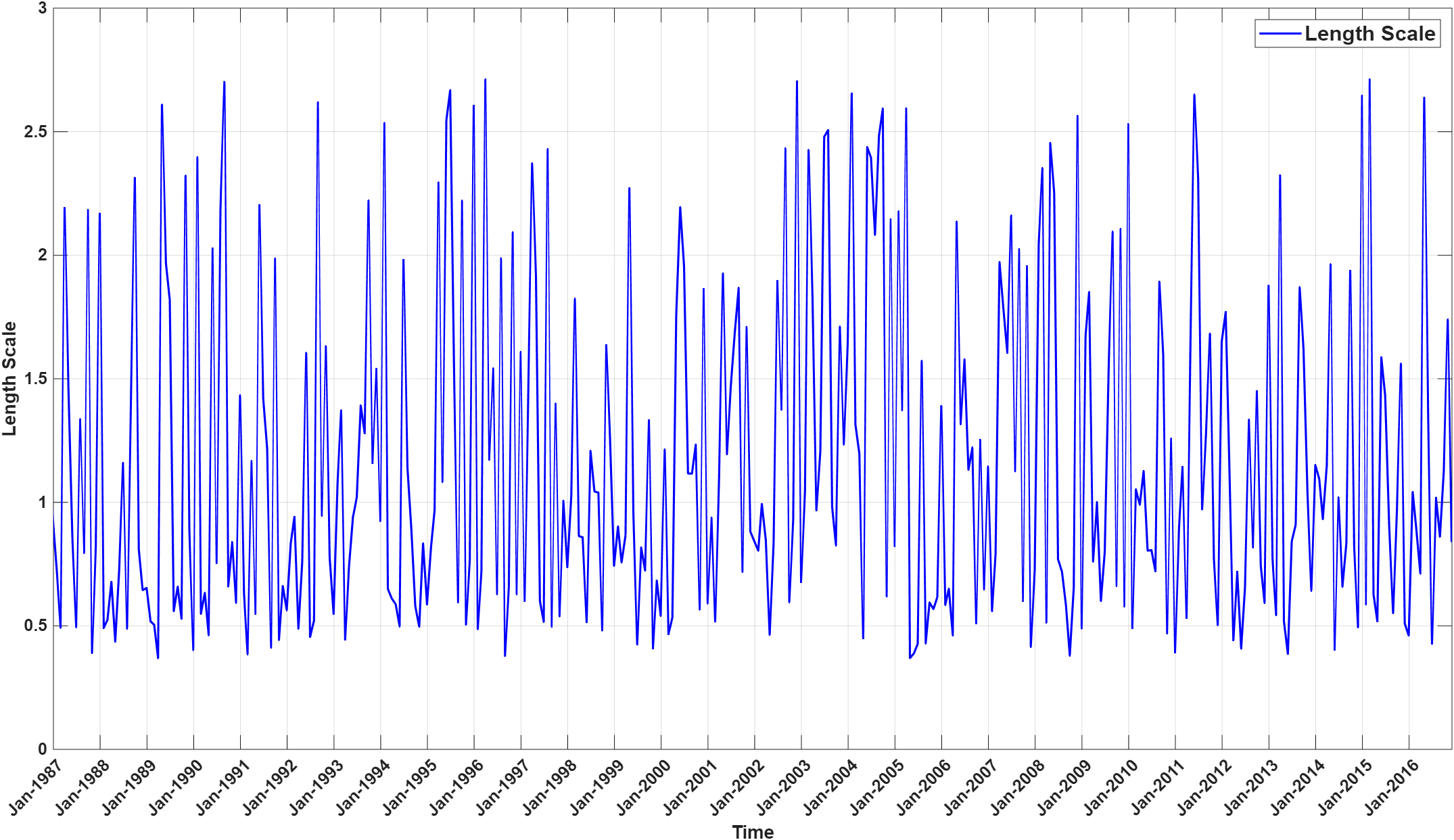}
\label{lengthscale}}
\quad
\subfigure[Power $\gamma$]{%
\includegraphics[height=4.5cm,width=12cm]{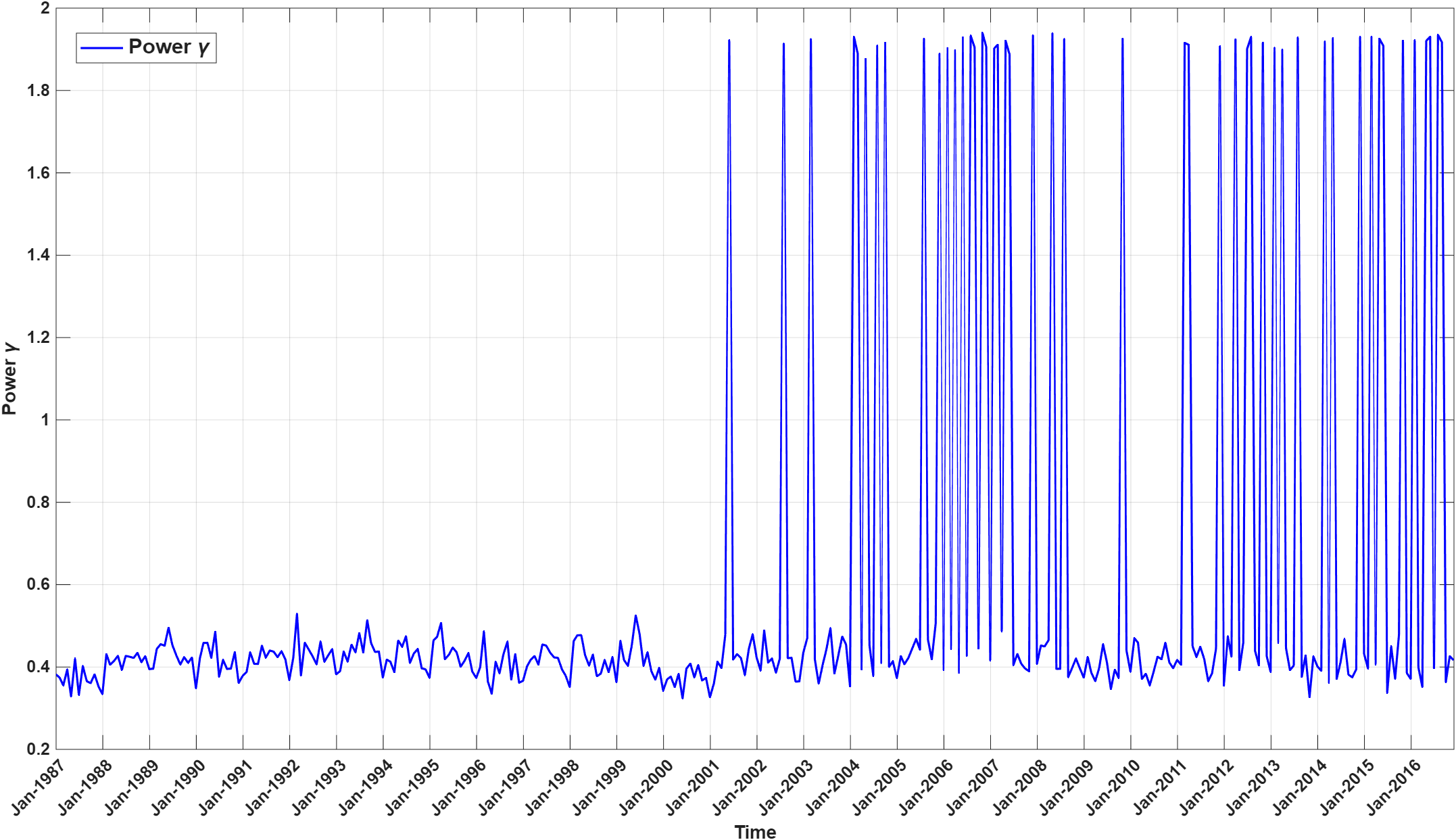}
\label{wb4}}
\caption{{This figure shows the estimated hyper-parameters of the monthly GPR models for the $\gamma$-exponential kernel.}}
\label{hyperparameters}
\end{center}
\end{figure}

\begin{figure}[ht!]
    \centering
    \includegraphics[scale=0.40]{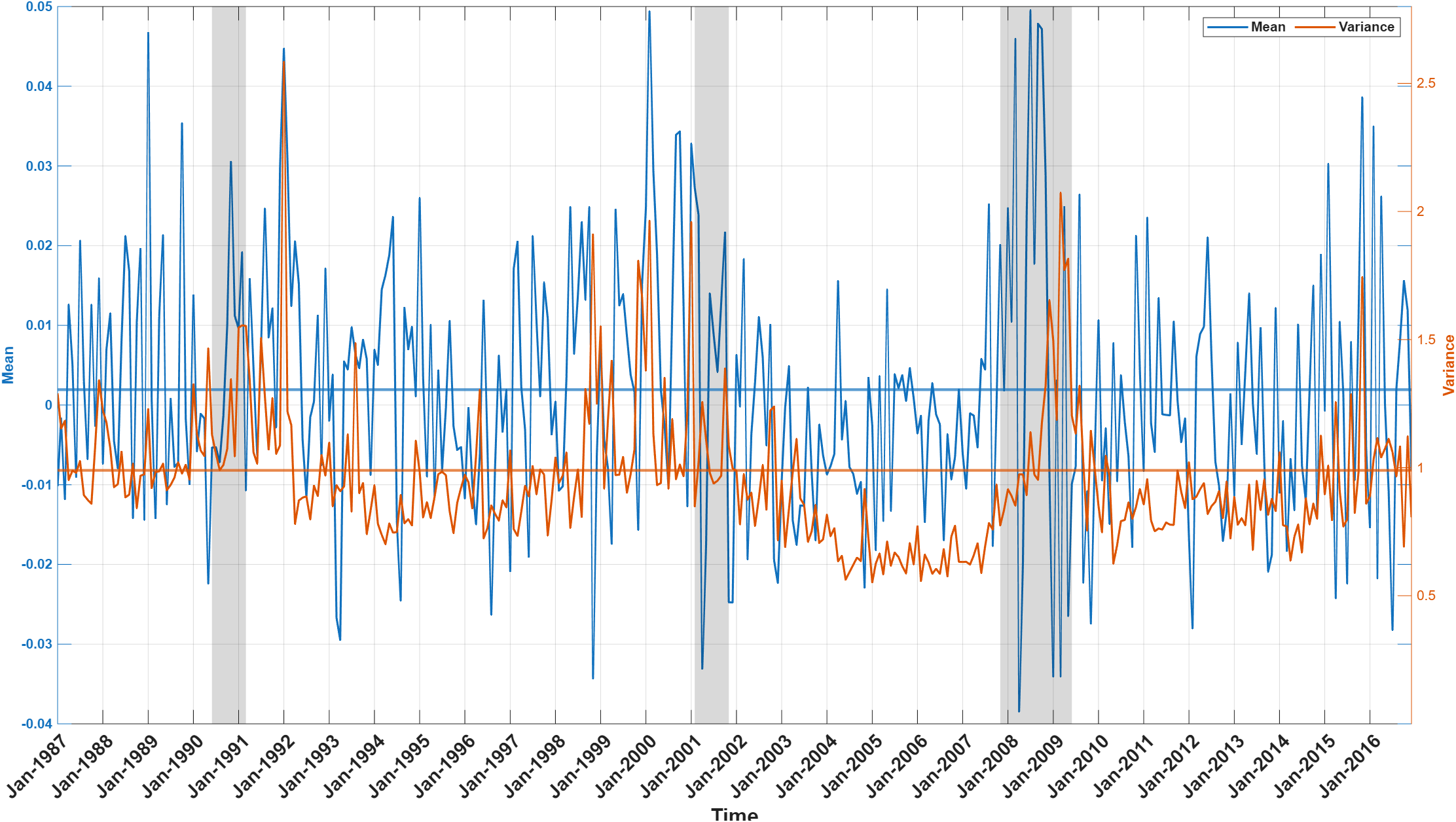}
    \caption{This figure shows the time-series of cross-sectional means and variances of realized standardized residuals in each test month. The solid horizontal lines represents the grand level mean and variance of the residuals.}
    \label{mean_var_residuals}
\end{figure}

\end{document}